\documentclass[preprint2]{aastex62}

\hypersetup{linkcolor=blue,citecolor=black,filecolor=cyan,urlcolor=blue}

\usepackage{natbib,layouts}
\usepackage[caption=false]{subfig}
\usepackage{multirow}
\usepackage{amsmath}
\usepackage{appendix}
\usepackage{courier}
\usepackage{ulem}
\newcommand{\1}[1]{\, \mathrm{#1}} % unit(y ;-)

\newcommand{\angstrom}{\mbox{\normalfont\AA}}
\newcommand{\co}[2]{\mathrm{CO}($#1 -- #2$)}

\def\kms    {\ifmmode{{\rm \ts km\ts s}^{-1}}\else{\ts km\ts s$^{-1}$}\fi}
\def\msol   {\ifmmode{{\rm M}_{\odot} }\else{M$_{\odot}$}\fi}
\def\lsol   {\ifmmode{L_{\odot}}\else{$L_{\odot}$}\fi}
\def\lfir   {\ifmmode{L_{\rm FIR}}\else{$L_{\rm FIR}$}\fi}
\def\lir   {\ifmmode{L_{\rm IR}}\else{$L_{\rm IR}$}\fi}
\def\zsol   {\ifmmode{{\rm Z}_{\odot}}\else{Z$_{\odot}$}\fi}
\def\etal   {{\rm et\ts al.}}
\def\tdust  {$\mathrm{T_{\rm dust}}$}
\def\lfir   {\ifmmode{L_{\rm FIR}}\else{$L_{\rm FIR}$}\fi}
\def\lir   {\ifmmode{L_{\rm IR}}\else{$L_{\rm IR}$}\fi}

\DeclareRobustCommand{\rchi}{{\mathpalette\irchi\relax}}
\newcommand{\irchi}[2]{\raisebox{\depth}{$#1\chi$}}

\def\ts     {\thinspace}
\def\ci     {\ifmmode{[{\rm C}{\rm \small I}]}\else{[C\ts {\scriptsize I}]}\fi}
\def\cii    {\ifmmode{[{\rm C}{\rm \small II}]}\else{[C\ts {\scriptsize II}]}\fi}
\def\nii    {\ifmmode{[{\rm N}{\rm \small II}]}\else{[N\ts {\scriptsize II}]}\fi}
\def\mue 	{\ifmmode{\mu{\rm m}}\else{$\mu$m}\fi}

\graphicspath{{./}{figures/}}

\submitjournal{ApJ}

\shorttitle{Complete Redshift Distribution of SPT galaxies}
\shortauthors{Reuter et al.}

\begin{document}

\title{The Complete Redshift Distribution of Dusty Star-forming Galaxies from the SPT-SZ Survey}

\correspondingauthor{Cassie Reuter}
\email{creuter@illinois.edu}

\author[0000-0001-7477-1586]{C. Reuter}
\affil{Department of Astronomy, University of Illinois, 1002 West Green St., Urbana, IL 61801, USA}

\author[0000-0001-7192-3871]{J.~D. Vieira}
\affil{Department of Astronomy, University of Illinois, 1002 West Green St., Urbana, IL 61801, USA}
\affil{Department of Physics, University of Illinois, 1110 West Green St., Urbana, IL 61801, USA}
\affil{National Center for Supercomputing Applications, University of Illinois, 1205 West Clark St., Urbana, IL 61801, USA}

\author[0000-0003-3256-5615]{J.~S. Spilker}
\affiliation{Department of Astronomy, University of Texas at Austin, 2515 Speedway, Stop C1400, Austin, TX 78712, USA}
\altaffiliation{NHFP Hubble Fellow}

\author[0000-0003-4678-3939]{A. Wei{\ss}}
\affil{Max-Planck-Institut f\"{u}r Radioastronomie, Auf dem H\"{u}gel 69, D-53121 Bonn, Germany
}

\author[0000-0002-6290-3198]{M. Aravena}
\affil{N\'{u}cleo de Astronom{\'i}a de la Facultad de Ingenier\'{i}a y Ciencias, Universidad Diego Portales, Av. Ej\'{e}rcito Libertador 441, Santiago, Chile}

\author[0000-0002-0517-9842]{M.~Archipley}
\affil{Department of Astronomy, University of Illinois, 1002 West Green St., Urbana, IL 61801, USA}

\author[0000-0002-3915-2015]{M. B{\'e}thermin}
\affil{Aix Marseille Univ., CNRS, CNES, LAM, Marseille, France}

\author{S.~C. Chapman}
\affil{Department of Physics and Astronomy, University of British Columbia, 6225 Agricultural Rd., Vancouver, V6T 1Z1, Canada}
\affil{National Research Council, Herzberg Astronomy and Astrophysics, 5071 West Saanich Rd., Victoria, V9E 2E7, Canada}
\affil{Department of Physics and Atmospheric Science, Dalhousie University, Halifax, Nova Scotia, Canada}

\author[0000-0002-6637-3315]{C. De~Breuck}
\affil{European Southern Observatory, Karl Schwarzschild Stra{\ss}e 2, 85748 Garching, Germany}

\author[0000-0002-5823-0349]{C. Dong}
\affil{Department of Astronomy, University of Florida, Bryant Space Sciences Center, Gainesville, FL 32611, USA
}

\author[0000-0002-5370-6651]{W.~B.~Everett}
\affil{Center for Astrophysics and Space Astronomy and Department of Astrophysical and Planetary Sciences, University of Colorado, Boulder, CO 80309, USA}

\author[0000-0002-3767-299X]{J.~Fu}
\affil{Department of Astronomy, University of Illinois, 1002 West Green St., Urbana, IL 61801, USA}

\author[0000-0002-2554-1837]{T.~R.~Greve}
\affil{Department of Physics and Astronomy, University College London, Gower Street, London WC1E 6BT, UK
}
\affil{Cosmic Dawn Center (DAWN), DTU-Space, Technical University of Denmark, Elektrovej 327, DK-2800 Kgs. Lyngby, Denmark}

\author[0000-0003-4073-3236]{C.~C. Hayward}
\affil{Center for Computational Astrophysics, Flatiron Institute, 162 Fifth Avenue, New York, NY, 10010, USA
}

\author{R. Hill}
\affil{Department of Physics and Astronomy, University of British Columbia, 6225 Agricultural Rd., Vancouver, V6T 1Z1, Canada}

\author[0000-0002-8669-5733]{Y.~Hezaveh}
\affil{D\'{e}partement de Physique, Universit\'{e} de Montr\'{e}al, Montreal, Quebec, H3T 1J4, Canada
}
\affil{Center for Computational Astrophysics, Flatiron Institute, 162 Fifth Avenue, New York, NY, 10010, USA
}

\author[0000-0002-5386-7076]{S. Jarugula}
\affil{Department of Astronomy, University of Illinois, 1002 West Green St., Urbana, IL 61801, USA}

\author[0000-0002-4208-3532]{K.~Litke}
\affil{Steward Observatory, University of Arizona, 933 North Cherry Avenue, Tucson, AZ 85721, USA
}

\author[0000-0001-6919-1237]{M. Malkan}
\affil{Department of Physics and Astronomy, University of California, Los Angeles, CA 90095-1547, USA
}

\author[0000-0002-2367-1080]{D.~P. Marrone}
\affil{Steward Observatory, University of Arizona, 933 North Cherry Avenue, Tucson, AZ 85721, USA
}

\author[0000-0002-7064-4309]{D. Narayanan}
\affil{Department of Astronomy, University of Florida, 211 Bryant Space Sciences Center, Gainesville, FL 32611, USA}
\affil{University of Florida Informatics Institute, 432 Newell Drive, CISE Bldg E251, Gainesville, FL 32611, USA}
\affil{Cosmic Dawn Center (DAWN), DTU-Space, Technical University of Denmark, Elektrovej 327, DK-2800 Kgs. Lyngby, Denmark}

\author[0000-0001-7946-557X]{K.~A. Phadke}
\affil{Department of Astronomy, University of Illinois, 1002 West Green St., Urbana, IL 61801, USA}

\author[0000-0002-2718-9996]{A.~A. Stark}
\affil{Harvard-Smithsonian Center for Astrophysics, 60 Garden Street, Cambridge, MA 02138, USA}

\author{M.~L.~Strandet}
\affil{Max-Planck-Institut f\"{u}r Radioastronomie, Auf dem H\"{u}gel 69, D-53121 Bonn, Germany
}

%%%%%%%%%%%%%%%%%%%%%%%%%%%%%%%%%%%%%
% ABSTRACT							%
\begin{abstract}					%
%%%%%%%%%%%%%%%%%%%%%%%%%%%%%%%%%%%%%
The South Pole Telescope (SPT) has systematically identified 81 high-redshift, strongly gravitationally lensed, dusty star-forming galaxies (DSFGs) in a 2500 square degree cosmological mm-wave survey.  We present the final spectroscopic redshift survey of this flux-limited ($S_{870\1{\mu m}} > 25\1{mJy}$) sample, initially selected at $1.4\1{mm}$.  The redshift survey was conducted with the Atacama Large Millimeter/submillimeter Array across the $3\1{mm}$ spectral window, targeting carbon monoxide line emission.  By combining these measurements with ancillary data, the SPT sample is now spectroscopically complete, with redshifts spanning $1.9$$<$$z$$<$$6.9$ and a median of $z=3.9 \pm 0.2$.  We present the mm through far-infrared photometry and spectral energy density fits for all sources, along with their inferred intrinsic properties.  

Comparing the properties of the SPT sources to the unlensed DSFG population, we demonstrate that the SPT-selected DSFGs represent the most extreme infrared-luminous galaxies, even after accounting for strong gravitational lensing.  The SPT sources have a median star formation rate of $2.3(2)\times 10^3\1{M_\odot yr^{-1}}$ and a median dust mass of $1.4(1)\times10^9\1{M_\odot}$.  However, the inferred gas depletion timescales of the SPT sources are comparable to those of unlensed DSFGs, once redshift is taken into account.  This SPT sample contains roughly half of the known spectroscopically confirmed DSFGs at $z$$>$$5$, making this the largest sample of high-redshift DSFGs to-date, and enabling the ``high-redshift tail" of extremely luminous DSFGs to be measured.  Though galaxy formation models struggle to account for the SPT redshift distribution, the larger sample statistics from this complete and well-defined survey will help inform future theoretical efforts.

\end{abstract}


%%%%%%%%%%%%%%%%%%%%%%%%%%%%%%%%%%%%%%%%%%%%%%%%%%%%%%%%%%
% SECTION 1												 %
% Introduction											 %
\section{Introduction} \label{sec:intro}		         %
%%%%%%%%%%%%%%%%%%%%%%%%%%%%%%%%%%%%%%%%%%%%%%%%%%%%%%%%%%
Millimeter (mm) and sub-millimeter (sub-mm) continuum observations have transformed our understanding of galaxy formation and evolution by demonstrating that luminous, dusty galaxies were a thousand times more abundant in the early Universe than they are today (see reviews by \citealt{blain02} and \citealt{casey14}).    
The most intense star formation in the universe takes place in these high-redshift ($z$$>$$1$) dusty star-forming galaxies (DSFGs), which form new stars at rates of $>$$100$$-$$1000$$\1{\msol{} yr^{-1}}$ behind dense shrouds of dust. DSFGs are thought to be the progenitors of the massive elliptical galaxies seen in the present-day universe~\citep{blain04}.  The exact details of how these galaxies form stars at such prodigious rates is still an open question~\citep{narayanan15}, though galaxy mergers likely play a role (e.g.~\citealp{tacconi08, engel10, narayanan10, hayward11, bothwell13b, hayward13, ma16}).  
While the first surveys of the redshift distributions of these DSFGs suggested that the population peaks at $z$$\sim$$2$~\citep{chapman03, chapman05}, modern redshift surveys suggest that the DSFG population peaks at higher redshifts (2.5-2.9; \citealp{simpson14, dacunha15, danielson17, dudzeviciute20}).  These updated surveys are also finding objects at increasingly higher redshift, extending past $z$$>$$6$ (e.g.~\citealp{riechers13, fudamoto17, strandet17, zavala18, marrone18}).  The presence of these extremely high-redshift DSFGs challenges our understanding of the underlying distribution of these objects and their role in the cosmic star formation history.

Dust emission at high redshift ($z$$>$$1$) exhibits dimming from increased cosmological distance, which is counteracted by a steep rise on the Rayleigh-Jeans side of the spectral energy distribution (SED) at a fixed observing wavelength, which leads to the so-called ``negative K-correction" in the sub-mm~\citep{blain93}.  Because of this effect, fluxes at mm and sub-mm wavelengths are roughly constant with redshift, facilitating the discovery of high-redshift DSFGs.  However, the dust-obscured nature of DSFGs suppresses emission at optical/UV wavelengths, making robust redshifts difficult to obtain, especially at high redshift.  Technological advances in correlator bandwidth have enabled  ``blind" spectroscopic surveys to be conducted at facilities such as the Atacama Large Millimeter/submillimeter Array (ALMA) and the Plateau de Bure Interferometer (PdBI/NOEMA).  These spectroscopic surveys search for molecular emission at millimeter wavelengths and can be conducted without prior optical/near-IR spectroscopy.  Because the molecular emission can be unambiguously related to the continuum emission, these surveys provide a direct and unbiased way to derive the redshifts of DSFGs (e.g.~\citealt{scott11,weiss09}).  Carbon monoxide (CO) is the second most abundant molecule in the Universe after H$_2$ and its rotational transitions are frequently targeted in blind searches.  These rotational transitions are spaced evenly every $115\1{GHz}$ and are among the brightest lines in the millimeter spectrum.  The CO line brightness is due to its abundance.  The low energy required to excite its rotational states, and the fact that the resulting emission lines are mostly accessible at frequencies of high atmospheric transmission make it ideal for line surveys with ALMA.  

One of the first millimeter wave redshift searches with ALMA identified the redshifts of 26 gravitationally lensed DSFGs selected with the South Pole Telescope (SPT; \citealp{vieira13,weiss13}).  Because gravitationally lensed sources (with magnifications $\mu$$\sim$$10$) are apparently brighter than unlensed sources, they require significantly less on-source time to survey ($t_{\rm obs}\propto\mu^{-2}$).  Blind CO surveys can be conducted with comparative ease and enable larger spectroscopic surveys \citep{weiss13, strandet16, neri20}.  While some unlensed blind CO surveys have been conducted~\citep{chapman15}, the majority of unlensed redshifts were obtained through optical spectroscopic searches~\citep{chapman03, chapman05, casey12, koprowski14, danielson17,brisbin17} or photometry~\citep{simpson14, simpson17,michalowski17}.  The discrepancies between the peaks of unlensed and lensed distributions can be explained by a combination of the selection wavelength, survey depth, and the redshift-dependent probability of strong lensing (e.g.~\citealt{bethermin15b, strandet16}).

Wide-area surveys conducted with the Atacama Cosmology Telescope (ATCA; \citealp{marsden14}), \textit{Herschel}~\citep{eales10, oliver12},  \textit{Planck}~\citep{planck15}, and the SPT~\citep{carlstrom11,vieira10, vieira13} span hundreds to thousands of square degrees and have enabled the discovery of hundreds of gravitationally lensed DSFGs at millimeter and sub-millimeter wavelengths.  For DSFGs at high-redshift, the rest-wavelength peak in the spectral energy distribution (SED) at $\sim$$100\1{\mu m}$ is shifted into the observing bands of these mm and sub-mm instruments.  Most of the brighter sources are gravitationally lensed as well (e.g. \citealt{blain96,negrello10, wardlow13, spilker16}).  The few intrinsically bright \textit{unlensed} sources in these samples are typically major mergers of DSFGs (e.g. \citealt{fu13,ivison13}) or protoclusters (e.g.~\citealt{overzier16, casey16, miller18, oteo18, lewis18, hill20}).  The magnification due to gravitational lensing enables us to obtain spectroscopic redshifts for a complete sample of DSFGs, measure the redshift distribution of DSFGs, and ascertain the prevalence of the highest-redshift DSFGs.  Samples of lensed DSFGs also afford the opportunity to study fainter observational diagnostics, in greater detail, than would otherwise be possible~(e.g. \citealt{bothwell17,bethermin18,spilker18,zhang18, litke19,dong19,jarugula19,cunningham20}).  

In this paper, we finalize the SPT-selected ALMA redshift survey, which started in~\citet{weiss13} and continued in~\citet{strandet16}.  The final catalog contains spectroscopic redshifts for all sources, making it the largest and most complete catalog of its kind to date.  In Sec.~\ref{sec:observations}, we present the $3\1{mm}$ line scans obtained from ALMA (Sec.~\ref{sec:specobs}), including 40 new sources.  We also present the photometry from SPT, ALMA, APEX, and \textit{Herschel} (Sec.~\ref{sec:photobs}).  In Sec.~\ref{sec:sed}, we present a methodology for fitting spectral energy distributions (SEDs) and deriving intrinsic source properties. Section~\ref{sec:results} is divided into two parts: spectroscopic results (Sec.~\ref{sec:specresults}) and photometric results (Sec.~\ref{sec:photoresults}).  In Sec.~\ref{sec:specresults}, we present the $3\1{mm}$ spectra and the resulting spectroscopic redshifts.  In Sec.~\ref{sec:photoresults}, we describe the fitted SEDs and the resulting intrinsic properties.  In Sec.~\ref{sec:discussion}, we discuss the redshift distribution of the complete SPT sample (Sec.~\ref{dis:zdist}), the possibility of temperature evolution (Sec.~\ref{sec:tdevolution}), the extreme nature of the SPT sources (Sec.~\ref{sec:ulirgs}) and the resulting high redshift tail of this distribution (Sec.~\ref{sec:highztail}).

For this paper, we adopt a flat $\Lambda$CDM cosmology, with $\Omega_{\Lambda}$$=$$0.696$ and $H_0=68.1\1{km s^{-1} Mpc^{-1}}$ \citep{planck16cosmo}.

%%%%%%%%%%%%%%%%%%%%%%%%%%%%%%%%%%%%%%%%%%%%%%%%%%%%%%%%%
% SECTION 2												%
% Observations											%
\section{Observations and Methods} \label{sec:observations}%
%%%%%%%%%%%%%%%%%%%%%%%%%%%%%%%%%%%%%%%%%%%%%%%%%%%%%%%%%

%========================================================
% Section 2.1 Sample selection
\subsection{Sample selection} \label{sec:sampleselect}
%========================================================
% Sample selection
The SPT-selected DSFG catalog is a flux-limited sample comprised of 81 bright sources, selected at $1.4\1{mm}$ from the $2500\1{deg^2}$ of the SPT-SZ survey~\citep{vieira10,mocanu13,everett20}.  The sources were selected with $S_{1.4\1{mm}}$$>$$20\1{mJy}$, corresponding to a signal to noise ratio of $>$$4.5$.  The relatively coarse SPT positions (beam size of $1'.05$ at $1.4\1{mm}$) were refined with observations with the Large Apex BOlometer CAmera (LABOCA) at $870\1{\mu m}$~\citep{siringo09}.  Given the smaller beam size ($20\arcsec$) and higher signal-to-noise ratio (typically $\sim$$2.5$ higher) with the LABOCA observations, a final flux density cut was performed to select sources with S$_{870\1{\mu m}}$$>$$25\1{mJy}$.  The complete source catalog and their positions are detailed in Ap.~\ref{ap:srcpos}.  

\begin{figure*}[htb]
	\centering 
	\includegraphics[width=\textwidth]{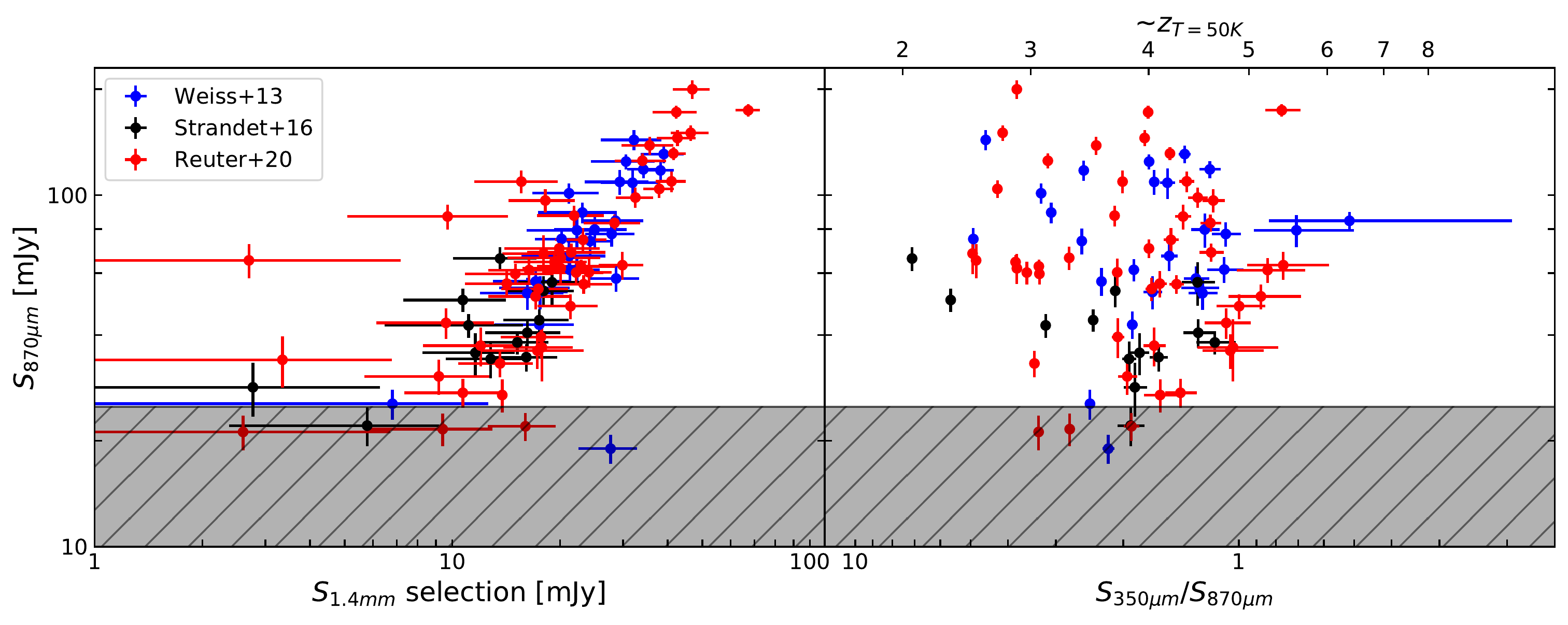}
	\caption{Flux density and color plots for all sources in the SPT-selected DSFG catalog, with 26 sources from \citet{weiss13} (\textit{blue}), 15 sources from \citet{strandet16} (\textit{black}) and the remaining 40 sources from this work (\textit{red}).  All of the source names and positions are detailed in Ap.~\ref{ap:srcpos}.  \textbf{Left:} APEX/LABOCA $870\1{\mu m}$ flux density versus the SPT~$1.4\1{mm}$ flux selection.  The initial selection included sources with a signal-to-noise ratio of 4.5, and were followed up with LABOCA.  The sample is defined such that sources have $S_{870\1{\mu m}}$$>$$25\1{mJy}$, excluding the grey hatched region.  Sources below the $S_{870\1{\mu m}}$ flux selection were not retained in the final sample.  \textbf{Right:} The APEX/LABOCA $870\1{\mu m}$ flux density versus the ratio of the \textit{Herschel}/SPIRE $350\1{\mu m}$ flux density to APEX/LABOCA $870\1{\mu m}$ flux density. The $S_{350\1{\mu m}}/S_{870\1{\mu m}}$ color corresponds to $\mathrm{T_{dust}}/(1+z)$.  At a fixed $\mathrm{T_{dust}}$, this color can be used as a crude proxy for redshift.}
	\label{fig:color_v_flux}
\end{figure*}

The spectroscopic survey of the SPT sample presented here is complete for $S_{870\1{\mu m}}$$>$$25\1{mJy}$.  The $1.4\1{mm}$ SPT and $S_{870\1{\mu m}}$ LABOCA fluxes of the final sample are shown in Fig.~\ref{fig:color_v_flux}.  The $\mathrm{S_{350\1{\mu m}}/S_{870\1{\mu m}}}$ color can be used as a rough indicator of redshift, shown in the right panel of Fig.~\ref{fig:color_v_flux}, assuming a constant dust temperature of $50\1{K}$ (see Sec.~\ref{sec:sedresults} for details).  Due to their extreme brightness, most of the sources were suspected to be gravitationally lensed by foreground galaxies, groups, or clusters~\citep{negrello07}.  High resolution $870\1{\mu m}$ observations~\citep{hezaveh13, spilker16} demonstrated that at least $70\%$ of the sample was strongly lensed.

%---------------------------------------------
% FIGURE 1 Sample Selection 
%---------------------------------------------

%========================================================
% Section 2.2 Spectroscopic Observations
\subsection{Spectroscopic Observations} \label{sec:specobs}
%========================================================

%========================================================
% Section 2.2.1 ALMA 3mm scans
\subsubsection{ALMA $3\1{mm}$ blind scans} \label{sec:alma3mm}
%========================================================
In order to obtain redshifts for the SPT sample, a blind spectroscopic redshift search was started in ALMA Cycle 0 (project ID: 2011.0.00957.S).  This program resulted in a $90\%$ line detection rate for the 26 sources surveyed~\citep{weiss13}.  An updated distribution was presented in \citet{strandet16} with an additional 15 sources observed in ALMA Cycle 1 (project ID: 2012.1.00844.S).  This work represents the conclusion of the SPT blind redshift survey, and presents spectroscopic scans for the remaining 40 sources from ALMA Cycles 3, 4 and 7 and 41 new spectroscopic redshifts.  The individual $3\1{mm}$ scans for all sources can be found in Ap.~\ref{ap:3mmspec}.  

\begin{figure}[!ht]
\begin{center}
\includegraphics[width=\columnwidth]{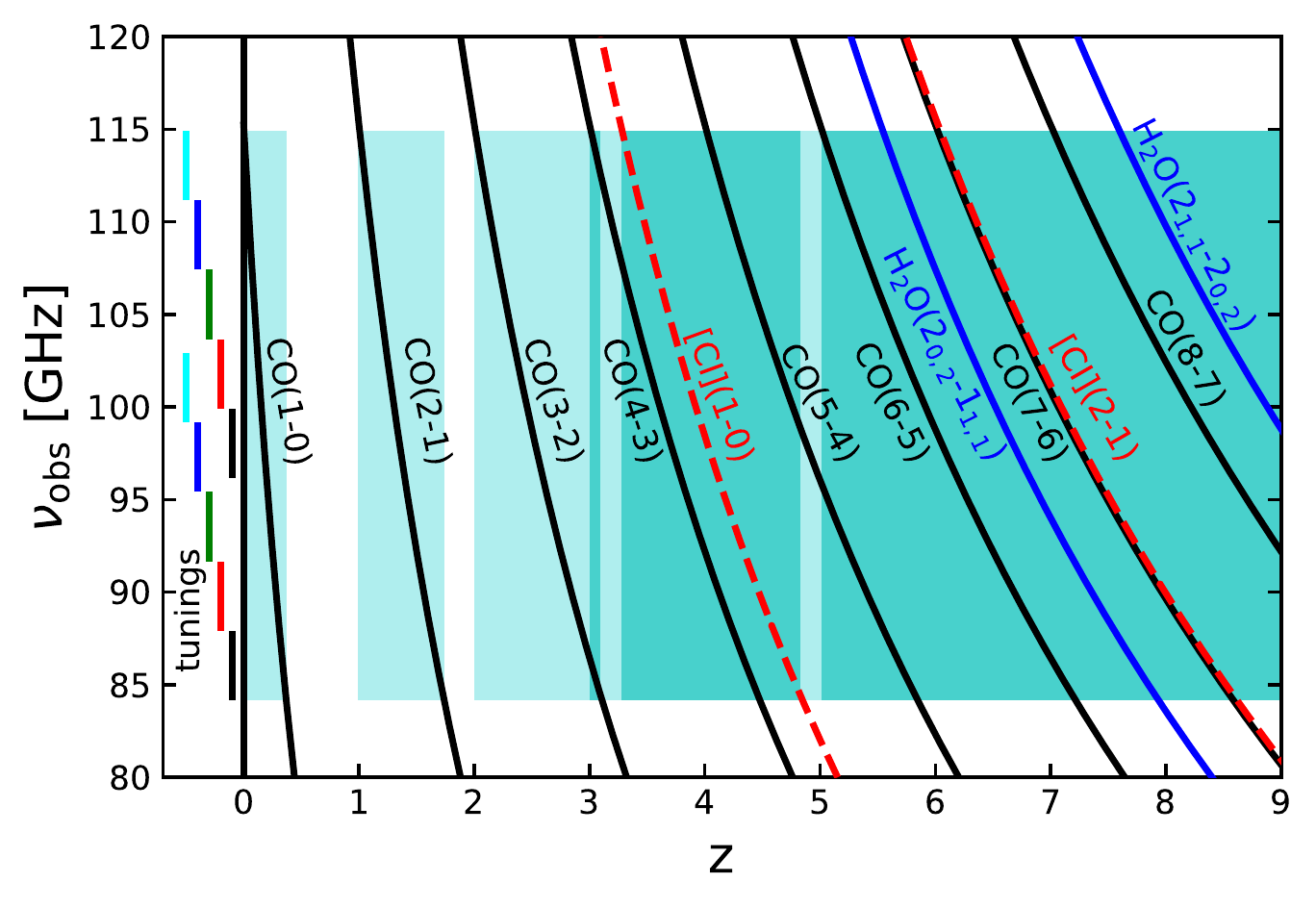}
\caption{Spectral coverage of the important CO, [CI], and $\mathrm{H_{2}O}$ emission lines as a function of redshift. The darker teal shaded region designates the redshift range in which two or more strong lines are expected to be detected, which would provide an unambiguous redshift for a given observation.  The lighter teal region marks redshift range where only a single line is detectable, and an ancillary spectroscopic observation or photometric redshift would be required to identify the correct redshift. The five frequency tunings used in the $3\1{mm}$ line scans are shown in the left panel\label{fig:almab3}}
\end{center}
\end{figure}

The blind spectroscopic search was conducted using ALMA's Band 3 receiver, which operates between $84 - 116\1{GHz}$.  The correlator has a total bandwidth of $7.5\1{GHz}$, which is split across two side bands.  The ALMA Band 3 $3\1{mm}$ atmospheric transmission window can be covered in five tunings, as shown in the left panel of Fig.~\ref{fig:almab3}.  This configuration results in overlapping coverage in the $96.2$$-$$102.8\1{GHz}$ region.  ALMA's primary beam ranges from $45\arcsec - 61\arcsec$ over the entire scanned frequency range.

Fig.~\ref{fig:almab3} also demonstrates this search's sensitivity to CO lines between the $\co{1}{0}$ and $\co{8}{7}$ transitions.  Scanning this region results in redshift coverage of $0.0$$<$$z$$<$$0.4$ and $1.0$$<$$z$$<8.6$ with a narrow redshift desert at $1.74$$<$$z$$<$$2.00$.  As previously shown in \citet{weiss13} and \citet{strandet16}, the SPT sources have a median redshift of $z$$\sim$$4$, enabling more than one emission line to be observed in each spectrum.  Notably, the [CI](1-0) line is predicted in the range $3.3$$<$$z$$<$$4.8$ and has been shown in \citet{weiss13} and \citet{strandet16} to be bright enough to provide secure spectroscopic redshifts at the observed sensitivities.
The H$_2$O(2$_{0,2}$-1$_{1,1}$) and [CI](2-1) lines are accessible at the highest redshifts (Figs.~\ref{fig:almab3} and~\ref{fig:zshifted_spec}) and are not typically observed.

 The Cycle 3 ALMA observations were conducted from December 2015 - August 2016 (project ID: 2015.1.00504.S).  During Cycle 3, between 34 and 41 antennas were employed, and resulted in typical synthesized beams of $3.9\arcsec$$\times$$4.7\arcsec$ to $3.3\arcsec$$\times$$ 4.1\arcsec$ (FWHM) from the low to high frequency ends of the band. Each target was observed for roughly 6~minutes on-source.
 Sources were grouped such that at least four targets were observed in a single execution block, so as to minimize overheads.  This resulted in total observation times of $25-45\1{minutes}$ per source.  Typical system temperatures were measured to be  $\mathrm{T_{sys}} = 55-84\1{K}$. Flux calibration was performed on Uranus, Neptune, Ganymede, J0519-4546, and J0538-4405.  Bandpass and phase calibration were determined from nearby quasars. 
 
The data were processed using the Common Astronomy Software Application (\texttt{CASA}; \citealp{mcmullin07,petry12}).  Calibrated data cubes were constructed using \texttt{CASA}'s \texttt{TCLEAN} package.  The cubes have a channel width of $62.5\1{MHz}$ ($\sim$$220\1{km s^{-1}}$).  Observations from Cycle 3 had a typical noise per channel of $0.5-0.8\1{mJy/beam}$ over the $37\1{GHz}$ bandwidth.  The \texttt{TCLEAN}ed continuum images have typical noise levels of $50\1{\mu Jy/beam}$.

The Cycle 4 observations were conducted from November 2016 - May 2017 (project ID: 2016.1.00672.S).  In Cycle 4, each scan utilized between 38 and 46 antennas, resulting in typical synthesized beams of $4.1\arcsec \times 5.0\arcsec$ to $3.5\arcsec \times 4.3\arcsec$ (FWHM) from the low to high frequency ends of the band. Each target was observed for $\sim$$12-15\1{minutes}$ on-source for each source, not including overheads.  Sources were again grouped such that each execution block contained multiple targets.  However, because groupings were not possible for every source, the total observation time ranged between $38-90\1{minutes}$ per source, including overheads.  Typical system temperatures were measured to be $\mathrm{T_{sys}} = 60-89\1{K}$. Flux calibration was performed on Mars, Uranus, Neptune, J0334-4008, J0538-4405, J2056-4714, and J0519-4546.  Bandpass and phase calibration were determined from nearby quasars. Because more antennas were available in Cycle 4, the typical noise per channel decreased to $0.4-0.6\1{mJy/beam}$.  The \texttt{TCLEAN}ed continuum images from Cycle 4 had typical noise levels of $40\1{\mu Jy/beam}$.

Because a total of three sources did not exhibit lines in their initial $3\1{mm}$ line scans (SPT0112-55, SPT0457-49 and SPT2340-59; more detail in Sec.~\ref{sec:specresults}), additional deeper $3\1{mm}$ scans for these sources were conducted in Cycle 7.  These observations were conducted from November 2019 - January 2020 (project ID: 2019.1.00486.S) with the aim of observing possible line features with fluxes of $1-2\1{mJy}$ identified in the earlier scans.  Each scan utilized between 42 and 49 antennas, resulting in minimum and maximum angular resolutions of $2.0\arcsec-4.0\arcsec$ from the high to low frequency ends of the observed band.  Each target was observed for $45-91\1{minutes}$ on-source, for a total time of $130-300\1{minutes}$ when including overheads. Typical system temperatures were $\mathrm{T_{sys}} = 62-72\1{K}$ resulting in a noise per channel of $0.2-0.3\1{mJy/beam}$.  Flux calibration was performed on J0006-0623, J0519-4546 and J0238+1636.  Bandpass and phase calibration were determined from nearby quasars. The continua for the three sources re-observed in Cycle 7 had typical noise levels of $10-70\1{\mu Jy/beam}$.  
The decreased noise in Cycle 7 is to due a combination of longer integration times and more antennas being available.

%---------------------------------------------
% FIGURE 2 ALMA Band 3 Configuration
%---------------------------------------------

%========================================================
% Section 2.2.2 Spectroscopic Confirmations
\subsubsection{Additional Spectroscopic Observations} \label{sec:zconfirm}
%========================================================
Because many observations from the blind $3\1{mm}$ line scans described in Sec.~\ref{sec:specobs} contain single CO lines, additional observations were required in order to break degeneracies between redshift solutions and obtain unambiguous spectroscopic redshifts.  In addition to the $3\1{mm}$ ALMA survey conducted in this work, we have also conducted surveys of the [CII] emission line and low-J CO lines.  The [CII] emission line survey was conducted using the First Light APEX Submillimetre Heterodyne receiver (FLASH; \citealt{heyminck06}).  A subset of these observations were published in~\citet{gullberg15} and used to confirm three redshifts from \citet{strandet16}.  More recently conducted [CII] observations were used to confirm four spectroscopic redshifts in this work.  The Australia Telescope Compact Array (ATCA) was used to conduct a survey of CO(1-0) and CO(2-1) and confirmed a subset of observations~\citep{aravena16}.  One of these observations was obtained after the publication of \citet{aravena16} is used to confirm one additional redshift in this work.  The details for all of these observations can be found in Ap.~\ref{ap:specfollowup}.

An additional 13 redshifts were confirmed through targeted CO line searches in the ALMA $2\1{mm}$ Band.  CO emission lines were targeted using the possible redshift solutions from the observed $3\1{mm}$ emission line.  To further guide which of the degenerate solutions should be targeted, the photometric redshift obtained from SED fitting was used to guide the target selection.  The ALMA correlator sidebands were configured such that at least one CO emission line would be observed for a given redshift.  The expected line strength was determined using the SPT DSFG CO SLED from~\citet{spilker14}.  These observations were carried out in ALMA Cycles 6 and 7 (project IDs: 2018.1.01254.S and 2019.1.00486.S) and all sources targeted yielded CO emission detections at sufficient significance levels to confirm the redshifts.  Details on these observations and the obtained spectra can be found in Ap.~\ref{sec:almazconfirm}.  

Of the spectroscopic redshifts presented, only two are based on a single line and are still awaiting additional observations to confirm the redshift (see Sec.~\ref{sec:1liners} for a detailed discussion).

%========================================================
% Section 2.3 Photometry
\subsection{Photometry and SED Fitting} \label{sec:photometry}
%========================================================
%========================================================
% Section 2.3.1 Photometric Observations
\subsubsection{Photometry} \label{sec:photobs}
%========================================================
The SPT DSFGs have superb FIR--mm photometric coverage, with flux densities measured at $3\1{mm}$ (ALMA), $2\1{mm}$, $1.4\1{mm}$ (SPT), $870\1{\mu m}$ (APEX/LABOCA), $500\1{\mu m}$, $350\1{\mu m}$, and $250\1{\mu m}$  (\textit{Herschel}/SPIRE) for all sources.  Additional \textit{Herschel}/PACS $160\1{\mu m}$ and $100\1{\mu m}$ observations were obtained for a subset of 65 sources.  Despite the large range in redshifts ($1.9$$<$$\mathrm{z}$$<$$6.9$), the photometry is complete between $71\1{\mu m}$$<$$\lambda_{\mathrm{rest}}$$<$$380\1{\mu m}$, and the peak of the FIR SED at $\sim$$100{\mu m}$ is always well constrained.   The flux densities for all photometric points can be found in Ap.~\ref{ap:fml}.  The absolute calibration uncertainties of $10\%$ for \textit{Herschel}/PACS, $7\%$ for \textit{Herschel}/SPIRE data, $12\%$ for APEX/LABOCA, $7\%$ for SPT, and $10\%$ for ALMA data added in quadrature to the errors quoted in Ap.~\ref{ap:fml}.

\paragraph{ALMA}
The ALMA $3\1{mm}$ continuum maps were obtained as a result of the observations described in Sec.~\ref{sec:alma3mm}.  The continuum images were created using \texttt{CASA}'s \texttt{TCLEAN} procedure with the full observed bandwidth ($84.2$$-$$114.9\1{GHz}$) with natural weighting in order to optimize sensitivity.  In cases where the source is unresolved, meaning $>$$90\%$ of the total flux detected was contained within one beam, we extract flux from the brightest pixel detected from the continuum to obtain a spectrum. The error on flux density was calculated using the RMS of the residual map produced by the \texttt{TCLEAN} procedure.  However, half of the SPT-selected sources are marginally resolved ($> 80 \%$ of the source's flux is contained within one beam).  In order to obtain spectra for these sources, \texttt{CASA}'s \texttt{imfit} routine is used to fit a 2D Gaussian to the source in each datacube slice and extract the flux and associated error.  

\paragraph{SPT}
The SPT $1.4\1{mm}$ and $2.0\1{mm}$ flux densities were extracted from CMB maps acquired from the first survey, SPT-SZ.  This survey was completed in November 2011 and covered $2500\1{deg^2}$ of the southern sky in three frequency bands, 95, 150, and $220\1{GHz}$ (corresponding to 3.2, 2.0 and $1.4\1{mm}$, respectively) with arcminute angular resolution.  Absolute calibration for both the 1.4 and $2.0\1{mm}$ bands is derived from the CMB and the calibration uncertainty is $\lessapprox$$10\%$.  The data were extracted and deboosted according to the procedure described in~\citet{everett20}.  

\paragraph{APEX}
The sources were observed at $870\1{\mu m}$ with LABOCA at APEX and the flux densities were extracted.  LABOCA is a 295-element bolometer array with an $11.4\arcmin$ field-of-view and a measured angular resolution of $19.7\arcsec$ (FWHM).  The center frequency of LABOCA is $345\1{GHz}$ ($870\1{\mu m}$) with a passband FWHM of $\sim 60\1{GHz}$. The measured noise performance for these observations was $60\1{mJy\,s^{1/2}}$.  These observations were performed between September 2010 and October 2013 (project IDs: M-085.F-0008-2010, M-087.F-0015-2011, E-087.A-0968B-2011, M-089.F-0009-2012, E-089.A-0906A-2012, M-091.F-0031-2013, E-091.A-0835B-2013, M-092.F-0021-2013). For more details on the observations, see \citet{strandet16}.

APEX/LABOCA maps were created for each source using the Bolometer Array analysis software (\texttt{BoA}; \citealp{schuller10}).  The resulting time-ordered data undergo various calibration, noise removal, and flagging procedures detailed fully in~\citet{greve12}.  The data is then gridded and individual maps are co-added with inverse variance weighting.  The flux densities were either extracted from the peak flux density, in case of point-like sources or by integrating over the emission region, in cases where LABOCA resolves the emission.

\paragraph{\textit{Herschel}{\normalfont /SPIRE}}
Flux densities at $250\1{\mu m}$, $350\1{\mu m}$ and $500\1{\mu m}$ for all sources were measured by the Spectral and Photometric Imaging Receiver (SPIRE)~\citep{griffin10} onboard the \textit{Herschel Space Observatory}.  The data were observed in two programs (project IDs: OT1\_jvieira\_4 and OT2\_jvieira\_5) conducted between August 2012 to March 2013.  The \textit{Herschel}/SPIRE data consists of triple repetition maps, with coverage complete to a radius of $5\arcmin$ from the nominal SPT position.  The maps were produced using the standard reduction pipeline \texttt{HIPE} v9.0~\citep{balm12,  ott11}. Flux densities were extracted by fitting a Gaussian profile to the SPIRE counterpart of the SPT detection and the noise was estimated by taking the RMS in the central $5\arcmin$ of the map.

\paragraph{\textit{Herschel}{\normalfont/PACS}}
Additional data were obtained for a subsample of 65 sources at $100$ and $160\1{\mu m}$ using the Photodetector Array Camera \& Spectrometer (PACS) onboard \textit{Herschel} (project IDs: OT1\_jvieira\_4, OT1\_dmarrone\_1, OT2\_jvieira\_5 and DDT\_mstrande\_1).  The additional data ensures that the thermal peak is well-sampled and has been constrained for all SPT sources with $z$$<$$2.5$.  The data were acquired using approximately orthogonal scans centered on the target at medium speed (i.e., with the telescope tracking at $20\arcmin s^{-1}$), spending a total of $180\1{s}$ on source per program.  Each scan was composed of ten separate $3\arcmin$ strips, each offset orthogonally by $4\arcsec$ and both wavelengths were observed simultaneously. The scans were co-added and weighted by coverage. The data were then handled by a variant of the reduction pipeline presented in~\citet{ibar10}.  The resulting noise levels were calculated using random aperture photometry and were found to be $\sigma$$\approx$$4$ and $7\1{mJy}$ at $100$ and $160\1{\mu m}$ respectively.  

%========================================================
% Section 2.3.2 SED Fitting
\subsubsection{SED Fitting} \label{sec:sed}
%========================================================
We fit each source in each sample with a modified blackbody law (e.g. \citealt{blain03,casey12}) given by 

\begin{equation}
    f_\nu \propto \big[ 1 - \exp (-\nu / \nu_0)^{\beta}]\big] \big[ B_\nu (T_{\mathrm{dust}}) - B_\nu(T_{\mathrm{CMB}})\big]
\end{equation}

\noindent where $B_\nu$ is the Planck function for a value of \tdust{} or the CMB temperature, $T_{\mathrm{CMB}}$.  In order to reduce the number of free parameters and mitigate the degeneracies between redshift and \tdust, we fix the Rayleigh-Jeans spectral slope, $\beta$.  Empirically, the value $\beta$$=$$2$ was well-matched to the data, so we fix this parameter for our modified blackbody fits in a similar fashion to what was done in \citet{greve12}.  However, rather than fix $\nu_0$$\approx$$3000\1{GHz}$ ($\lambda_0$$\approx$$100\1{\mu m}$), we use the empirical relationship between $\lambda_{0}$ and \tdust{} given in Eq.~2 of~\citet{spilker16} to constrain $\lambda_{0}$.  Using this relation provides a better alternative to assuming a single value for $\lambda_0$ when an independent estimate of the size of the emission region is not available.  We find that the introduction of this dependency between \tdust{} and $\lambda_{0}$ improves both the reduced $\rchi^2$ value and photometric redshift.  It should be noted, however, that this procedure tends to increase the value of the dust temperature by $\sim$$20\%$.  The only free parameters in this SED fit are the overall SED normalization, dust temperature, and redshift.  

Because a modified blackbody fit alone does not typically describe the mid-IR excess found in the Wien side ($<\lambda_{\mathrm{rest}}$$=$$50\1{\mu m}$) of the thermal emission peak, we perform another fit including an additional power law component~\citep{blain03}.  The power law component introduces another free parameter,  $\alpha$, which is the power law slope.  The combined modified blackbody and power law fit empirically describes all of the available photometry, including the data on the Wien side of the thermal emission peak.  In this work, we use this SED fit to define the frequency at which thermal emission peaks, $\lambda_{\mathrm{peak}}$, obtain a best fit to $\beta$ for the $\mathrm{M_{d}}$ calculation, and to determine total \lir.  

In order to fit the data, we employ a Markov Chain Monte Carlo (MCMC) algorithm using the \texttt{emcee} package~\citep{foremanmackey13} to sample the posterior probability function.  To ensure uniform photometric coverage in the fit region, we mask data shortward of $\lambda_{\mathrm{rest}}$$=$$50\1{\mu m}$~\citep{greve12} for the modified blackbody fits.  Fitting done with an additional power law included all available photometry points. The results of this fitting procedure are described in Sec.~\ref{sec:photoresults}.  

%========================================================
% Section 2.3.3 Calculating intrinsic source properties
\subsubsection{Calculating Intrinsic Source Properties} \label{sec:srcprops}
%========================================================
In this section, we describe the intrinsic source properties, which are calculated using the SED fits.  Because we constrain $\lambda_0$ as a function of \tdust{} (as described in Sec.~\ref{sec:sed}), we are able to better understand the \tdust{} distribution of the sample.  The apparent FIR luminosity (\lfir) is calculated by integrating the fitted SED over the wavelength range $42.5-122.5\1{\mu m}$~\citep{helou88}.  In order to obtain the IR luminosity (\lir), we integrate the modified blackbody function with an additional power law over the $8-1000\1{\mu m}$ range.

With FIR luminosity and \tdust{} values, we derive star formation rates and dust masses for each source.  The dust masses are calculated according to: 

\begin{equation} \label{eq:mdust}
    M_{d} = \mu^{-1} \frac{D_{L}^2 S_{\nu}}{(1+z)\kappa(\nu)} [B_{\nu_r}(T_d) - B_{\nu_r}(T_{\mathrm{CMB}}(z))]^{-1}
\end{equation}

\noindent
where $S_{\nu}$ is the flux density at $345\1{GHz}$ in the rest frame, determined from our SED fit.  $D_L$ is defined as the luminosity distance, $T_{\mathrm{CMB}}(z)$ is the cosmic microwave background temperature at redshift $z$, and $\mu$ is the magnification factor.  We adopt $\kappa(\nu)/\1{m^2 kg^{-1}} = 0.015 \times
(\nu_r/250\1{GHz})^{\beta}$ \citep{weingartner01,dunne03}, where $\beta$$=$$2.0$ is the dust emissivity index.  

To derive total star formation rates (SFR), we use the following conversion from~\citet{murphy11}:

\begin{equation} \label{eq:sfr}
    \frac{\mathrm{SFR}}{\msol \1{yr}^{-1}} = 1.49 \times 10^{-10} \mu^{-1} \frac{\mathrm{L_{IR}}[8-1000\1{\mu m}]}{\lsol}
\end{equation}

\noindent
where the infrared luminosity, \lir, is calculated from the $8-1000\1{\mu m}$ range using the modified blackbody fit with an additional power law to describe the mid-IR excess.  This conversion was calculated using Starburst99~\citep{leitherer99} for a Kroupa initial mass function~\citep{kroupa01}.  

In order to calculate the intrinsic properties of these sources, the magnifications presented in~\citet{spilker16} are used when available.  \citet{spilker16} surveyed a sample of 47 SPT DSFGs, and constructed gravitational lens models.  However, because of the final flux density cut ($S_{870\1{\mu m}}$$>$$25\1{mJy}$) used to define the sample presented in this work, not all sources modeled in \citet{spilker16} were retained.  As a result, 39 of the sources presented in this work have detailed lens modeling.  For sources with multiple components, a flux weighted average is used as the magnification.  For the remaining sources without lens models, the median magnification ($\big< \mu_{870\1{\mu m}} \big>$$=$$5.5$) of the modeled sources is adopted. 

%%%%%%%%%%%%%%%%%%%%%%%%%%%%%%%%%%%%%%%%%%%%%%%%%%%%%
% SECTION 3																
% Results																
\section{Results} \label{sec:results}									
%%%%%%%%%%%%%%%%%%%%%%%%%%%%%%%%%%%%%%%%%%%%%%%%%%%%%

%========================================================
% Section 3.1 Spectroscopy Results
\subsection{Spectroscopy Results} \label{sec:specresults}
%========================================================
Building on the work of~\citet{weiss13} and~\citet{strandet16}, the final SPT-selected DSFG sample is composed of 81 sources.  We have obtained spectroscopic redshifts for the complete SPT-selected sample, making our catalog the largest and most complete redshift survey of high redshift DSFGs to date.  We begin by presenting the final blind $3\1{mm}$ CO line scans obtained by ALMA and the resulting spectroscopic redshifts.  For the cases where only a single CO line was detected, we discuss any ancillary spectroscopic data used to confirm the redshift in Ap.~\ref{ap:1liners}.  We also discuss the three $3\1{mm}$ spectra where no spectroscopic lines were present in the initial $3\1{mm}$ scans.  These sources were re-observed and deeper scans enabled secure redshifts to be obtained.  Finally, we discuss the two single line spectra where no ancillary data has yet been obtained, and discuss the most probable redshift.

\begin{figure*}[t]
	\centering
	\includegraphics[width=\textwidth]{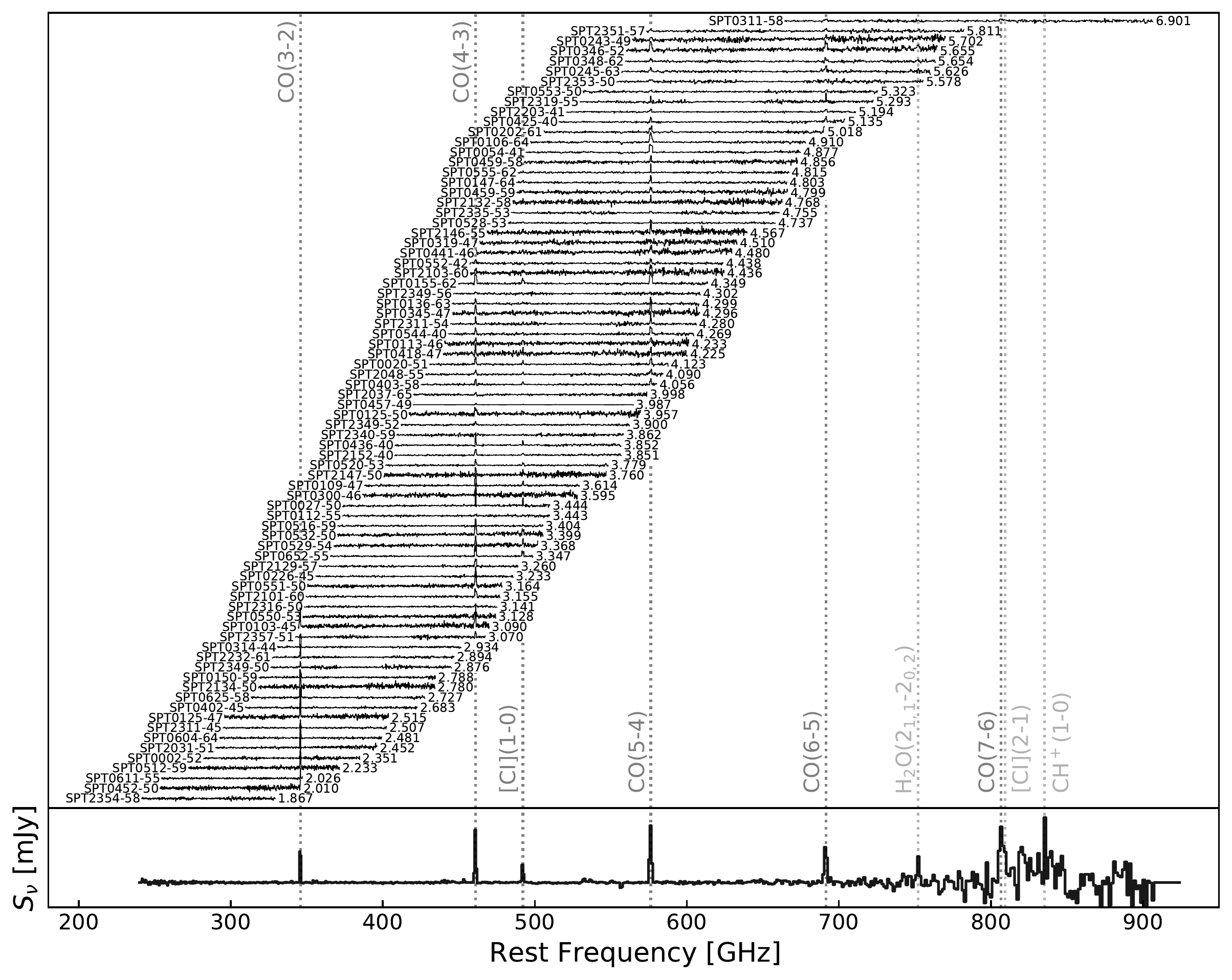}
	\caption{Top panel: All obtained ALMA $3\1{mm}$ spectra in the rest frame, with channel widths of $62.5\1{MHz}$.  The associated redshift is shown on the right of each spectrum.  Not shown: three DSFGs which do not have ALMA $3\1{mm}$ data, but were confirmed through other programs (SPT0538-50, SPT0551-48, SPT2332-53) and published in \citet{greve12} and \citet{strandet16}. By performing a $1.4\1{mm}$ flux-weighted average of the observed continuum-subtracted rest frame spectra, we obtain the composite spectrum presented in the bottom panel.  This was first done in \citet{spilker14}, but has been updated to include the final SPT-selected sample.  Emission lines used for redshift confirmation are shown in \textit{dark grey}, while other important emission lines are shown in \textit{light grey}.}
	\label{fig:zshifted_spec}
\end{figure*}

All of the SPT-selected DSFG spectra, including those originally published in~\citet{weiss13} and~\citet{strandet16}, are summarized in Fig.~\ref{fig:zshifted_spec}.  A complete summary of the spectroscopic lines detected for each source can be found in Ap.~\ref{ap:3mm}.  In this work, we detect 62 strong line features from $^{12}$CO and [CI] with integrated SNR $>$$5$$\sigma$.  We detect an additional 29 weaker features ($>$$3$$\sigma$), which include HCN, HCO$^{+}$, H$_{2}$O, $^{13}$CO, and CN.  

We detect $3\1{mm}$ continuum emission for all the 40 previously unpublished SPT-selected DSFGs presented in this work.  The positions for these sources were obtained by fitting Gaussian profiles to the ALMA~$3\1{mm}$ data and listed in Ap.~\ref{ap:srcpos}.  The continuum flux densities were also obtained in these fits and are given with the other photometric observations in Ap.~\ref{ap:fml}.  

%---------------------------------------------
% FIGURE 3 All spectra ordered by z
%---------------------------------------------

%========================================================
% Section 3.1.1 Unambiguous cases
\subsubsection{Unambiguous Cases}
%========================================================
We detect two or more line features in the $3\1{mm}$ spectra for $\sim$$46\%$ of the SPT-selected catalog.  Because of the unique distances between the CO rotational states, these redshifts can be related to rest frame spectra unambiguously.  The redshifts are derived by averaging the redshifts for individual line detections, which typically differ at the $< .1\%$ level.  Because these line profiles were fitted using a Markov-Chain Monte Carlo (MCMC) to sample to posterior probability, the values differ slightly from the values published in \citet{weiss13}, \citet{strandet16} and \citet{strandet17}.  However, the redshifts with previously published values also differ at the $< .1\%$ level and agree within their stated error bars.  Tab.~\ref{tab:lines} summarizes all of detected line features for the full SPT DSFG catalog and their derived redshifts. 

%========================================================
% Section 3.1.2 Single line detections
\subsubsection{Single Line Detections} \label{sec:1liners}
%========================================================
Spectroscopic redshifts can also be calculated from spectra with a single line feature.  However, such redshifts have multiple degenerate solutions, as the observed transition cannot be unambiguously identified.  Ancillary spectroscopic observations are required in order to break the degeneracy.  Line scans specifically targeting CO transitions and [CII] at the expected redshift solutions have been obtained with ALMA, APEX and ATCA.  These observations are described in detail in Ap.~\ref{ap:specfollowup} and are used in order to confirm an additional 15 redshifts.  Any ancillary data obtained is noted in the comments section of Tab.~\ref{tab:lines}.  

However, there are still two sources (SPT0150-59 and SPT0314-44) with a single $3\1{mm}$ feature that do not yet have any associated ancillary spectroscopy.  These sources are identified in Tab.~\ref{tab:lines} as the bolded sources.  In both cases, these spectra cannot result from CO transitions of J=4-3 or higher because these lines would be accompanied by another line within the observing band (see Fig.~\ref{fig:almab3}).  For more detail, see Sec.~\ref{sec:sedresults}.  Additionally, we can use the available photometry to determine the most probable redshift solution.  Both methods indicate that CO(3-2) is the most probable identification, but follow up spectroscopy is needed for redshift confirmation.  

%========================================================
% Section 3.1.3 No line detections
\subsubsection{No Line Detections}
%========================================================
Combining the results from our two previous redshift papers, there are three sources where no line could be identified in the $3\1{mm}$ window: SPT0128-51, SPT0457-49 and SPT2344-51.  In this work, we find one additional source, SPT0112-55, where we do not detect a line in our initially shallow Band 3 observations.  SPT0128-51 and SPT2344-51 failed to meet our $S_{870\1{\mu m}}$$>$$25\1{mJy}$ cut and are not retained in the final flux-limited sample.  We obtained deeper $3\1{mm}$ scans in ALMA Cycle 7 for SPT0112-55 and SPT0457-49.  We also re-observed SPT2340-59, which was originally published with a tentative single line detection in \citet{strandet16} to secure its redshift.  All three showed $>$$5\sigma$ detections of CO(4-3) and [CI](1-0) and the spectra are shown in Figs.~\ref{fig:3mmthumbs_SPT0002-SPT0243}-\ref{fig:3mmthumbs_SPT2349-52-SPT2357}.  While SPT0112-55 does not have obvious multiplicity, both SPT0457-49 and SPT2340-59 had flux split between two sources.  Optical imaging reveals that SPT2340-59 is almost certainly lensed, while SPT0457-49 is a protocluster candidate.  

%========================================================
% Section 3.1.4 summary section
\subsubsection{Summary of Spectroscopic Results} \label{sec:specsum}
%========================================================
In summary, all of the combined observational efforts have yielded secure redshifts for the complete flux-limited sample of 81 sources from the $2500\1{deg^2}$ SPT survey.  Of these, 79 sources had multiple spectroscopic lines, detected either solely from the $3\1{mm}$ window or with ancillary spectroscopic observations.  Only two of the redshifts are based on a single line, but they have no other possible redshift solution and agree with our photometric redshift from the distribution of dust temperatures.  Altogether, this is the largest and most complete collection of spectroscopic redshifts for high-redshift DSFGs obtained so far at mm wavelengths.

%========================================================
% Section 3.2 Photometric Results
\subsection{Photometric Results} \label{sec:photoresults}
%========================================================
% section outline
In this section, we first discuss the results of fitting the available photometry with the procedure outlined in Sec.~\ref{sec:sed}.  We present the fits, along with the derived dust temperatures and photometric redshifts in Sec.~\ref{sec:sedresults}.  With the SED fits in hand, we calculate the intrinsic source properties for the sample, and present them in Sec.~\ref{sec:indv_src_props}.  

%========================================================
% Section 3.2.1 SED fits
\subsubsection{SED Fits, Dust Temperature and Photometric Redshifts} \label{sec:sedresults}
\begin{figure*}[t]
	\centering
	\includegraphics[width=\textwidth]{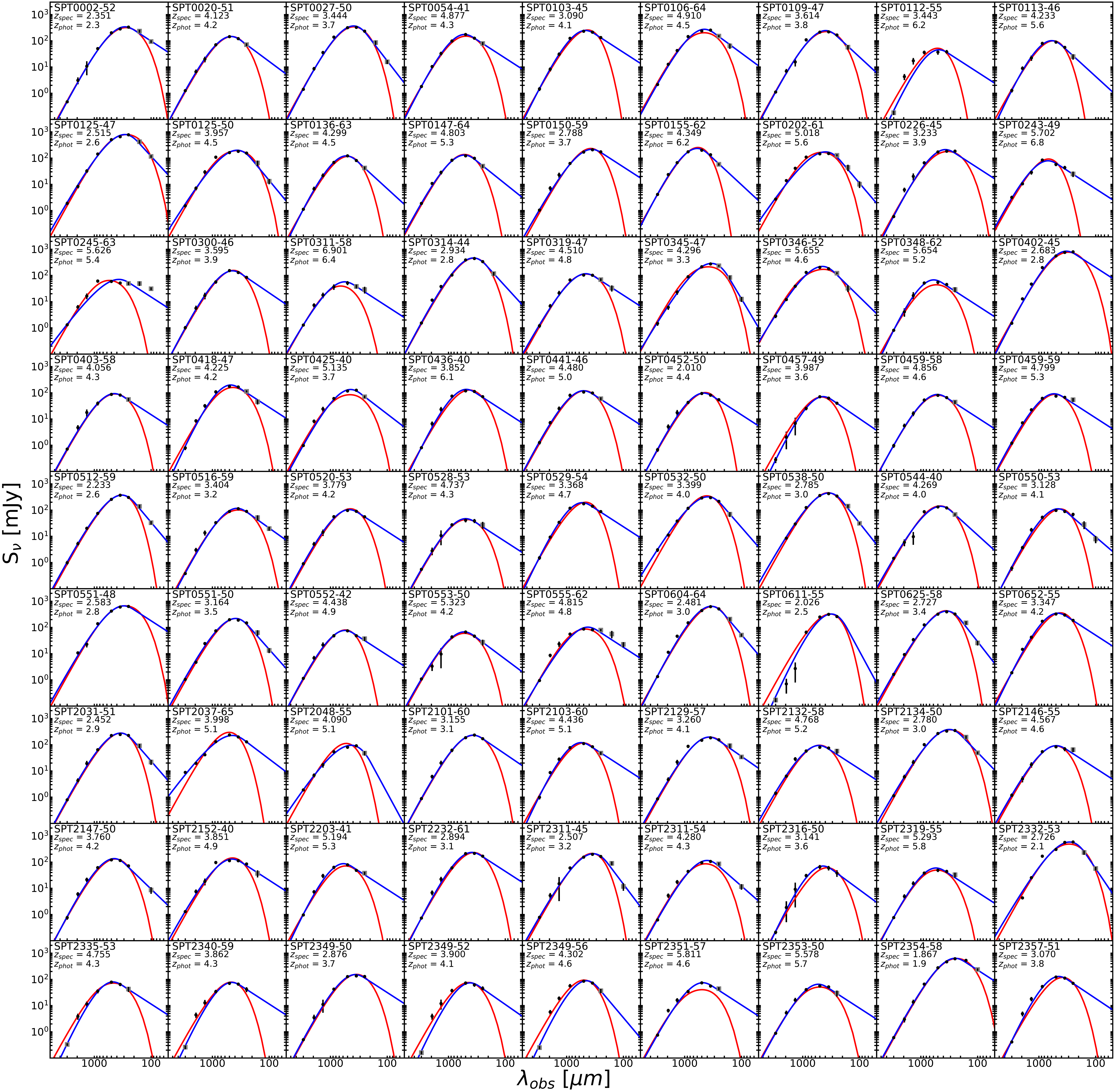}
	\caption{Spectral energy distribution fits for all 81 of the SPT-selected DSFGs.  The modified blackbody model (\textit{red}) was fit by masking data shortward of $\mathrm{\lambda_{rest}} < 50\1{\mu m}$, such that all sources have roughly uniform photometric coverage and excess emission on the Wein side of the blackbody does not artificially drive the dust temperature towards higher values.  Sources which exhibit multiple components in $3\1{mm}$ are fitted by masking the $3\1{mm}$ point.  Any photometry points which were masked for the modified blackbody fit are represented by the \textit{grey} squares.  In order to account for the mid-IR excess, a power law can be added on to the modified blackbody function~\citep{blain03} to better describe all available data (\textit{blue}). }
	\label{fig:sedthumbs}
\end{figure*}

%========================================================
More information can be obtained by fitting the photometry using the process outlined in Sec.~\ref{sec:sed}.  We first fit a modified blackbody to the FIR thermal emission peak for rest wavelengths $>$$50\1{\mu m}$.  We also fix the redshift parameter to the spectroscopic values obtained in Sec.~\ref{sec:specresults}, shown in Fig.~\ref{fig:sedthumbs} in red.  This model describes the FIR thermal emission peak well, with a median reduced $\rchi^2$ value of 1.8.  We also perform another fit with an additional power law component in order to fit the complete \lir wavelength range (shown in Fig.~\ref{fig:sedthumbs} in blue), which gives a median reduced $\rchi^2$ of 0.7.  

Sources with large $\rchi^2$ values generally exhibit a discrepancy between the $3\1{mm}$ flux and the value predicted by the modified blackbody.  Because of the discrepancy in beam sizes between ALMA and SPT, objects which are broken into multiple components are sometimes below the ALMA $3\1{mm}$ detection threshold, leading to an underestimation of the total $3\1{mm}$ flux.  Using ALMA~$870\1{\mu m}$ imaging~\citep{spilker16}, we verify that the source is split into multiple components.  In cases of multiple components, the $3\1{mm}$ point is then masked when fitting the thermal emission peak.  One of these sources, SPT2349-56, has already been identified as a protocluster \citep{miller18, hill20}.  This discrepancy in $3\1{mm}$ flux from over-resolving the emission may be a good way to separate unlensed protoclusters from lensed DSFGs. The remaining sources are being investigated as potential protocluster candidates.  As previously stated, any photometry $<$$50\1{\mu m}$ in the rest frame was masked for the modified blackbody fit.  Any masked photometry points are shown in Fig.~\ref{fig:sedthumbs} in grey.  

%---------------------------------------------
% FIGURE 4 All SED fits
%---------------------------------------------

\begin{figure}[t]
	\centering
	\includegraphics[width=\columnwidth]{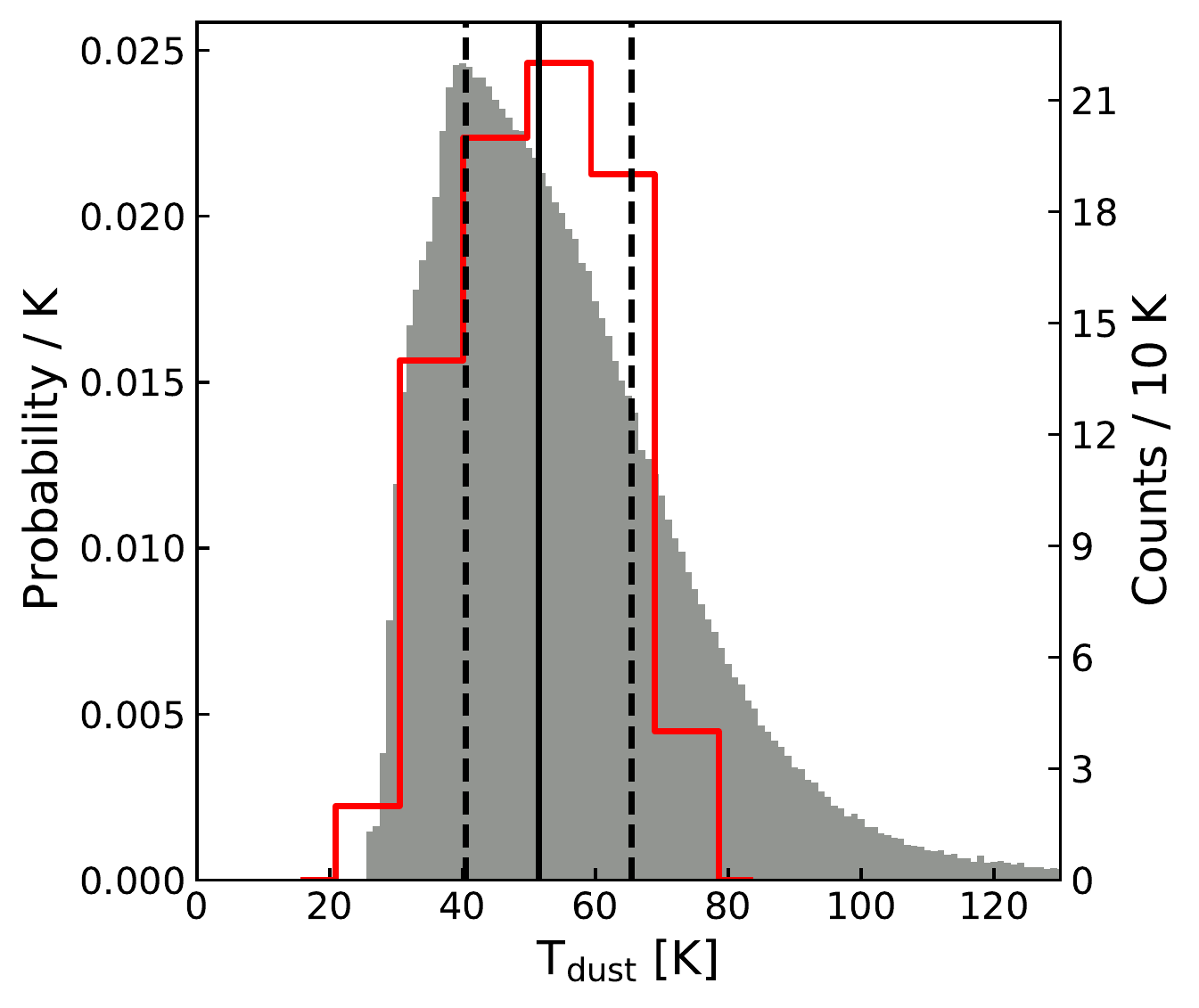}
	\caption{The probability distribution of the dust temperature for all sources in the SPT-DSFG sample. Though this probability distribution peaks at $40\1{K}$, the distribution is skewed towards warmer sources.  The median of this distribution is at $52.4\1{K}$ (solid line) and the inner quartiles of the distribution are given by the dashed lines.  {The adopted median for each fitted \tdust{} (Tab.~\ref{tab:srcprops}) is shown for comparison in \textit{red}, and is described by the right axis.}  }
	\label{fig:td_dist}
\end{figure}

By extracting a dust temperature from the SED fit, a dust temperature probability distribution was created by sampling each source's dust temperatures $10^3$ times using a Monte Carlo procedure, shown in Fig.~\ref{fig:td_dist}.  Though the distribution peaks near \tdust$=$$40\1{K}$, the median of the distribution is significantly higher at $T_{\mathrm{dust}}$$=$$52.4\pm2.3$K, with a tail extending past $\sim$$100\1{K}$.  Given the relationship between $\lambda_0$ and \tdust{} discussed in Sec.~\ref{sec:sed}, the implied median of the $\lambda_0$ distribution is $155 \pm 7 \1{\mu m}$.  These warm dust temperatures suggest that there \textit{could} be a correlation between redshift and dust temperature, which we discuss in Sec.~\ref{sec:tdevolution}.

%---------------------------------------------
% FIGURE 5 T_d distribution
%---------------------------------------------

\begin{figure}[h]
	\centering
	\includegraphics[width=\columnwidth]{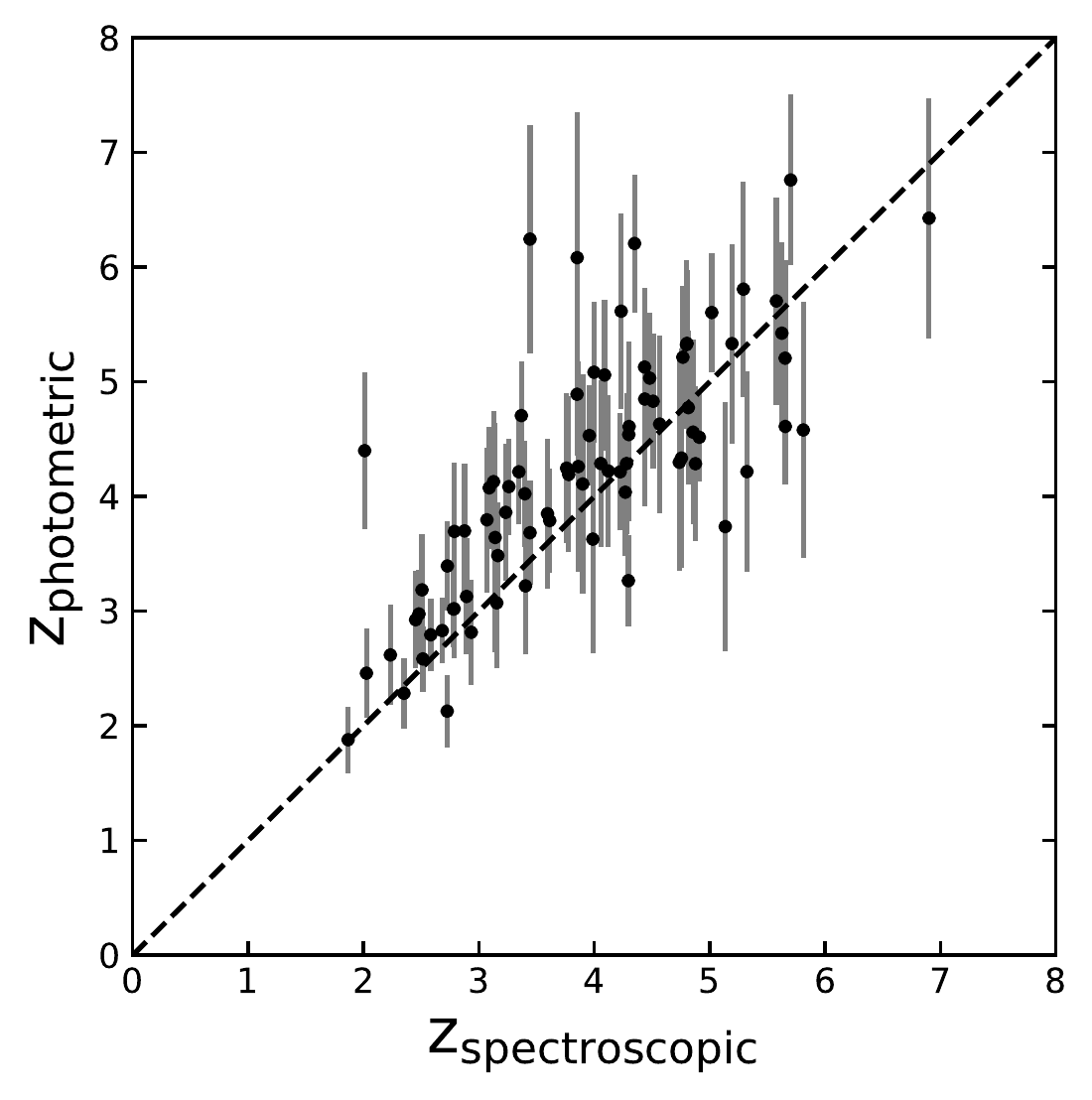}
	\caption{Photometric redshift compared with spectrosocpic redshift.   We find good agreement between photometric and spectroscopic redshifts, though photometric redshift has large associated errors.  Unity is shown as the dashed line.}  
	\label{fig:zphot_v_zspec}
\end{figure}

Although spectroscopic redshifts have been obtained for all sources, the photometric redshift can still be used in order to break degenerate redshift solutions in the two cases where a single CO transition was detected, as well as inform future spectroscopic and photometric surveys.  In order to find the photometric redshift, we use the dust temperature distribution given in Fig.~\ref{fig:td_dist} as a prior and simultaneously fit the \tdust{} and redshift parameters.  While this method exhibits good agreement with the spectroscopic redshift, as seen in Fig.~\ref{fig:zphot_v_zspec}, the associated errors are large.  However, this method can be used to ascertain a rough estimate of redshift in the absence of observationally expensive spectroscopic data. (See e.g.~\citealt{casey20} for a similar method and comparison to template fitting.).  

This photometric redshift fitting can also be used to break degeneracies for the two single $3\1{mm}$ line sources (SPT0150-59 and \mbox{SPT0314-44}).  For a given degenerate redshift solution, we fix the redshift parameter to that solution and use the SED fitting procedure to find the associated dust temperature.  Using the dust temperature distribution given in Fig.~\ref{fig:td_dist} as a prior, we assign a likelihood of the source being at each possible redshift solution (see Ap.~\ref{ap:1liners} and Fig.~\ref{fig:1linerprobs} for details).  For both single-line sources, CO(3-2) is the most probable identification.  If the detected line was in fact the higher-J transition (e.g. CO(4-3)), an additional CO line should have been detected.  Taken together, these two independent pieces of evidence indicate that there is no ambiguity in the redshift of these sources.  

%---------------------------------------------
% FIGURE 6 z_phot v z_spec
%---------------------------------------------

\begin{figure}[h]
	\centering
	\includegraphics[width=\columnwidth]{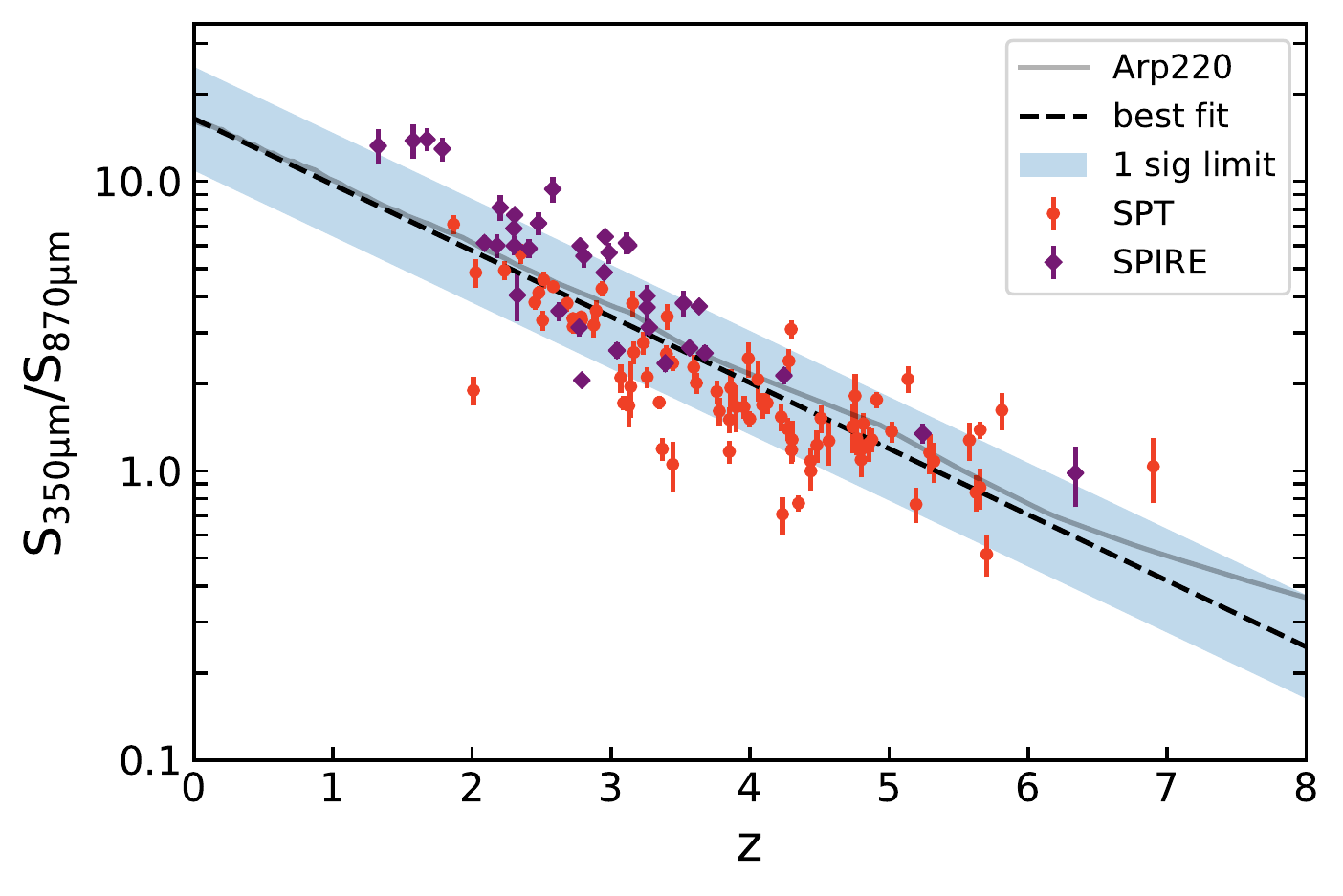}
	\caption{The $S_{350\1{\mu m}}/S_{870\1{\mu m}}$ color serves as a rough indicator for redshift.  The SPT catalog is compared to \textit{Herschel}/SPIRE sources and is well-described by an exponential fit with a term to describe the intrinsic scatter.  The Arp220 colors were obtained by artificially observing the Arp220 SED~\citep{silva98} for a range of redshifts. }
	\label{fig:redness}
\end{figure}

Though we are complete in spectroscopic redshifts, photometry provides insights that inform future surveys.  Because the SPT DSFGs have excellent FIR-mm coverage, it is possible to use single flux density ratios to obtain information about the redshift.  For instance, the $S_{350\1{\mu m}}/S_{870\1{\mu m}}$ ratio can be used as a rough indicator of redshift.  We find relatively good agreement between our catalog, \textit{Herschel}/SPIRE sources~\citep{negrello10, conley11,omont11, harris12, wardlow12, bussmann13, gladders12, riechers13, messias14} and the redshifted SED of Arp220~\citep{silva98}, shown in Fig.~\ref{fig:redness}.  We perform a maximum likelihood estimation assuming the $350-870\1{\mu m}$ flux density ratios for the \textit{Herschel}/SPIRE and SPT sources can be described by an exponential model with an extra Gaussian variance term to account for intrinsic scatter.  We find the following exponential fit best describes the available data:
\begin{equation}
    z = (5.55 \pm 0.3) - \\
     (4.55 \pm 0.2) \times \log_{10} \Bigg( \frac{S_{350\1{\mu m}}}{S_{870\1{\mu m}}} \Bigg)
\end{equation}
\noindent which is shown with a $1\sigma$ limit in Fig.~\ref{fig:redness}.  In performing this fit, we find an intrinsic logarithmic scatter of $4.2 \pm 1.1$.  While a redshift determination using this method would have large uncertainties, it is nevertheless useful in instances where limited IR or mm photometry is available.   

%---------------------------------------------
% FIGURE 7 SMG Redness
%---------------------------------------------
%========================================================
% Section 3.2.2 Individual source properties
\subsubsection{Individual Source Properties} \label{sec:indv_src_props}
%========================================================
To demonstrate the extreme star-forming nature of the SPT DSFGs, we compare our sources to a flux-limited blank-field sample throughout this work.  The original LABOCA ECDFS submillimetre survey (LESS) mapped the full Extended Chandra Deep Field South \citep[ECDFS;][]{weiss09}.  The survey area encompassed $0.5$$\times$$0.5\1{deg^2}$ and detected a total of 126 DSFGs above a significance level of $3.7\sigma$. The individual source detections were followed up with ALMA in Cycle 0 and the sample was thus called ALESS \citep{hodge13,swinbank14}.  Photometry was obtained at $1.4\1{GHz}$, 870, 500, 350, 250 160, 100 and 70$\1{\mu m}$ in order to obtain the photometric redshifts~\citep{swinbank14}.  The \texttt{MAGPHYS} SED modeling code was used to obtain photometric properties~\citep{dacunha15}.  Spectroscopic redshifts were also obtained for a subset of these sources~\citep{danielson17}.  Because the ALESS sample is complete for sufficiently bright sources, we use it as an example of the classic unlensed $850\1{\mu m}$-selected DSFGs with a flux-limited selection.  Unless otherwise stated, the same formalism used for the SPT analysis was also applied to the ALESS sources throughout this work.

\begin{figure*}[tbh]
	\centering
	\includegraphics[width=\textwidth]{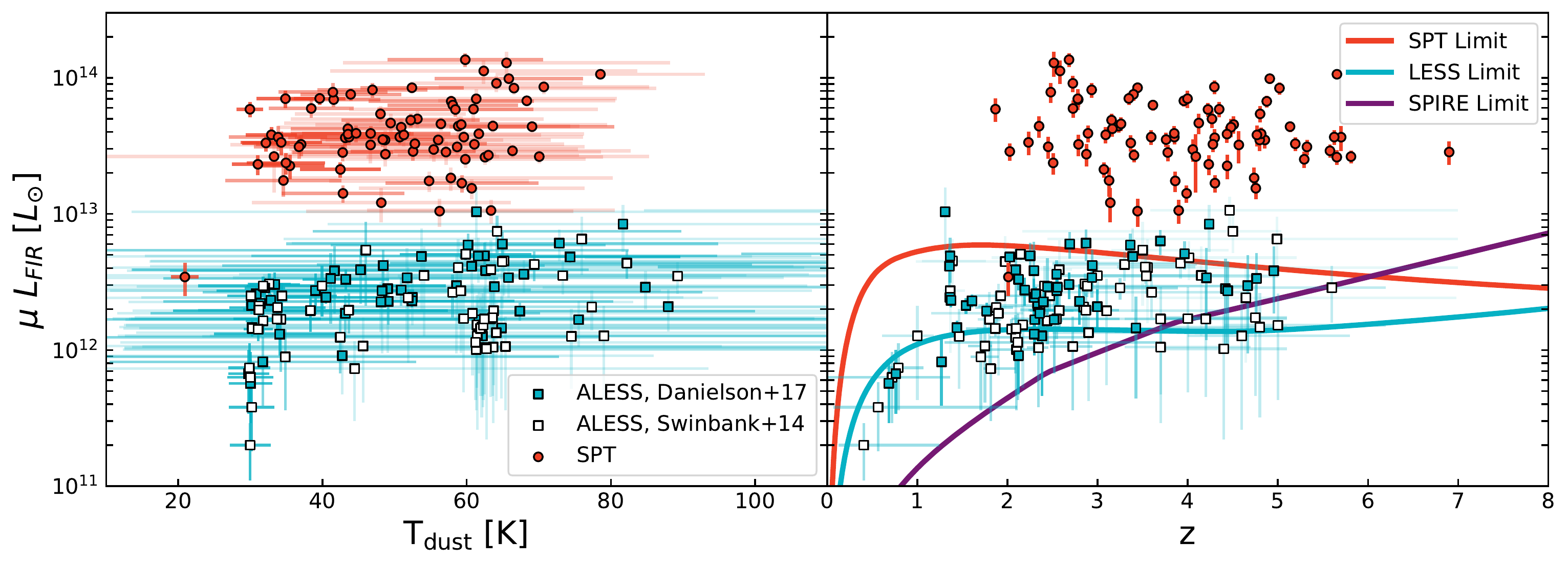}
	\caption{\textbf{Left:} Apparent $\mathrm{L_{FIR}}$ vs. \tdust{} for the unlensed ALESS sources (\textit{teal}) presented in~\citet{swinbank14, danielson17} and the lensed SPT sources presented in this work (\textit{red}).  The offset in apparent $\mathrm{L_{FIR}}$ between the lensed (SPT) and unlensed (ALESS) sources is due to gravitational magnification.  It should be noted that for many of the ALESS sources, the photometry is not well constrained, which is reflected in the size of the associated error bars.  \textbf{Right:} The FIR luminosity as a function of redshift for the same sources.  The teal, purple, and red curves represent the limiting FIR luminosity as a function of redshift corresponding to: $3$$\times$ the LESS limit (r.m.s. $\sim$$1\1{mJy}$), $3$$\times$ the SPIRE $500\1{\mu m}$ confusion limit ($30\1{mJy}$), and the $3\sigma$ SPT $2.0\1{mm}$ survey limit ($3.9\1{mJy}$), given the SED model of this paper and assuming a $35\1{K}$ dust temperature.  Many of the sources discovered in the LESS survey split into multiple components, and were treated as individual sources in the ALESS survey and consequently fall below the plotted LESS survey limit.  The SPT survey, with its longer wavelength selection, is more sensitive to sources at the highest redshifts ($z$$>$$5$) than \textit{Herschel}.} 
	\label{fig:lfir_v_zTd}
\end{figure*}

The distributions of \tdust, apparent FIR luminosity, and redshift are shown for both the SPT sample and the ALESS sample~\citep{dacunha15} in Fig.~\ref{fig:lfir_v_zTd}.  The FIR luminosity and \tdust{} values for both samples are derived from the SED fitting procedure described in Sec.~\ref{sec:sed}.  Because many of the ALESS sources do not have robust FIR detections due to their relative faintness and the \textit{Herschel}/SPIRE confusion limit \citep{nguyen10}, their fitted SED values have large associated uncertainties when using the same SED fitting routine described in Sec.~\ref{sec:sed}.  In the left panel of Fig.~\ref{fig:lfir_v_zTd}, the lensed SPT sources are offset from the largely unlensed ALESS sample, as expected from gravitational lensing.  Gravitational lensing randomly samples sources from the background and increases the solid angle they subtend.  This effect makes the lensed sources appear more luminous at a given \tdust{} than an unlensed source.  It would take a median magnification of $\sim$$18$ to make the median apparent \lfir{} for the SPT sources without lens models ($3.88$$\times$$10^{13}\1{L_{\odot}}$) on average intrinsically identical to the median \lfir{} for the ALESS sources ($2.3(2)$$\times$$10^{12}\1{L_\odot}$).

On the right panel of Fig.~\ref{fig:lfir_v_zTd}, both samples exhibit a roughly constant FIR luminosity as a function of redshift, demonstrating the negative K-correction inherent in sub/mm surveys of DSFGs.  The approximate detection limits for both surveys and the \textit{Herschel}/SPIRE confusion limit are also shown, assuming a standard template from \citet{chary01}.  These limits illustrate the importance of selection wavelength and survey depth on the resultant redshift distribution of the sample. Though SPT is only sensitive to the most luminous sources, the detection threshold at the selection wavelength of $1.4\1{mm}$ corresponds to a decreasing (apparent) luminosity at higher redshift. This is in contrast to the selection curve for the \textit{Herschel} $500\1{\mu m}$ selection, which corresponds to an increasing luminosity at higher redshift because $500\1{\mu m}$ corresponds to rest wavelengths near or beyond the peak of the dust SED at higher redshifts.

%---------------------------------------------
% FIGURE 8 LFIR v. z and T_d
%---------------------------------------------

\begin{figure*}[tb]
	\centering
	\includegraphics[width=\textwidth]{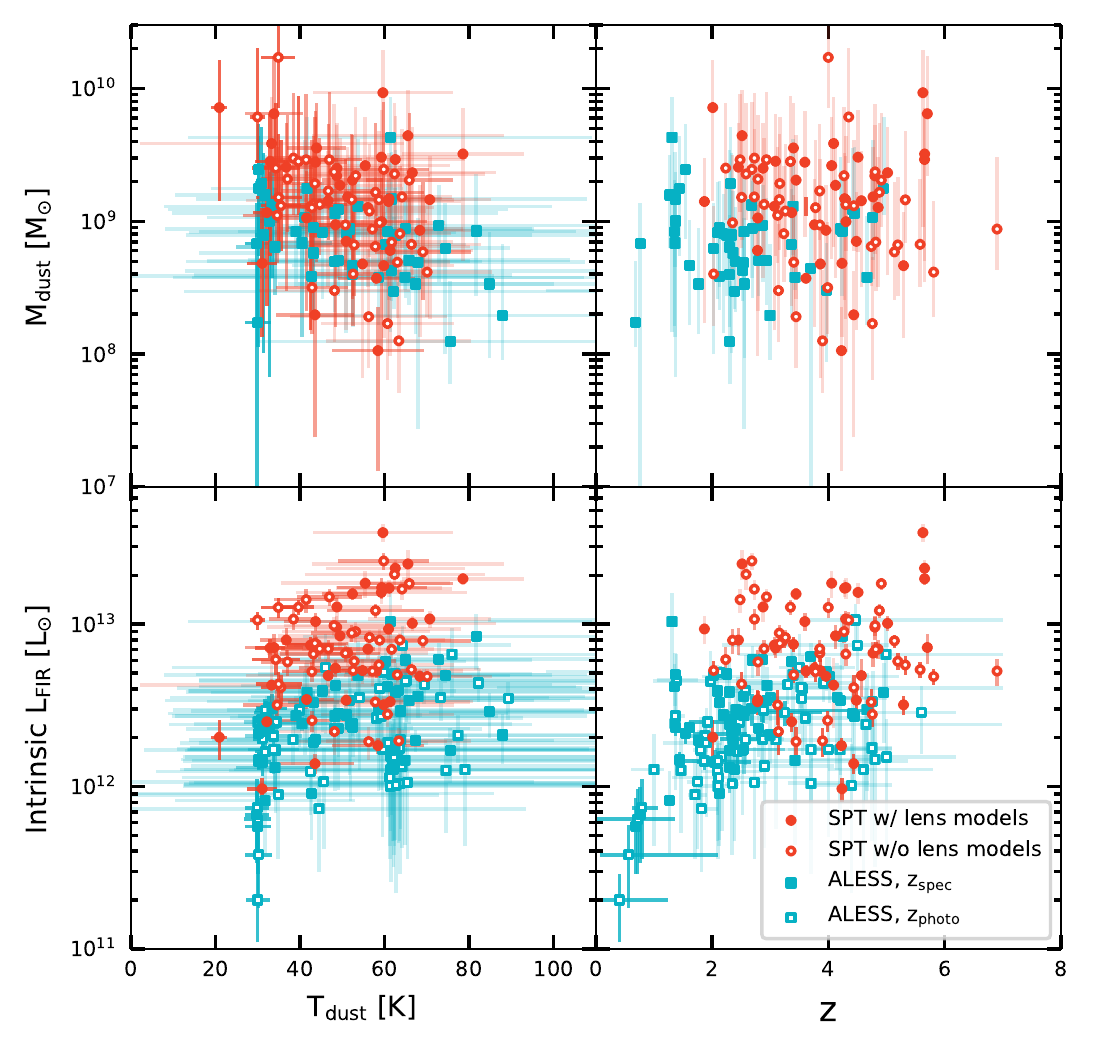}
	\caption{Derived intrinsic source properties for the SPT sample, compared with the ALESS sample.  \textbf{Top rows:} Dust mass ($\mathrm{M_{dust}}$) as a function of \tdust{} (left) and redshift (right).  \textbf{Bottom rows:} Intrinsic FIR luminosity ($\mathrm{L_{FIR}}$) as a function of \tdust{} (left) and redshift (right).  The SPT sample is on average hotter, has higher intrinsic luminosities, and skews towards higher redshift than the ALESS sample. ALESS sources (\textit{teal}) with spectroscopic redshifts are shown with \textit{filled} symbols while the ALESS sources with photometric redshifts are shown with \textit{open} symbols. SPT sources (\textit{red}) with lens models are shown with \textit{filled} symbols while the SPT sources without lens models are shown in \textit{open} symbols, and we adopted $\big< \mu \big>$$=$$5.5$.  }
	\label{fig:td_md_z_lfir_4panel}
\end{figure*}

In order to determine the intrinsic properties for each source, as in Fig.~\ref{fig:td_md_z_lfir_4panel}, lens models are needed to calculate the magnification factor.  Using the 39 SPT-DSFGs with lens models~\citep{spilker16}, the magnification factors for individual sources are used to calculate the intrinsic luminosities, dust mass and SFR values, as described in Sec.~\ref{sec:srcprops}.  All of the derived properties can be found in Ap.~\ref{ap:srcpropstable}.  It should be noted that for the remaining sources without lens models, the median magnification ($\big< \mu_{870\1{\mu m}} \big>$$=$$5.5$) is adopted.  This is a reasonable assumption, given that SPT sources with lens models from high resolution ALMA imaging presented in \citet{spilker16} were effectively drawn at random from the larger SPT sample.  However, given that the median magnifications are merely an estimate, the intrinsic properties for these sources should be also treated as estimates.  At the time of this publication, ALMA $870\1{\mu m}$ imaging has been obtained for all sources and the construction of a complete lens model catalog is currently underway. 

%---------------------------------------------
% FIGURE 9 intrinsic source property comparisons to ALESS
%---------------------------------------------

The intrinsic dust masses and FIR luminosities are shown in Fig.~\ref{fig:td_md_z_lfir_4panel} compared with \tdust{} and redshift.   We find that the SPT sources have a median intrinsic dust mass of $1.4(1)$$\times$$10^{9}\1{M_\odot}$, while ALESS sources have a median of $7.4(9)$$\times$$10^{8}\1{M_\odot}$.  The median SPT intrinsic FIR luminosity is $7.1(5)$$\times$$ 10^{12}\1{L_\odot}$ while ALESS sources exhibit a lower median $\mathrm{L_{FIR}} = 2.37(2) $$\times$$10^{12}\1{L_\odot}$. Thus, the SPT sources exhibit both higher luminosities as well as dust masses compared to the ALESS sources.  These results are discussed in detail in Sec.~\ref{sec:ulirgs}.

%%%%%%%%%%%%%%%%%%%%%%%%%%%%%%%%%%%%%
% SECTION 4 											      %
% Discussion											      %
\section{Discussion} \label{sec:discussion}				      %
%%%%%%%%%%%%%%%%%%%%%%%%%%%%%%%%%%%%% 
In this section, we begin by discussing the complete spectroscopic redshift distribution in Sec.~\ref{dis:zdist}, along with the various selection effects.  We explore the dust temperature evolution as a function of redshift in Sec.~\ref{sec:tdevolution}.  We then discuss the extreme nature indicated by the intrinsic source properties of the sample in Sec.~\ref{sec:ulirgs}.  Finally, because our sample is at higher redshift than most DSFG samples, we discuss the implications for the space density of high redshift DSFGs in Sec.~\ref{sec:highztail}. 

%========================================================
% Section 4.1 Redshift Distribution
\subsection{Redshift Distribution}\label{dis:zdist}
\begin{figure}[h]
	\centering
	\includegraphics[width=\columnwidth]{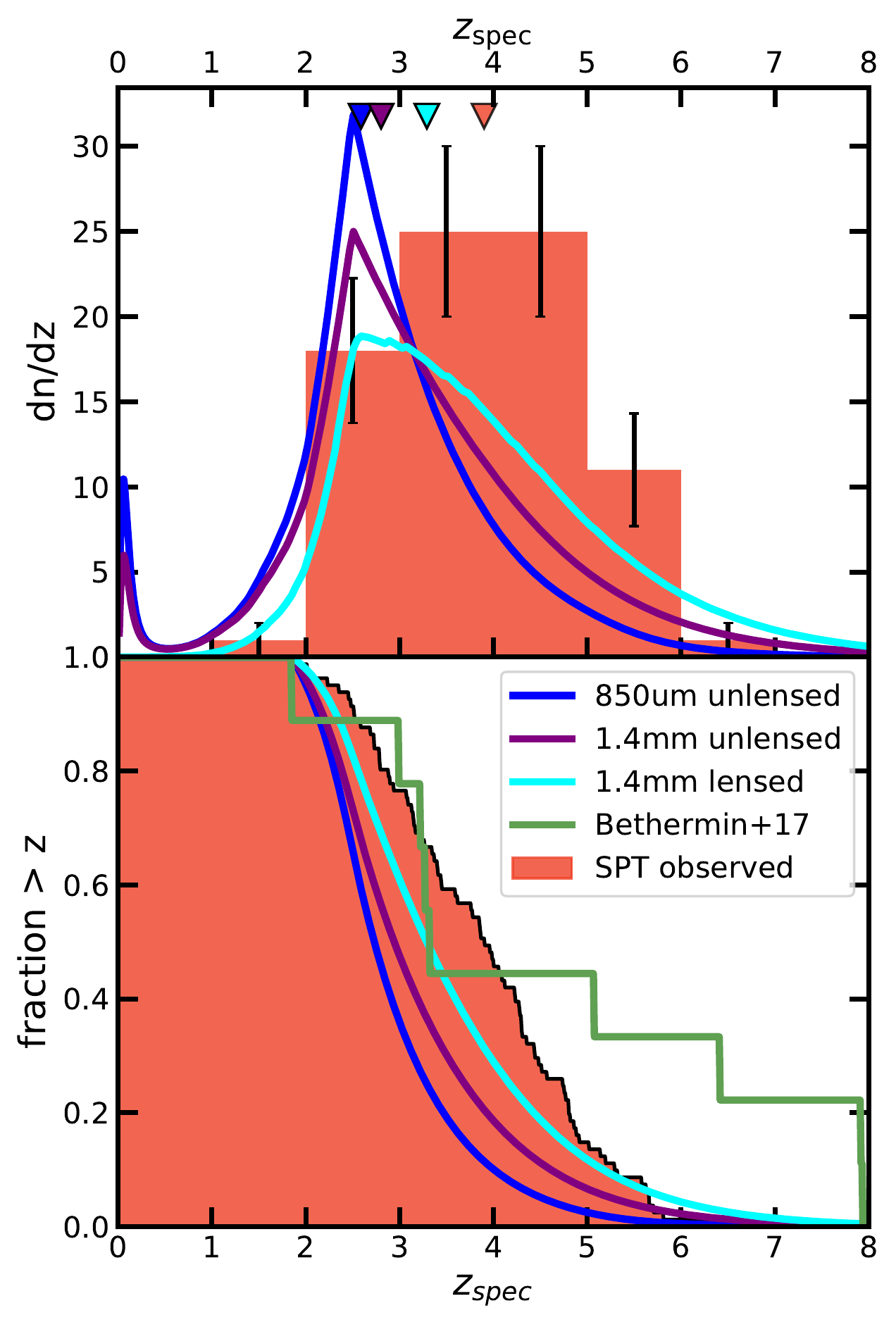}
	\caption{The observed SPT-DSFG redshift distribution is shown overlaid with with modeled source catalog predictions from~\citet{bethermin12b,bethermin15b} and \citet{bethermin17}.  \textbf{Top:} The observed redshift distribution compared with modeled source catalogs with selections of $850\1{\mu m}$ (\textit{blue}), $1.4\1{mm}$ (\textit{purple}) and $1.4\1{mm}$ with gravitational lensing taken into account (\textit{cyan}).  The filled triangles at the top of the panel represent the median redshifts of their respective samples.   \textbf{Bottom:}  The cumulative source density as a function of redshift. }
	\label{fig:z_dist}
\end{figure}

%========================================================
The redshift distribution of the full SPT-DSFG sample is shown in Fig.~\ref{fig:z_dist}.  The median of this distribution is $z_{\mathrm{median}} = 3.9 \pm 0.2$, which is unchanged from the previously published values of $z_{\mathrm{median}} = 3.9 \pm 0.4$ from~\citet{strandet16}.  Given that the new sources presented in this work represent an effectively random sampling of the original sample published in \citet{weiss13} and \citet{strandet16} (shown in Fig.~\ref{fig:color_v_flux}), it was expected that the median of the distribution would not shift significantly.  The final distribution peaks between $3$$<$$z$$<$$5$, with a large fraction ($76\%$) of the sample at $z>3$, shown in the top panel of Fig.~\ref{fig:z_dist}.  The SPT sources peak at significantly higher redshift than their unlensed counterparts, largely selected at shorter wavelengths ($z$$\sim$$2.3$$-$$2.9$; \citealt{simpson14, koprowski14, miettinen15, danielson17, simpson17, michalowski17, brisbin17}). 

%---------------------------------------------
% FIGURE 10 dn/dz and dn/dz cumulative (2 panel)
%---------------------------------------------

%---------------------------------------------
% Table 1 1 dn/dz binned counts
%---------------------------------------------
\begin{table}
\begin{center}
\begin{tabular}{ | l | l | }
%\tablecaption{Redshift distribution }
\hline
redshift bin  & dN$/$dz \\
\hline
$1$$<$$z$$<$$2$	& 1 \\
$2$$<$$z$$<$$3$	& 18 \\
$3$$<$$z$$<$$4$	& 25 \\
$4$$<$$z$$<$$5$	& 25 \\
$5$$<$$z$$<$$6$	& 11 \\
$6$$<$$z$$<$$7$	& 1 \\
\hline
\end{tabular}
\caption{Empirical counts in each redshift bin for the final SPT $dN/dz$ distribution. \label{tab:dndz}}
\end{center}
\end{table}

%========================================================
% Section 4.1.1 Source Multiplicity
\subsubsection{Sources with Multiplicity}\label{dis:multiplicity}
%========================================================
Due to the large ($\gtrapprox$$1\arcmin$) beam size of the initial SPT selection, the SPT-DSFG catalog contains multiple sources which break up into multiple individual galaxies at the same redshift when observed at higher resolution ($\lessapprox$$1\arcmin$).  As previously discussed in Sec.~\ref{sec:photoresults}, the differences between the ALMA and SPT beam sizes can result in over-resolving the emission, causing the $3\1{mm}$ flux to be underestimated in SED fitting.  Though we treat these systems as individual DSFGs in this work, at least two of these sources are known to break into multiple components.  SPT0311-58~\citep{strandet16} was shown in \citet{marrone18} to be composed of two individual galaxies at $z$$=$$6.900$, a western source with an intrinsic $\lir = 33\times10^{12}\lsol$ and an eastern source with an intrinsic $\lir = 4.6$$\times$$10^{12}\lsol$.  SPT2349-56 is among the most actively star-forming high-redshift protoclusters known \citep{miller18} at z=4.3.  This system is composed of at least 29 individual galaxies, six of which are at $\lir$$>$$3$$\times$$10^{12}\lsol$ ~\citep{hill20}.  Additional APEX/LABOCA observations show SPT0348-62, SPT0457-49, SPT0553-50, and SPT2335-53, also have significant overdensities and lie at high redshift ($3$$<$$z$$<$$7$, Wang \etal~\textit{in prep.}).  The multiplicities in these systems would potentially change the overall shape of the observed SPT-DSFG redshift distribution, favoring even higher redshifts. For the purpose of simplicity and homogeneity, we treat each of these systems as a single source in this analysis. 

%========================================================
% Section 4.1.2 Selection effects
\subsubsection{Selection Effects}\label{dis:selection}
%========================================================
The two major selection effects that impact our distribution are due to the long wavelength selection ($1.4\1{mm}$) and the high $S_{870\1{\mu m}}$ flux cut, which selects primarily gravitationally lensed sources.  Both of these effects are known to bias a redshift distribution towards higher redshifts.  In this section, we compare the model of \citet{bethermin15b} with our observed distribution.  This model was selected because it best described the distribution in \citet{weiss13} and takes into account both selection wavelength and lensing.  These effects were previously shown to account for the redshift distribution for the subset of the SPT DSFG sample published in \citet{strandet16} and \citet{bethermin15b}. 
Some works~\citep{blain02, greve08, dacunha13, staguhn14} suggest that the CMB could make cold DSFGs at high redshifts difficult to detect.  However, since the SPT-selected DSFGs are warm, with a median dust temperature of \tdust$=52.4\1{K}$ and minimum of \tdust$=20.9\1{K}$, this effect only becomes relevant at very high redshifts ($z$$>$$6-18$).  

In Fig.~\ref{fig:z_dist}, we examine the effect of a long wavelength selection on the observed redshift distribution using the \citet{bethermin15b} model.  By applying a different selection wavelength on a realistic population of galaxies, we see the median shift from $z$$\sim$$2.6$ at $850\1{\mu m}$ to $z$$\sim$$2.8$ at $1.4\1{mm}$.  A Kolmogorov-Smirnov (K-S) test reveals that the observed SPT-DSFG distribution is distinct from the modeled data populations at levels of $7.7\sigma$ and $5.4\sigma$ for the $850\1{\mu m}$ and $1.4\1{mm}$ wavelength selections, respectively.  Long wavelength selection alone does not explain the difference between the \citet{bethermin15b} model and the observed SPT data.  

Based on models of the high redshift DSFG population (e.g.~\citealt{baugh05, lacey10, bethermin12a, hayward13,bethermin17, lagos19,lovell20}), very few sources would be intrinsically bright enough to exceed our adopted flux density threshold at $870\1{\mu m}$ ($>$$25\1{mJy}$).  Because of the high apparent luminosities, the SPT DSFG sample was expected to consist almost solely of gravitationally lensed sources~\citep{blain96, negrello07}.  Indeed, the lens models provided in \citet{spilker16} demonstrate that at least ${\sim}$$74\%$ of the flux-limited SPT sample is strongly lensed, while a small fraction of sources are unlensed ($\sim$$5\%$) and the remaining sources are weakly lensed.  Strong lensing preferentially selects sources at high redshift, biasing an intrinsic redshift distribution to higher redshifts. The probability of sources at $z$$<$$1.5$ undergoing strong lensing is heavily suppressed relative to sources at higher redshifts ($z$$>$$4$) (e.g. \citealt{hezaveh11}).  The $1.4\1{mm}$ lensed model in Fig.~\ref{fig:z_dist} demonstrates the combined effect of the lensing potential on the observed redshift distribution and long wavelength selection.  Though this model brings the median of the \citet{bethermin15b} mock catalog to $z$$\sim$$3.2$, the observed SPT DSFG redshift distribution is still higher than this model and a KS test rules it out at a level of $3.2\sigma$.  A possible reason for discrepancy between the \citet{bethermin15b} model and the SPT redshift distribution is that the model was fit to data from previous surveys, which typically peaked at lower redshift.  The original \citet{bethermin12a} model is based on the stellar mass function and evolution of the main sequence and relies on SED template fitting.  As these parameters become better constrained with more observational data from the SPT catalog and others, more realistic models will be possible in the future.

The lensing analysis in \citet{bethermin15b} assumes that the sources do not undergo significant size evolution over cosmic time.  Compact sources have a higher probability of being highly magnified than more diffuse sources~\citep{hezaveh12a}.  Thus, size evolution verses redshift coupled with gravitational lensing could potentially bias a redshift distribution towards higher redshift, as discussed in \citet{weiss13}.  The fitted values of $\mathrm{L_{FIR}}$ and \tdust{} can then be used to determine the effective blackbody radius of the source via the modified blackbody function (e.g. Eq.~1 of \citealt{spilker16}).  The median effective blackbody radius for SPT-selected DSFGs is $1.0$$\pm$$0.1 \1{kpc}$, which is consistent with the median radius found using lens modeling ($1.04$$\pm$$0.07 \1{kpc}$; \citealt{spilker16}).  The sizes of the SPT DSFGs are also compatible with the sizes ($0.3$$-$$3\1{kpc}$) observed in unlensed DSFGs ~\citep{ikarashi15, simpson15, hodge16, ikarashi17}.  None of these observations show evidence of size evolution with redshift above $z$$>$$1$.  

%========================================================
% Section 4.2 Dust temperature evolution
\subsection{Dust Temperature Evolution}\label{sec:tdevolution}
%========================================================
\begin{figure}[!h]
	\centering
	\includegraphics[width=\columnwidth]{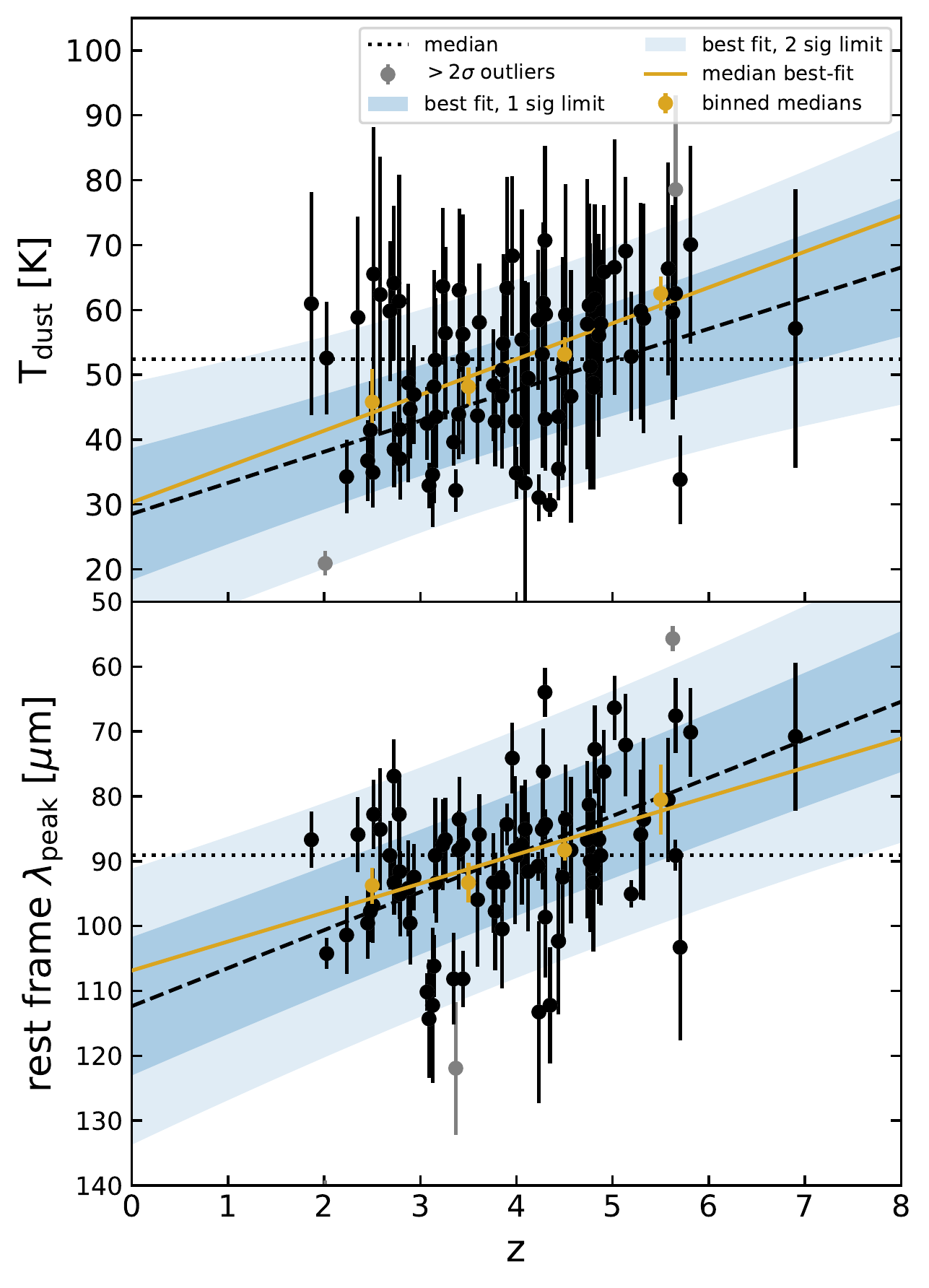}
	\caption{\textbf{Top:} Fitting the dust temperature distribution (\textit{blue}) yields a non-zero linear slope.  However, taking into account the intrinsic scatter weakens the signficance of such a slope, as reflected in the confidence intervals shown.  Outliers ($>$$2\sigma$) are shown in grey, and removing them does not appreciably change the fitted slope.  Binning the dust temperatures with bin widths $z$$=$$1$ also yields a linear dependence (\textit{yellow}).  \textbf{Bottom:} Peak wavelength ($\lambda_{\mathrm{peak}}$) versus redshift.  The wide sampling of the SED makes the $\lambda_{\mathrm{peak}}$ constraint insensitive to the SED fitting procedure.  The same analysis was repeated and the evolution with redshift is weaker, but still produces a non-zero slope.}
	\label{fig:Td_v_z}
\end{figure}
 
With the first spectroscopically complete sample of DSFGs in hand and excellent FIR photometric coverage for all sources, we are now in a position to investigate whether the typical FIR properties of DSFGs evolve with redshift. For example, observations of `normal' star-forming galaxies~\citep{magdis12, magnelli14} and backwards evolution modeling~\citep{bethermin12a, schreiber18} imply that higher redshift sources should exhibit progressively warmer dust temperatures.   One possible explanation for this effect would be that \tdust{} is proportional to the $\lir/M_{\rm dust}$ ratio, which is proportional to the specific star formation rate~\citep{narayanan18,liang19,ma19}.  While these studies focused on dust temperature evolution in lower-SFR objects than our sample, determining whether DSFGs show temperature evolution has been difficult due to the highly uncertain photometric redshifts, confusion limits, and possible selection effects towards more luminous (and therefore, potentially warmer) galaxies in flux-limited surveys.  

We examine this relationship by considering the dust temperatures from Sec.~\ref{sec:sedresults} versus the spectroscopic redshifts measured in Sec.~\ref{sec:specresults}, shown in Fig.~\ref{fig:Td_v_z}.  In order to fit the data, we perform a maximum likelihood estimation assuming a linear model with an extra Gaussian variance term to account for intrinsic scatter.  Using this method, the best fit line is given in Fig.~\ref{fig:Td_v_z} as the dashed line, with a slope of $4.7$$\pm$$1.5\1{K/z}$.  However, the fitted intrinsic scatter is substantial ($8.4$$\pm$$1.3\1{K}$).  While there is evidence of temperature evolution, the large amount of intrinsic scatter implies the fit is also consistent with very shallow to no evolution with redshift.  

Performing a least squares fit without the intrinsic scatter term will also give a non-zero linear dependence.  However, such a fit is largely driven by the presence of significant outliers (e.g. SPT0452-50 with \tdust$=$$21\1{K}$ and SPT0346-52 with \tdust$=$$79\1{K}$).  One way to mitigate the presence of outliers in this approach is to consider the binned temperature distribution.  Sources were binned with a bin width of $\Delta z$$=$$1$ and the median dust temperature was adopted, with bootstrapped errors representing the error on \tdust.  Though fitting these points again presents a positive slope of $1.9 \pm 0.5\1{K/z}$, a one-sided reduced $\rchi^2$ test yields a p-value of $0.275$, which is not statistically significant ($<$$1 \sigma$).  Moreover, hypothesis testing reveals that a line with no slope (p-value $=$$0.184$) is favored.  

Since the dust temperature is highly sensitive to the exact SED fit function, we also examine the rest frame $\lambda_{\mathrm{peak}}$.  The wide sampling of the SED makes the $\lambda_{\mathrm{peak}}$ constraint insensitive to the SED fitting function.  The $\lambda_{\mathrm{peak}}$ values used were extracted by fitting a modified blackbody fit with an added power law (individual fits shown in Fig.~\ref{fig:sedthumbs} in blue).  Repeating the linear fit with maximum likelihood estimation, we find that $\lambda_{\mathrm{peak}}$ also exhibits a non-zero slope of $-5.9$$\pm$$1.3\1{\mu m/z}$.  However, like dust temperature, any evidence of evolution with redshift is obscured by the $9.5$$\pm$$1.1\1{\mu m}$ intrinsic scatter.  Like dust temperature, the presence of outliers (e.g. SPT0245-63, SPT0529-54 and SPT0452-50) could affect the significance of a slope, so we performed the same binned analysis.  While the binned medians yielded a negative slope of $-3$$\pm$$2\1{\mu m/z}$, it again had a low significance ($<$$1\sigma$).  

Taken together, SPT sample's dust temperature and peak wavelengths show evidence of the thermal emission peak evolving with redshift, proportional to $(1$$+$$z)$, though it is not strong evidence.  The intensely star forming nature of our sample implies that our sources are starburst galaxies.  Unlike the `normal' main sequence galaxies, starbursts have been shown to have almost constant radiation field intensities from $z$$=$$0$ to $z$$\sim$$2$~\citep{bethermin15a, schreiber18}.  However, recent observations of GN20~\citep{cortzen20} suggest that optically thick dust \textit{could} obscure temperature evolution in [CI] at high-redshift.  While it is possible that luminous starburst galaxies could evolve in dust temperature versus redshift, strong evidence for temperature evolution is obscured by the large scatter on \tdust.

%========================================================
% Section 4.3 SPT Sources are Extreme ULIRGs/HyLIRGs
\subsection{SPT Sources are Extreme ULIRGs/HyLIRGs} \label{sec:ulirgs}
%========================================================
A consequence of modifying our SED fitting function to have the optically thick transition wavelength ($\lambda_0$) vary as a function of dust temperature is that the fitted dust temperatures obtained are systematically higher than if $\lambda_0$ was fixed to $100\1{\mu m}$, as in~\citet{greve12}. Fitting the ALESS photometry with this same SED fit gives a median dust temperature of \tdust$=$$54$$\pm$$4\1{K}$.  While this dust temperature is significantly higher than what is commonly reported for DSFGs, which typically range between $30$$-$$40\1{K}$~\citep{chapman05, greve12, swinbank14, dacunha15}, this is consistent with the values we obtain for the SPT DSFGs (\tdust$=$$52$$\pm$$2\1{K}$).  

We use the procedure outlined in Sec.~\ref{sec:srcprops} to calculate dust mass for the SPT sources and show the demagnified, intrinsic values throughout this section.  Molecular gas mass was obtained by assuming a constant gas-to-dust ratio (taken to be $100$; e.g.~\citealt{casey14}).  Because the FIR peak for the ALESS sample is not always well-constrained, fitting the full IR range as described in Sec.~\ref{sec:sed} is not always possible.  We instead adopt the values derived from \texttt{MAGPHYS} for comparison purposes~\citep{dacunha15}, which takes into account the available photometry, as detailed in \citep{swinbank14}.  When a direct SED fit is possible, the two methods produce results that agree within the stated errors, with no bias.  We derive a median SPT dust mass of $1.4$$\times$$10^{9}\1{M_{\odot}}$, compared with a median ALESS dust mass of $7.3$$\times$$10^{8}\1{M_{\odot}}$.  This implies that the SPT-selected sources have a factor of two higher dust reservoirs than the ALESS sample, despite being earlier in cosmic history (Fig.~\ref{fig:td_md_z_lfir_4panel}).  

% SPT sources are brighter
The SPT sources exhibit higher FIR luminosities than their unlensed analogs.  Because of the relatively shallow SPT survey limit, shown in Fig.~\ref{fig:lfir_v_zTd}, the SPT is sensitive to all but the brightest sources.  Generally, these sources are lensed and have apparent luminosities that are an order of magnitude larger than unlensed sources. However, even after correcting for the gravitational magnification, the intrinsic SPT FIR luminosities range from $1.0$$-$$37$$\times$$10^{12}\1{\lsol}$ with a median of $7.1(5)$$\times$$10^{12}\1{\lsol}$, compared to the median ALESS value of $\mathrm{L_{FIR}}$$=$$2.3$$\times$$10^{12}\1{\lsol}$.  If the median magnification of the SPT sources lacking lens models was raised from $\langle \mu_{870\1{\mu m}} \rangle$$=$$5.5$ to $\sim$$18$, these discrepancies would also be resolved. We find this possibility unlikely, however, as the SPT sources with lens models from high resolution ALMA imaging presented in \citet{spilker16} were effectively drawn at random from the larger SPT sample of 81 sources presented here. Thus, SPT sources, in addition to being strongly lensed, are also intrinsically more luminous than standard DSFGs found in blank-field surveys.

%---------------------------------------------
% FIGURE 12 Mgas v SFR
%---------------------------------------------
\begin{figure}[!h]
	\centering
	\includegraphics[width=\columnwidth]{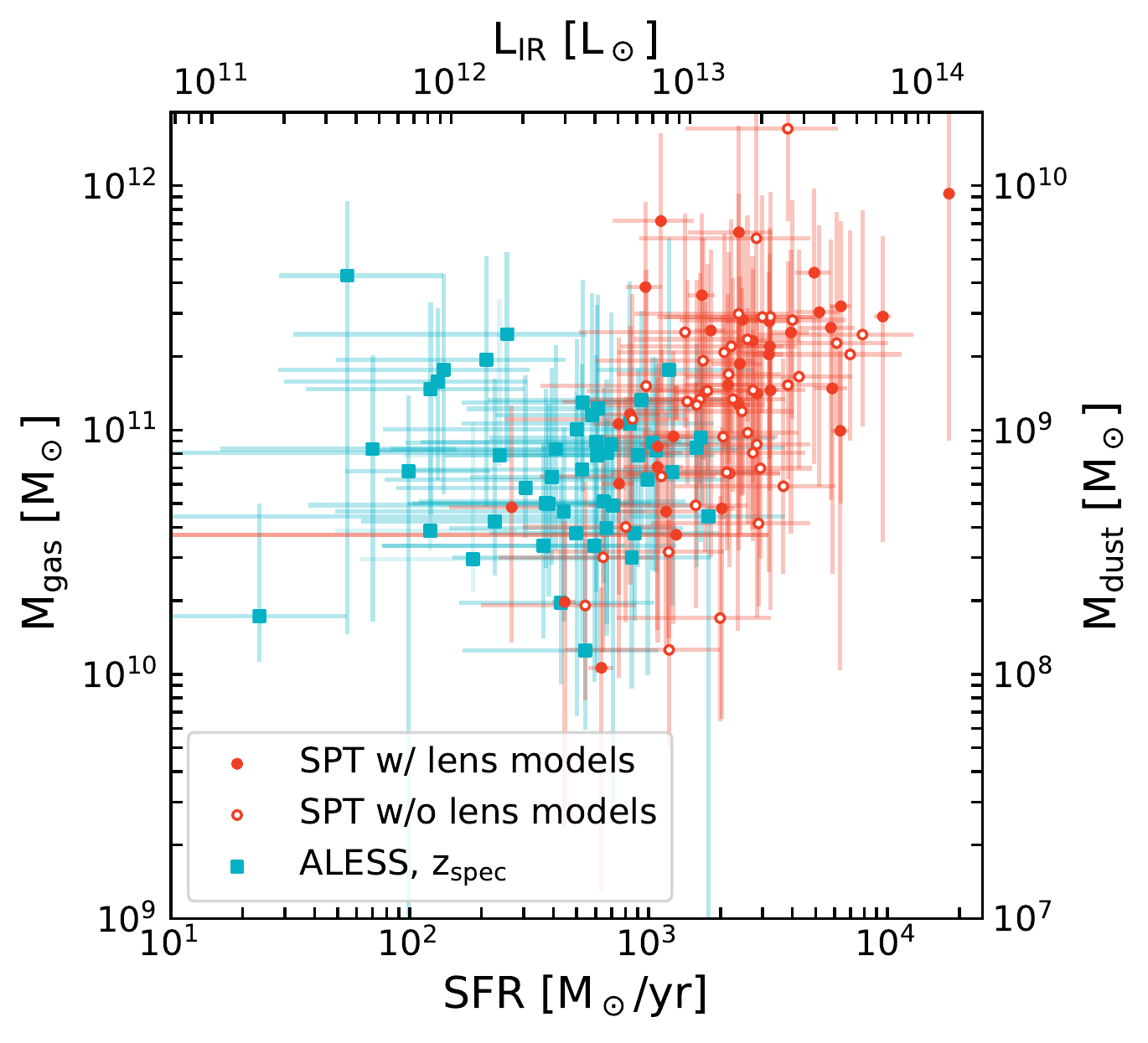}
	\caption{Gas mass versus SFR for the demagnified SPT-selected sources and the ALESS sample.  Though the SPT-selected sources have comparable amounts of gas mass to the ALESS sample, the SPT sources also experience higher star formation rates. }
	\label{fig:mgas_v_sfr}
\end{figure}

\begin{figure}[!h]
	\centering
	\includegraphics[width=\columnwidth]{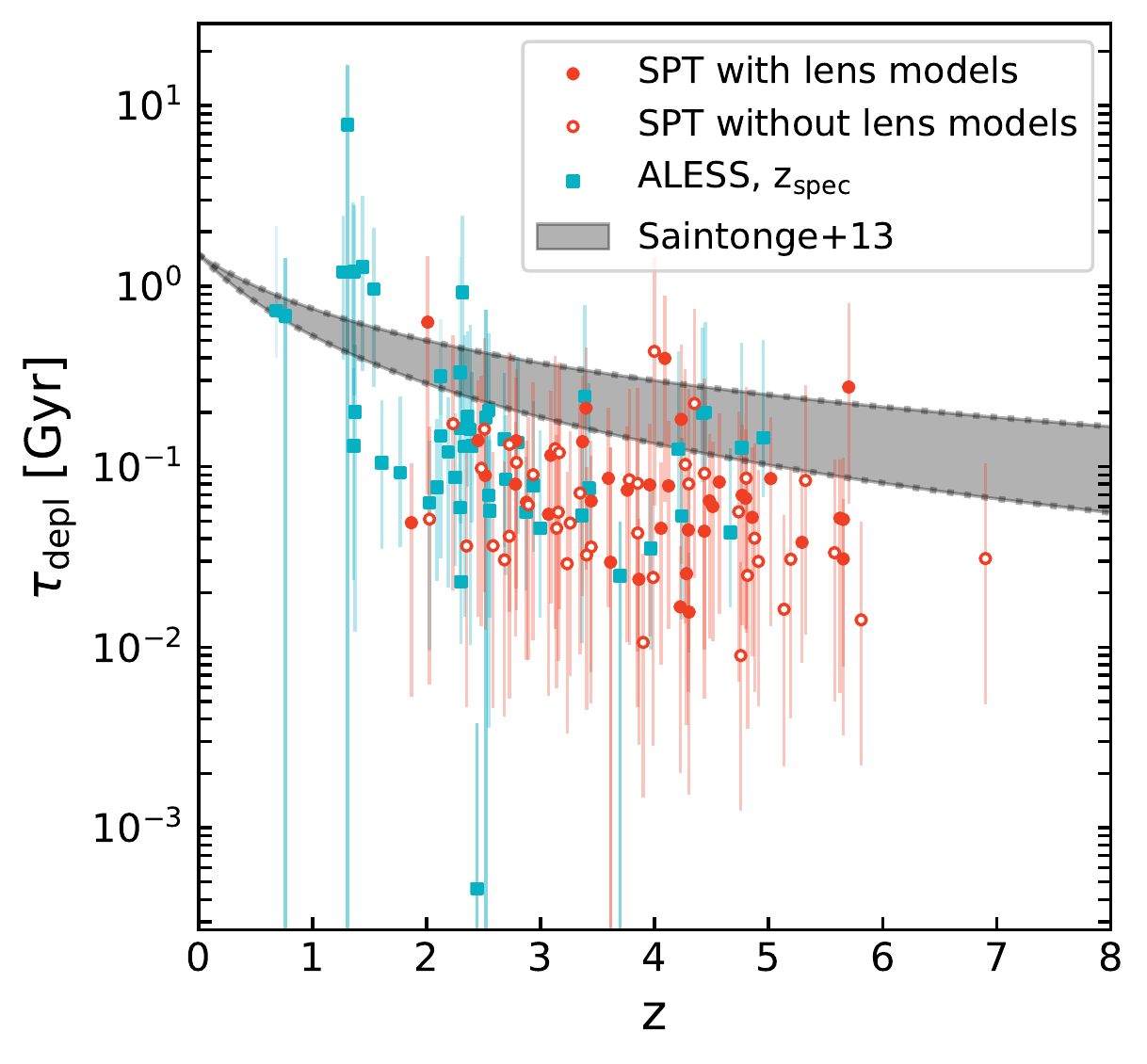}
	\caption{Depletion time ($\tau_{\mathrm{depl}}$) versus redshift for the SPT-selected sample and ALESS sample. The extreme properties of the SPT sample result in short depletion times, which imply that the SPT sources are experiencing a rapid burst of star formation.  The typical values presented in \citet{saintonge13} for main sequence galaxies are shown in grey.  
	}
	\label{fig:Tdepl_v_z}
\end{figure}

Taking into account the full IR range of the SED, we find a median $L_{\rm IR}$$=$$1.5(1)$$\times$$10^{13}\1{\lsol}$, which corresponds to a median SFR of $2.3(2)\times 10^3\1{\msol yr^{-1}}$ for the SPT sample. Fig.~\ref{fig:mgas_v_sfr} shows the gas mass (derived from $M_{\rm dust}$) verses SFR (derived from $L_{\rm IR}$). The SPT sample exhibits a correlation between the two quantities.  By taking the ratio of gas mass to SFR, we calculate the gas depletion times ($\tau_{\mathrm{depl}}$) shown in Fig.~\ref{fig:Tdepl_v_z}.  Because the magnification correction enters into both the SFR and gas mass calculations in the same way, it cancels out, assuming no differential magnification \citep{hezaveh12a}.  While ($\tau_{\mathrm{depl}}$) in `normal' main sequence galaxies are known to evolve with redshift~\citep{saintonge13, genzel15, tacconi18}, both the ALESS and SPT samples show comparable depletion times below the main sequence.  We also find an evolution in $\tau_{\mathrm{depl}}$ with redshift, which was already observed for main sequence galaxies.  Alternatively, this effect could also be the result of a strong redshift evolution of metallicity, with the highest redshift galaxies being the most metal-poor.  This would cause an underestimation of gas mass, which would also explain the lower depletion times.  In order to address this properly, however, we would need stellar masses for our systems.  Because of the high-redshift dust obscured nature of these sources, this requires the use of future facilities such as the \textit{James Webb Space Telescope}.  

Though the dust temperatures of the SPT and ALESS sources are similarly distributed, with medians of \tdust$=$$52(2)\1{K}$ and \tdust$=$$54(4)\1{K}$ respectively, SPT sources are on average more luminous than their unlensed analogs, even after accounting for the gravitational lensing.  The simple blackbody relationship would imply that these sources \textit{should} exhibit similar luminosities, if they were the same sizes.  The SPT sources have a median radius of $1.0(1)\1{kpc}$, while ALESS sources have a median of $0.46(4)\1{kpc}$.  The higher intrinsic luminosities could therefore stem from larger sizes, as discussed in Sec.~\ref{dis:selection}.  This fact, coupled with the larger dust masses, implies that the SPT sources could have larger gas reservoirs, which is supported by their systematically higher dust masses.

%---------------------------------------------
% FIGURE 13 Depletion time
%---------------------------------------------

%========================================================
% Section 4.4 High-z tail
\subsection{High-z Tail}\label{sec:highztail}
%========================================================
As previously discussed, $850\1{\mu m}$-selected samples of DSFG from blank fields (e.g. \citealt{chapman05, swinbank14}) peak at a lower redshift than the SPT sample, typically between $z$$\sim$$2.3$$-$$2.9$.  This discrepancy is attributed to selection effects, as discussed in the previous section (Sec.~\ref{dis:selection}).
 While the SPT sample includes some sources equivalent to those selected in the blank-field surveys, it also includes the previously undiscovered high redshift and extreme tail of DSFGs.  
In addition to discovering the highest redshift DSFG to-date (SPT0311-58 at $z=6.9$; \citealp{strandet17, marrone17}), and with 12 sources at $z$$>$$5$, the SPT DSFG catalog contains roughly half of the DSFGs discovered at $z$$>$$5$~\citep{oteo17, asboth16, walter12, rawle14, riechers13, dowell14, strandet17, riechers14,  riechers17, pavesi18, jin19, fudamoto17, zavala18, riechers20}. 
%\footnote{Note that SPT0311-58 consists of two sources \citep{marrone17} and SPT2349-56 contains 6 sources with L$_{\rm FIR}>3\times10^{12}\lsol$ \citep{hill20}}

%---------------------------------------------
% FIGURE 14 demagnified L v. z and T_d
%---------------------------------------------
\begin{figure}[h]
	\centering
	\includegraphics[width=\columnwidth]{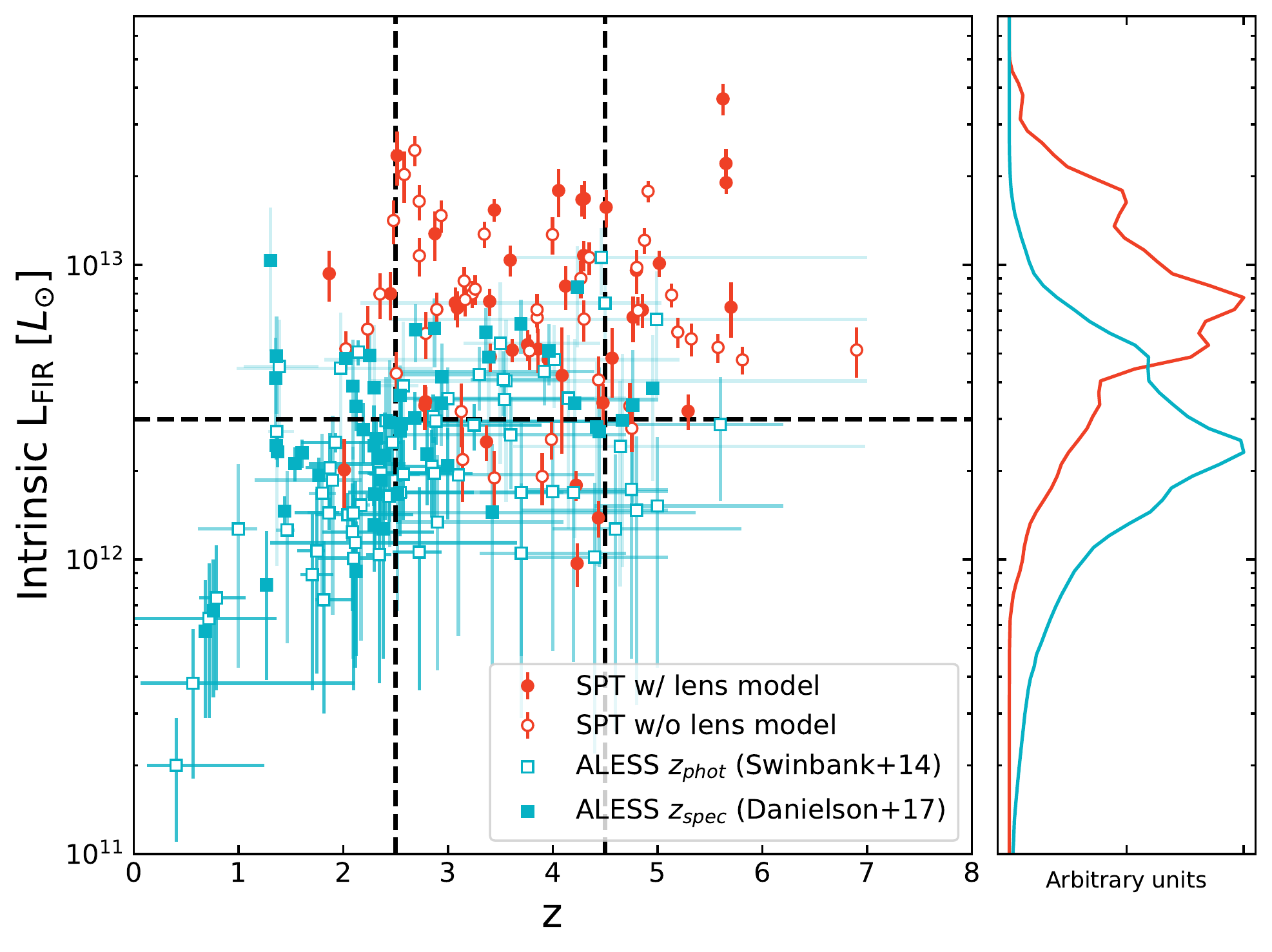}
	\caption{Intrinsic luminosity as a function of redshift for DSFGs from SPT and ALESS.  SPT sources are demagnified using lens models from~\citet{spilker16} where available.  The dashed horizontal line represents the $\mathrm{L_{FIR}}$ above which the SPT and ALESS samples are complete, while the vertical lines correspond to $2.5$$<$$z$$<$$4.5$, where the surveys have equal source density.}
	\label{fig:lfir_v_zTd_demag}
\end{figure}

\begin{figure}[t]
	\centering
	\includegraphics[width=\columnwidth]{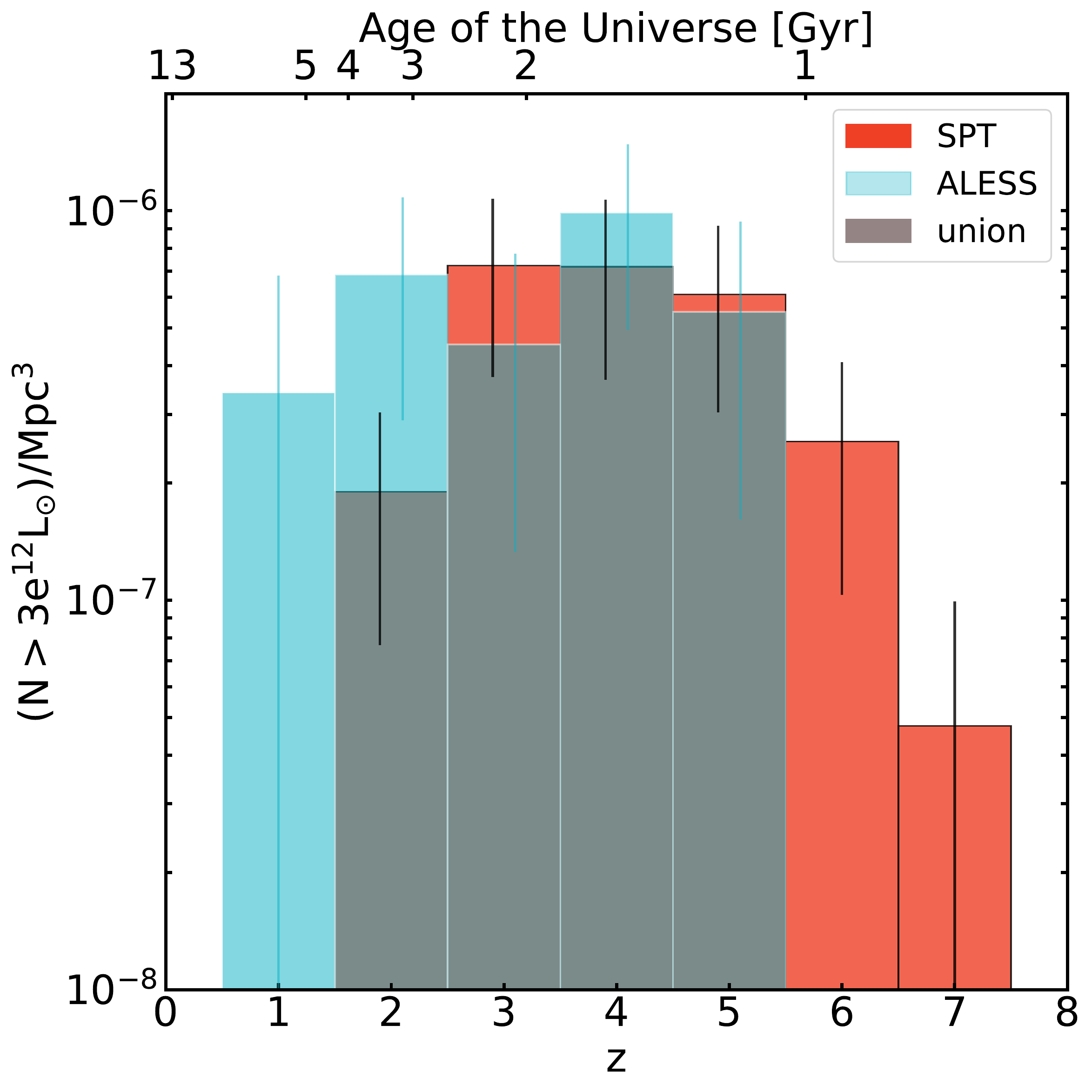}
	\caption{The spatial density of ALESS and SPT sources.  Because the ALESS survey has a well-defined survey volume, we use it to correct for the unknown SPT survey volume.  We assume both surveys are complete above a FIR luminosity threshold of \lfir$>3\times 10^{12}\1{L_\odot}$ and use the region where both surveys have equal source density ($2.5 < z < 4.5$) to scale the SPT source density.  In this plot, union simply represents the overlapping source density regions of both surveys.}
	\label{fig:spt_aless_rho}
\end{figure}

A way to assess the relative rarity of SPT-selected sources is to consider the space density.  However, because the majority SPT-selected sources are lensed, the volume probed changes as a function of lensing and the effective volume is difficult to calculate.  In order to estimate the survey volume, we consider the ALESS survey, which has a well-defined survey area and can be considered complete above a luminosity threshold of $>$$3$$\times$$10^{12}\1{\lsol{}}$, as shown in Fig.~\ref{fig:lfir_v_zTd_demag}.  This particular luminosity threshold was selected because it retains most of the SPT sources, while ensuring there are sufficiently many ALESS sources to constitute an accurate comparison sample.  We assume the two surveys have equal source density in the region $2.5$$<$$z$$<$$4.5$ for $>$$3$$\times$$10^{12}\1{\lsol{}}$.  We use this region to estimate the volume for the SPT survey and compare the two redshift distributions directly in Fig.~\ref{fig:spt_aless_rho}.  The redshift region $2.5$$<$$z$$<$$4.5$ was chosen because it is the largest region covered by both surveys, which helps mitigate the effects of cosmic variance.  The error bars for SPT are taken to the statistical errors from SPT added in quadrature with the ALESS errors in order to account for the scaling.  Error estimation using bootstrapping produced similar results to the errors shown.  Various combinations of luminosity threshold and redshift scaling region were tested, and the one chosen proved to be the best compromise between catalog completeness and constraining power.

\begin{figure}[t]
	\centering
	\includegraphics[width=\columnwidth]{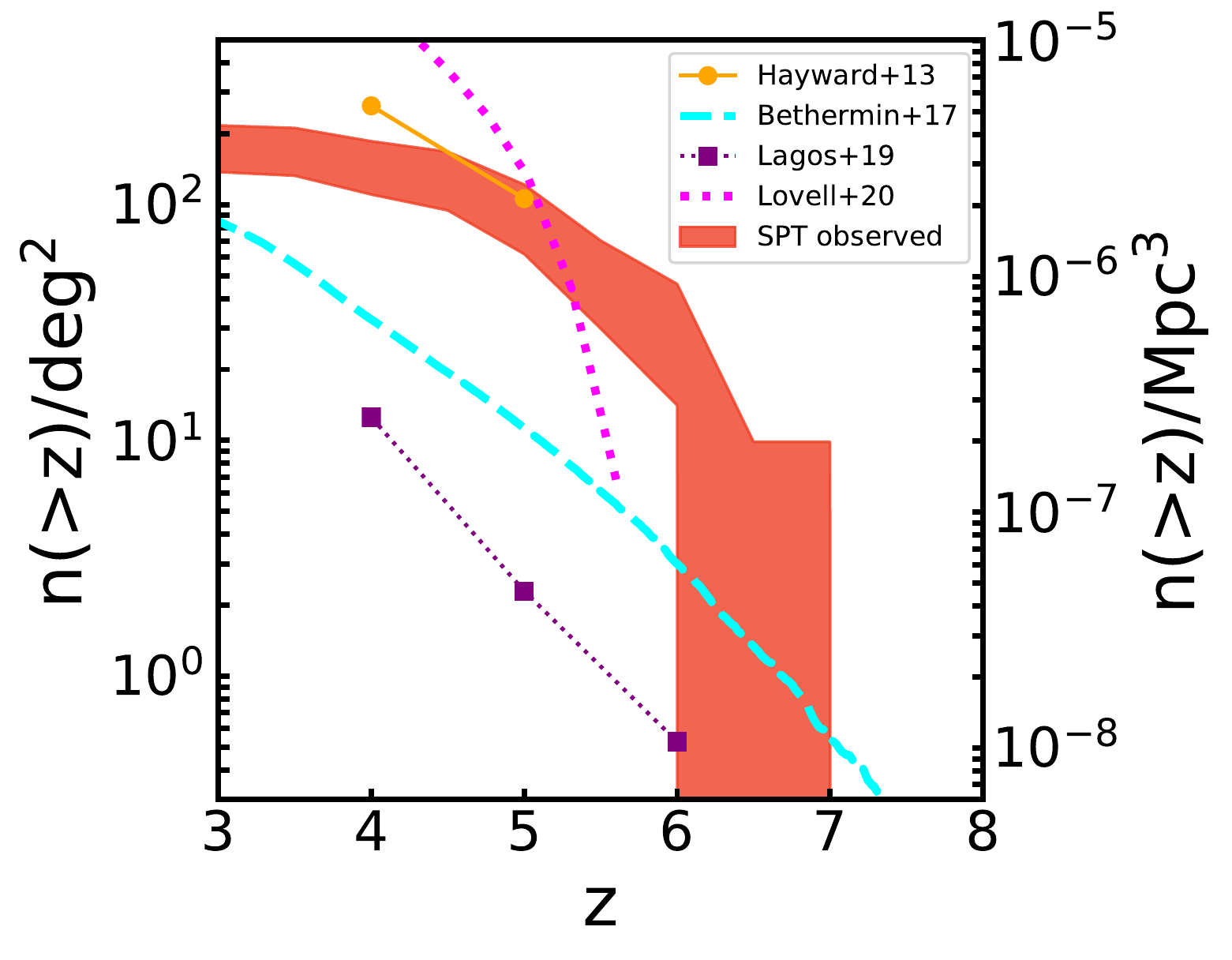}
	\caption{Spatial densities of high-redshift luminous galaxies with $\lfir>3\times10^{12}\lsol$ measured in this work from the SPT survey compared to model predictions from \citet{hayward13, bethermin17,lagos19} and \citet{lovell20}.} 
	\label{fig:spt_rho}
\end{figure}

Because the SPT distribution is both selected at long wavelength and spectroscopically complete, we have measured the high redshift tail of the overall DSFG distribution.  By summing the contributions in Fig.~\ref{fig:spt_aless_rho}, we infer that the expected number of DSFGs above $z$$>$$4$ is $2.6(7)$$\times$$10^{-6}\1{/Mpc^3}$ above $3$$\times$$10^{12}\1{\lsol{}}$.  We can also predict the number of $z$$>$$6$ sources above $3$$\times$$10^{12}\1{\lsol{}}$, which we take to be $0.95(1.0)$$\times$$10^{-7}\1{/Mpc^3}$, but this number should be used with caution, as it is based upon just one source~\citep{marrone18}.  To place this number in context, we compare with other observations~\citep{coppin09, ivison16, cooke18} and theoretical predictions~\citep{hayward13, bethermin17,lagos19, lovell20} in Tab.~\ref{tab:density} and Fig.~\ref{fig:spt_rho}.  The stated number densities are taken to have an upper limit of $z$$=$$8$ and the errors are again calculated using the statistical errors from both the SPT and ALESS surveys in quadrature.  

The \citet{hayward13} model combines a semi-empirical model with 3D hydrodynamical simulations and a 3D dust radiative transfer.  Strong lensing is not included in the modeling and the model predicted $dn/dz$ is determined using sources with $S_{1\1{mm}}$$>$$1\1{mJy}$, consistent with the expected intrinsic flux densities of our sample.  While the \citet{hayward13} model over-predicts the number of $z$$>$$4$ and $z$$>$$5$ galaxies, it does not predict any galaxies at $z$$>$$6$, which is contrary to our observations.  

As previously discussed, \citet{bethermin15b} model is an empirical model, which starts from the observed FIR number counts.  This model also includes the effects of magnification by strong lensing, so it can directly predict the $dn/dz$ for the SPT sample.  We find that the updated \citet{bethermin17} model is, in general, in good agreement with the SPT measurement. 

The \texttt{Shark} semi-analytic model~\citet{lagos18} is also compared in Tab.~\ref{tab:density} and Fig.~\ref{fig:spt_rho}.  The number densities of galaxies with \lir$>$$3\times10^{12}\1{\lsol}$ were computed using the simulated lightcone with an area of $107.9\1{deg^2}$ and covers a redshift range $0$$\le$$z$$\le$$8$, and includes all galaxies in the model with stellar masses $>$$10^8\1{ M_{\odot}}$ and with a (dummy) AB magnitude (of mass-to-light-ratio = 1) $\le 32$~mags. Note that here, \lir{} corresponds to the total luminosity that is re-emitted by dust in Shark at wavelengths $\le$$1000\1{\mu m}$.  While this model under-predicts the number of sources at $z$$>$$4$ and $z$$>$$5$, it exactly matches at $z$$\sim$$6$.

\citet{lovell20} modelled the $S_{\rm 850\1{\mu m}}$ emission using a full radiative transfer code coupled with a full cosmological lightcone. To compare with our model we have selected sources at $S_{\rm 850\1{\mu m}}>3$mJy, which corresponds to roughly $\lfir\sim3\times10^{12}\lsol$. The model is in good agreement with the SPT measurement at $z$$\sim$$6$, but over-predicts the density of sources at $z$$\sim$$4$.

With SPT, we have measured the ``high-redshift tail" of the distribution of luminous DSFGs. The SPT measurement for $z$$>$$4$ luminous sources lies within the range of model predictions and previous measurements and can already constrain models of galaxy evolution. The next-generation mm-wave surveys, combined with spectroscopic followup, will further improve this measurement and extend the redshift constraints out to $z$$\sim$$8$. 

%---------------------------------------------
% FIGURE 15 ALESS/SPT SMG source density
%---------------------------------------------

%---------------------------------------------
% TABLE 2 Source densities of literature/theory values
%---------------------------------------------
\begin{table*}
\scriptsize
\begin{center}
\begin{tabular}{ |r c l l l l l l l| } 
\hline
Reference & Selection & \lfir{} limit & \multicolumn{2}{c}{N($4$$<$$z$$<8$)} & \multicolumn{2}{c}{N($5$$<$$z$$<8$)} & \multicolumn{2}{c|}{N($6$$<$$z$$<8$)}\\
                        & [$\1{\mu m}$]         & [$\times$$10^{12}$$\1{L_\odot}$]             &[$\times$$10^{-7}$$\1{Mpc^{-3}}$] &[$\1{deg^{-2}}$] &[$\times$$10^{-7}$$\1{Mpc^{-3}}$] &[$\1{deg^{-2}}$] &[$\times$$10^{-7}$$\1{Mpc^{-3}}$] &[$\1{deg^{-2}}$] \\
\hline
\citealt{coppin09}        & 850       & $>2.3 $ & $>$$1.5$          & $>$$5.6$   & -- & -- & -- & -- \\ 
\citealt{ivison16}        & 250-500   & $>2.3$  & $6$               & $22$      & -- & -- & -- & -- \\
\citealt{cooke18}$^*$     & 850       & $>3.2$  & $>$$50$$-$$60$    & $>$$187$$-$$225$        & -- & -- & -- & -- \\
\textit{this work}        & 1400      & $>3$    & $26(7)$           & $131(36)$  & $10(4)$ & $50(20)$ & $0.95(1.0)$ & $4.7(5.1)$ \\

\hline
\citealt{hayward13}      &  --        & $>1.4 $ & $70$      & $262$     & $40$   & $106$    & $0$      & $0$ \\ 
\citealt{bethermin17}    &  --        & $>3 $   & $8.66$    & $32.4$    & $4.29$ & $11.4$   & $1.80$    & $3.01$ \\ 
\citealt{lagos19}        &  --        & $>3$    & $3.34$    & $12.5$    & $0.87$ & $2.30$   & $0.32$  &  $0.53$ \\ 
\citealt{lovell20}       &  --        & $>3$    & $192$     & $524$     & $50.4$ & $138$    & $3.15$ & $8.60$ \\ 
\hline
\end{tabular}
\caption{Observed and modeled source densities of high-redshift luminous DSFGs.  
* In \citet{cooke18}, this estimate is for the range $4$$<$$z$$<$$5$.  }
\label{tab:density}
\end{center}
\end{table*}

%---------------------------------------------
% FIGURE 16 SPT SMG source density
%---------------------------------------------

%%%%%%%%%%%%%%%%%%%%%%%%%%%%%%%%%%%%%%%%%%%%%%%%%%%%%
% SECTION 5 										%
% Conclusions										%
\section{Conclusions} \label{sec:conclusions}		%
%%%%%%%%%%%%%%%%%%%%%%%%%%%%%%%%%%%%%%%%%%%%%%%%%%%%%

In this paper we present the final spectroscopic redshift distribution for the SPT-selected DSFG sample, as well as the complete FIR, sub-mm, and mm photometry. Our main results and conclusions are summarized as follows:

\begin{itemize}
\item Spectroscopic redshifts have been obtained for all 81 SPT-selected sources.  This survey has the highest spectroscopic completeness of any sample of high redshift galaxies to-date.  With the high redshift sources presented in this work, the SPT DSFGs represent roughly half of the $z$$>$$5$ DSFGs in literature to-date, which will be vital in constraining models of massive galaxy formation.  
\item Fitting the photometry by using the empirical relationship between $\lambda_{0}$(\tdust) derived in \citet{spilker16} improved the SED fitting and provided better photometric redshifts than the original method presented in \citet{greve12}.  In the absence of observationally expensive spectroscopic data, photometric redshifts can give a rough estimate of redshift.  They can also serve to identify the correct redshift when multiple spectroscopic solutions are available.  In instances of limited photometric data, we also demonstrated that the $S_{350\1{\mu m}}/S_{870\1{\mu m}}$ ratio can also serve as a rough redshift indicator.  Taken together, these results demonstrate the utility of photometric data in determining redshift and can serve to guide prioritization of targets for followup, even when limited IR/mm data is available.  
\item The dust temperatures obtained in Sec.~\ref{sec:sedresults} suggest a mild correlation between redshift and dust temperature (e.g.~\citealt{cooke18}), but this result is not yet statistically significant.  
\item Due to gravitational lensing, the apparent luminosity of SPT sources is an order-of-magnitude greater than the unlensed DSFG population.  However, the intrinsic luminosities are still significantly higher than the unlensed $850\1{\mu m}$-selected population, with higher dust masses and lower depletion times for the SPT sources. Thus, even after accounting for magnification due to strong gravitational lensing, SPT-selected DSFGs are more extreme than those typically found in blank-field surveys. 
\item Even though the SPT survey encompassed a contiguous 2500 deg$^2$ area and the sample is flux-limited, the gravitational lensing intrinsic to the SPT sample makes it difficult to estimate the total volume probed by the survey. Scaling the SPT sample to that of the ALESS sample allows us to estimate the source density of luminous DSFGs at high redshift. From our data and analysis, we estimate $4.7\pm5.1$$\1{deg^{-2}}$ and $0.95$$\pm$$1$$\times$$10^{-7}$ Mpc$^{-3}$ DSFGs with \lfir$>$$3$$\times$$10^{12}$$\1{L_\odot}$ at $z$$>6$.
\item While the $3\1{mm}$ data was obtained in this paper, it will serve as a starting point for several papers currently in preparation.  The $3\1{mm}$ line fluxes from high-J CO lines will be combined with the line fluxes from low-J CO observations with ATCA.  Spectral Line Energy Distributions will be fitted for all sources, which is currently in preparation.  The composite $3\1{mm}$ spectrum, shown in the bottom panel of Fig.~\ref{fig:zshifted_spec}, will be further examined in Reuter~\etal~in prep.  Finally, ALMA $870\1{\mu m}$ imaging has been obtained for the complete SPT sample, enabling lens models to be constructed for the complete catalog of SPT DSFGs.  The fitted lens models and their associated properties, including source size and magnifications, will be studied in Reuter~\etal~in prep.
\item Though this SPT-selected redshift catalog is now complete, the next generation of CMB experiments are already in operation.  SPT-3G ~\citep{benson14} has already begun the process of surveying $1500\1{deg^2}$ field with a sensitivity of $\sim$$5$ times the original SPT-SZ survey and will detect two orders-of-magnitude more sources, extending all the way to $z$$>$$8$.  CMB-S4~\citep{abazajian19}, still in the design phase, is projected to find an order-of-magnitude more DSFGs than SPT-3G.  The next generation of mm-wave surveys with spectroscopic followup will be key to uncovering the the earliest stages of the dust-obscured Universe.  Future upgrades (e.g. \citealt{mroczkowski19}) to significantly extend the instantaneous bandwidth of the ALMA receivers will be crucial for ascertaining the redshifts for the next generation of mm surveys.  
\end{itemize}

\software{\texttt{MAGPHYS}~\citep{dacunha15}, \texttt{CASA}~\citep{mcmullin07, petry12}, \texttt{Miriad}~\citep{sault95}, \texttt{BoA}~\citep{schuller10}, \texttt{HIPE}~\citep{balm12,  ott11}, \texttt{emcee}~\citep{foremanmackey13}}

\smallskip

%%%%%%%%%%%%%%%%%%%%%%%%%%%%%%%%%%%%%
% ACKNOWLEDGEMENTS									 %
\acknowledgments 											 %
%%%%%%%%%%%%%%%%%%%%%%%%%%%%%%%%%%%%%
We thank the anonymous referee whose careful reading and insightful suggestions greatly improved this paper.
We thank A.~M.~Swinbank, C.~Lagos and C.~Lovell for useful discussions.
The SPT is supported by the NSF through grant OPP-1852617.
D.P.M. J.D.V., K.C.L., K.P. and S.J. acknowledge support from the US NSF under grants AST-1715213 and AST-1716127.
S.J. and K.C.L acknowledge support from the US NSF NRAO under grants SOSPA5-001 and SOSPA4-007, respectively.
J.D.V. acknowledges support from an A. P. Sloan Foundation Fellowship.
J.S.S. acknowledges support through the NASA Hubble Fellowship grant \#HF2-51446  awarded  by  the  Space  Telescope  Science  Institute,  which  is  operated  by  the  Association  of  Universities  for  Research  in  Astronomy,  Inc.,  for  NASA,  under  contract  NAS5-26555.  
M.A. has been supported by the grant ``CONICYT+PCI+REDES 19019."
The National Radio Astronomy Observatory is a facility of the National Science Foundation operated under cooperative agreement by Associated Universities, Inc.
D.N. acknowledges support from the US NSF under grant 1715206 and Space Telescope Science Institute under grant AR-15043.0001 
This paper makes use of the following ALMA data: ADS/JAO.ALMA\#2011.0.00957.S, ADS/JAO.
\\ALMA\#2012.1.00844.S, ADS/JAO.ALMA \\
\#2015.1.00504.S, ADS/JAO.ALMA \#2016.
\\ 1.00672.S, ADS/JAO.ALMA\#2018.1.01254.S, \\
ADS/JAO.ALMA\#2019.1.00486.S.  ALMA is a partnership of ESO (representing its member states), NSF (USA) and NINS (Japan), together with NRC (Canada), MOST and ASIAA (Taiwan), and KASI (Republic of Korea), in cooperation with the Republic of Chile. The Joint ALMA Observatory is operated by ESO, AUI/NRAO and NAOJ.
This work is based in part on observations made with \textit{Herschel} under program IDs:  OT1\_jvieira\_4 and OT2\_jvieira\_5. \textit{Herschel} is a European Space Agency Cornerstone Mission with significant participation by NASA. 
We also use data from the Atacama Pathfinder Experiment under program IDs: E-086.A-0793A-2010, M-085.F-0008-2010, M-087.F- 0015-2011, M-091.F-0031-2013, E-094.A-0712A-2014, M-095.F-0028-2015, E-096.A-0939A-2015. APEX is a collaboration between the Max-Planck-Institut f{\"u}r Radioastronomie, the European Southern Observatory, and the Onsala Space Observatory. 
The Australia Telescope Compact Array is part of the Australia Telescope National Facility which is funded by the Australian Government for operation as a National Facility managed by CSIRO.

%%%%%%%%%%%%%%%%%%%%%%%%%%%%%%%%%%%%%
% APPENDIX							%
\appendix						    %
%%%%%%%%%%%%%%%%%%%%%%%%%%%%%%%%%%%%%

%========================================================
% Section A All SPT sources and 3mm positions
\section{SPT sources and positions}
\label{ap:srcpos}
%========================================================
\renewcommand{\theHtable}{A.\arabic{table}}%<---!!!!---
\renewcommand{\theHfigure}{A.\arabic{figure}}%<---!!!!---
\setcounter{table}{0}
\setcounter{figure}{0}
\renewcommand\thetable{\thesection.\arabic{table}}  

We present a total of 81 sources in this work.  Their positions are calculated using $3\1{mm}$ continuum images, and are given in Tab.~\ref{tab:smgpos}.

%========================================================
% Table A.1 All SPT sources and 3mm positions
%========================================================
\startlongtable
\begin{deluxetable*}{llll}
\tabletypesize{\footnotesize}
\tablecaption{Positions of SPT-selected DSFGs, obtained by fitting ALMA $3\1{mm}$ continuum images. \label{tab:smgpos}}
\tablewidth{0pc}
\tablecolumns{4}
\tablehead{
\colhead{Short Name} & \colhead{Source} & \colhead{R.A.} & \colhead{Dec.} \\
}
\startdata
\hline
SPT0002-52 & SPT-S J000223-5232.1 & 0:02:23.76 & -52:31:52.7 \\
SPT0020-51 & SPT-S J002023-5146.5 & 0:20:23.58 & -51:46:36.4 \\
SPT0027-50 & SPT-S J002706-5007.4& 0:27:06.54 & -50:07:19.8 \\
SPT0054-41 & SPT-S J005439-4151.9 & 0:54:39.52 & -41:51:53.2 \\
SPT0103-45 & SPT-S J010312-4538.8 & 1:03:11.50 & -45:38:53.8 \\
SPT0106-64 & SPT-S J010623-6413.08 & 1:06:23.86 & -64:12:50.0 \\
SPT0109-47 & SPT-S J010949-4702.02 & 1:09:49.66 & -47:02:11.4 \\
SPT0112-55 & SPT-S J011207-5516.15 & 1:12:09.03 & -55:16:36.0 \\ 
SPT0113-46 & SPT-S J011308-4617.7 & 1:13:09.01 & -46:17:56.1 \\
SPT0125-47 & SPT-S J012506-4723.7 & 1:25:07.08 & -47:23:56.0 \\
SPT0125-50 & SPT-S J012549-5038.2 & 1:25:48.45 & -50:38:21.0 \\
SPT0136-63 & SPT-S J013652-6307.4 & 1:36:50.28 & -63:07:26.7 \\
SPT0147-64 & SPT-S J014707-6458.9 & 1:47:07.07 & -64:58:52.1 \\
SPT0150-59 & SPT-S J015011-5924.1 & 1:50:09.26 & -59:23:57.1 \\
SPT0155-62 & SPT-S J015547-6250.9 & 1:55:47.75 & -62:50:50.0 \\
SPT0202-61 & SPT-S J020258-6121.1 & 2:02:58.75 & -61:21:11.0 \\
SPT0226-45 & SPT-S J022649-4515.9 & 2:26:49.46 & -45:15:39.0 \\
SPT0243-49 & SPT-S J024307-4915.5 & 2:43:08.81 & -49:15:34.9 \\
SPT0245-63 & SPT-S J024543-6320.7 & 2:45:44.08 & -63:20:38.7 \\
SPT0300-46 & SPT-S J030003-4621.3 & 3:00:04.37 & -46:21:23.8 \\
SPT0311-58 & SPT-S J031134-5823.4 & 3:11:33.14 & -58:23:33.4 \\
SPT0314-44 & SPT-S J031428-4452.2 & 3:14:28.33 & -44:52:22.8 \\
SPT0319-47 & SPT-S J031931-4724.6 & 3:19:31.88 & -47:24:33.6 \\
SPT0345-47 & SPT-S J034510-4725.6 & 3:45:10.77 & -47:25:39.5 \\
SPT0346-52 & SPT-S J034640-5204.9 & 3:46:41.13 & -52:05:01.9 \\
SPT0348-62 & SPT-S J034841-6220.9 & 3:48:42.10 & -62:20:50.9 \\
SPT0402-45 & SPT-S J040202-4553.3 & 4:02:02.19 & -45:53:14.0 \\
SPT0403-58 & SPT-S J040331-5850.0 & 4:03:32.69 & -58:50:08.8 \\
SPT0418-47 & SPT-S J041839-4751.8 & 4:18:39.67 & -47:51:52.5 \\
SPT0425-40 & SPT-S J042517-4036.7 & 4:25:17.38 & -40:36:49.4 \\
SPT0436-40 & SPT-S J043640-4047.1 & 4:36:41.30 & -40:47:08.7 \\
SPT0441-46 & SPT-S J044143-4605.3 & 4:41:44.08 & -46:05:25.6 \\
SPT0452-50 & SPT-S J045247-5018.6 & 4:52:45.83 & -50:18:42.4 \\
SPT0457-49 & SPT-S J045719-4932.0 & 4:57:17.52 & -49:31:51.7 \\
SPT0459-58 & SPT-S J045859-5805.1 & 4:58:59.80 & -58:05:14.1 \\
SPT0459-59 & SPT-S J045912-5942.4 & 4:59:12.34 & -59:42:20.3 \\
SPT0512-59 & SPT-S J051258-5935.6 & 5:12:57.98 & -59:35:42.0 \\
SPT0516-59 & SPT-S J051640-5920.4 & 5:16:37.98 & -59:20:32.1 \\
SPT0520-53 & SPT-S J052040-5329.6 & 5:20:40.14 & -53:29:51.3 \\
SPT0528-53 & SPT-S J052850-5300.2 & 5:28:50.30 & -53:00:20.7 \\
SPT0529-54 & SPT-S J052903-5436.5 & 5:29:03.09 & -54:36:40.2 \\
SPT0532-50 & SPT-S J053250-5047.1 & 5:32:51.04 & -50:47:07.6 \\
SPT0538-50 & SPT-S J053816-5030.8 & 5:38:16.50 & -50:30:52.5 \\
SPT0544-40 & SPT-S J054400-4036.2 & 5:44:01.12 & -40:36:31.2 \\
SPT0550-53 & SPT-S J055001-5356.5 & 5:50:00.56 & -53:56:41.4 \\
SPT0551-48 & SPT-S J055155-4825.2 & 5:51:54.65 & -48:25:01.8 \\
SPT0551-50 & SPT-S J055138-5058.0 & 5:51:39.42 & -50:58:02.0 \\
SPT0552-42 & SPT-S J055226-4243.9 & 5:52:26.52 & -42:44:12.7 \\
SPT0553-50 & SPT-S J055321-5006.9 & 5:53:20.39 & -50:07:11.8 \\
SPT0555-62 & SPT-S J055518-6218.6 & 5:55:16.00 & -62:18:50.2 \\
SPT0604-64 & SPT-S J060457-6447.4 & 6:04:57.57 & -64:47:22.0 \\
SPT0611-55 & SPT-S J061154-5513.9 & 6:11:57.88 & -55:14:09.5 \\
SPT0625-58 & SPT-S J062523-5835.2 & 6:25:22.18 & -58:35:20.0 \\
SPT0652-55 & SPT-S J065206-5516.0 & 6:52:07.24 & -55:16:00.1 \\
SPT2031-51 & SPT-S J203100-5112.1 & 20:30:58.87 & -51:12:25.2 \\
SPT2037-65 & SPT-S J203729-6513.3 & 20:37:31.98 & -65:13:16.8 \\
SPT2048-55 & SPT-S J204824-5520.7 & 20:48:22.87 & -55:20:41.3 \\
SPT2101-60 & SPT-S J210113-6048.9 & 21:01:13.77 & -60:48:56.2 \\
SPT2103-60 & SPT-S J210328-6032.6 & 21:03:30.90 & -60:32:39.9 \\
SPT2129-57 & SPT-S J212912-5702.1 & 21:29:12.53 & -57:01:54.3 \\
SPT2132-58 & SPT-S J213242-5802.9 & 21:32:43.23 & -58:02:46.4 \\
SPT2134-50 & SPT-S J213404-5013.2 & 21:34:03.34 & -50:13:25.2 \\
SPT2146-55 & SPT-S J214654-5507.8 & 21:46:54.02 & -55:07:54.7 \\
SPT2147-50 & SPT-S J214720-5035.9 & 21:47:19.05 & -50:35:53.5 \\
SPT2152-40 & SPT-S J215212-4036.6 & 21:52:12.61 & -40:36:29.5 \\
SPT2203-41 & SPT-S J220316-4133.5 & 22:03:16.85 & -41:33:26.2 \\
SPT2232-61 & SPT-S J223250-6114.7 & 22:32:51.14 & -61:14:44.8 \\
SPT2307-50 & SPT-S J230721-4930.4 & 23:07:25.16 & -50:03:35.4 \\
SPT2311-45 & SPT-S J231148-4546.6 & 23:11:50.52 & -45:46:44.8 \\
SPT2311-54 & SPT-S J231125-5450.5 & 23:11:23.97 & -54:50:30.1 \\
SPT2316-50 & SPT-S J231657-5036.7 & 23:16:58.90 & -50:36:31.7 \\
SPT2319-55 & SPT-S J231922-5557.9 & 23:19:21.64 & -55:57:57.8 \\
SPT2332-53 & SPT-S J233227-5358.5 & 23:32:26.50 & -53:58:39.8 \\
SPT2335-53 & SPT–S J233513-5324.0 & 23:35:13.96 & -53:24:21.0 \\
SPT2340-59 & SPT-S J234009-5943.1 & 23:40:08.96 & -59:43:32.3 \\
SPT2349-50 & SPT-S J234942-5053.5 & 23:49:42.20 & -50:53:30.9 \\
SPT2349-52 & SPT-S J234928-5246.7 & 23:49:29.42 & -52:46:48.6 \\
SPT2349-56 & SPT-S J234944-5638.3 & 23:49:42.78 & -56:38:23.2 \\
SPT2351-57 & SPT-S J235149-5722.2 & 23:51:50.80 & -57:22:18.3 \\
SPT2353-50 & SPT-S J235339-5010.1 & 23:53:39.26 & -50:10:08.1 \\
SPT2354-58 & SPT-S J235434-5815.1 & 23:54:34.31 & -58:15:08.3 \\
SPT2357-51 & SPT-S J235718-5153.6 & 23:57:16.83 & -51:53:52.8 \\
\enddata
\tablecomments{Long source names are based on positions measured with the SPT data.  Source positions are based on ALMA $3\1{mm}$ continuum data.  This table is available in machine-readable format at: \href{https://github.com/spt-smg/publicdata}{https://github.com/spt-smg/publicdata}}
\end{deluxetable*}

%========================================================
% Section B All 3mm spectroscopy
\section{ALMA $3\1{mm}$ Spectra}
\label{ap:3mmspec}
%========================================================
\renewcommand{\theHtable}{B.\arabic{table}}%<---!!!!---
\renewcommand{\theHfigure}{B.\arabic{figure}}%<---!!!!---
\setcounter{table}{0}
\setcounter{figure}{0}
\renewcommand\thetable{\thesection.\arabic{table}}  
\renewcommand\thefigure{\thesection.\arabic{figure}}  

The spectroscopic redshifts were obtained by scanning ALMA's Band 3, using the procedure outlined in Sec.~\ref{sec:specobs}.  The resulting $3\1{mm}$ line scans (project IDs: 2015.1.00504.S and 2016.1.00672.S) are shown in Figs.~\ref{fig:3mmthumbs_SPT0002-SPT0243} - \ref{fig:3mmthumbs_SPT2349-52-SPT2357}.  Each target was observed for roughly $6-15\1{minutes}$ on-source.  Originally, three spectra (SPT0112-55, SPT0457-49 and SPT2340-59) did not show any spectroscopic lines.  Additional deeper $3\1{mm}$ scans were obtained in 2019.1.00486.S and revealed two spectroscopic lines each.  

%---------------------------------------------
% FIGURE B.1 3mm thumbs
%---------------------------------------------
\begin{figure*}[h] 
	\includegraphics[width=\textwidth]{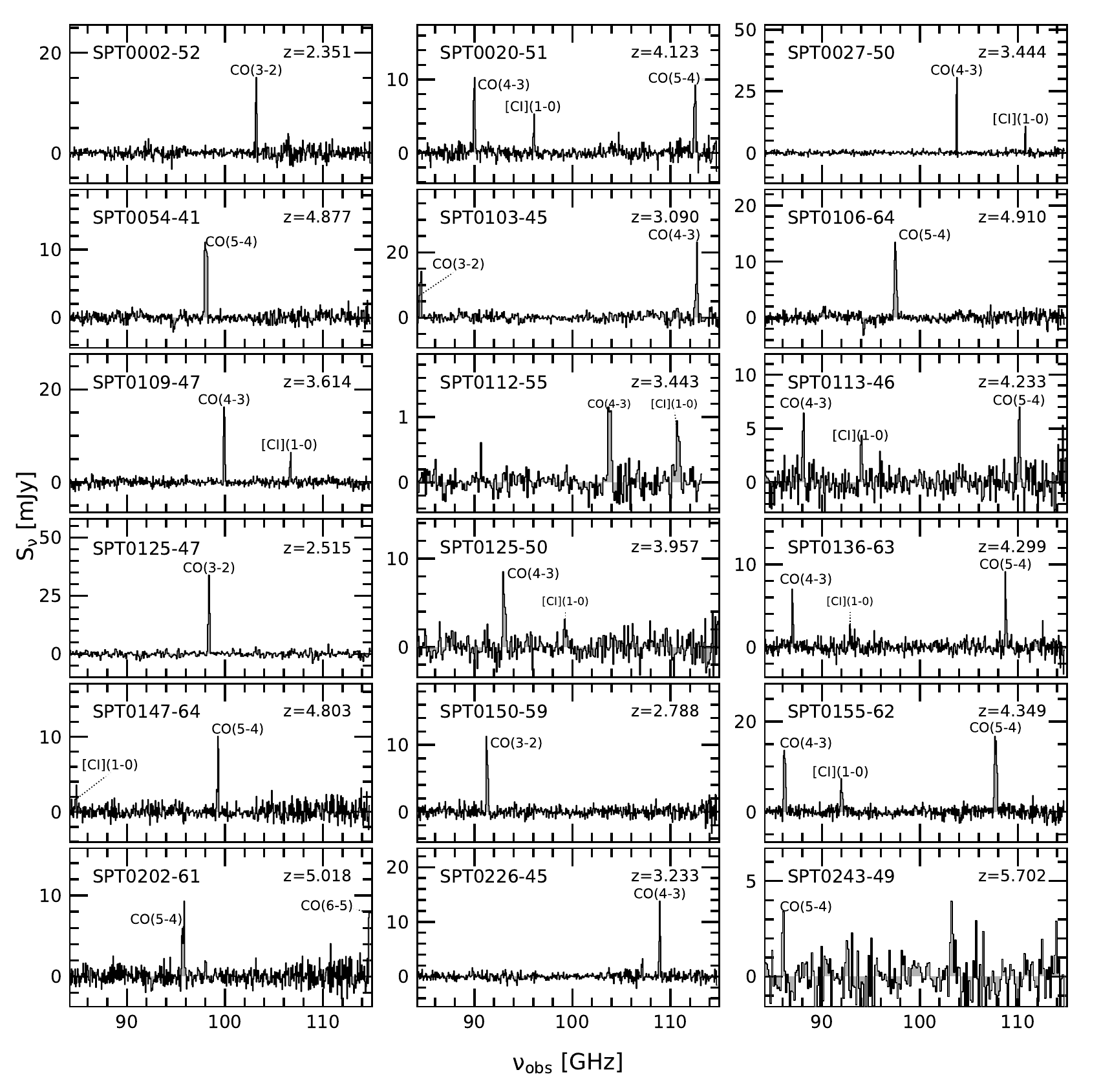}
	\caption{The $3\1{mm}$ ALMA spectra (spanning $84.2$$-$$114.9\1{GHz}$) of all SPT DSFGs. Marginal spectroscopic detections ($>$$3\sigma$) are designated by the smaller font and dashed line.  Though no spectroscopic lines were detected in SPT0112-55 initially, two lines were detected when the source was reobserved, as shown above.  }
	\label{fig:3mmthumbs_SPT0002-SPT0243}
\end{figure*}

\begin{figure*}[h] 
	\includegraphics[width=\textwidth]{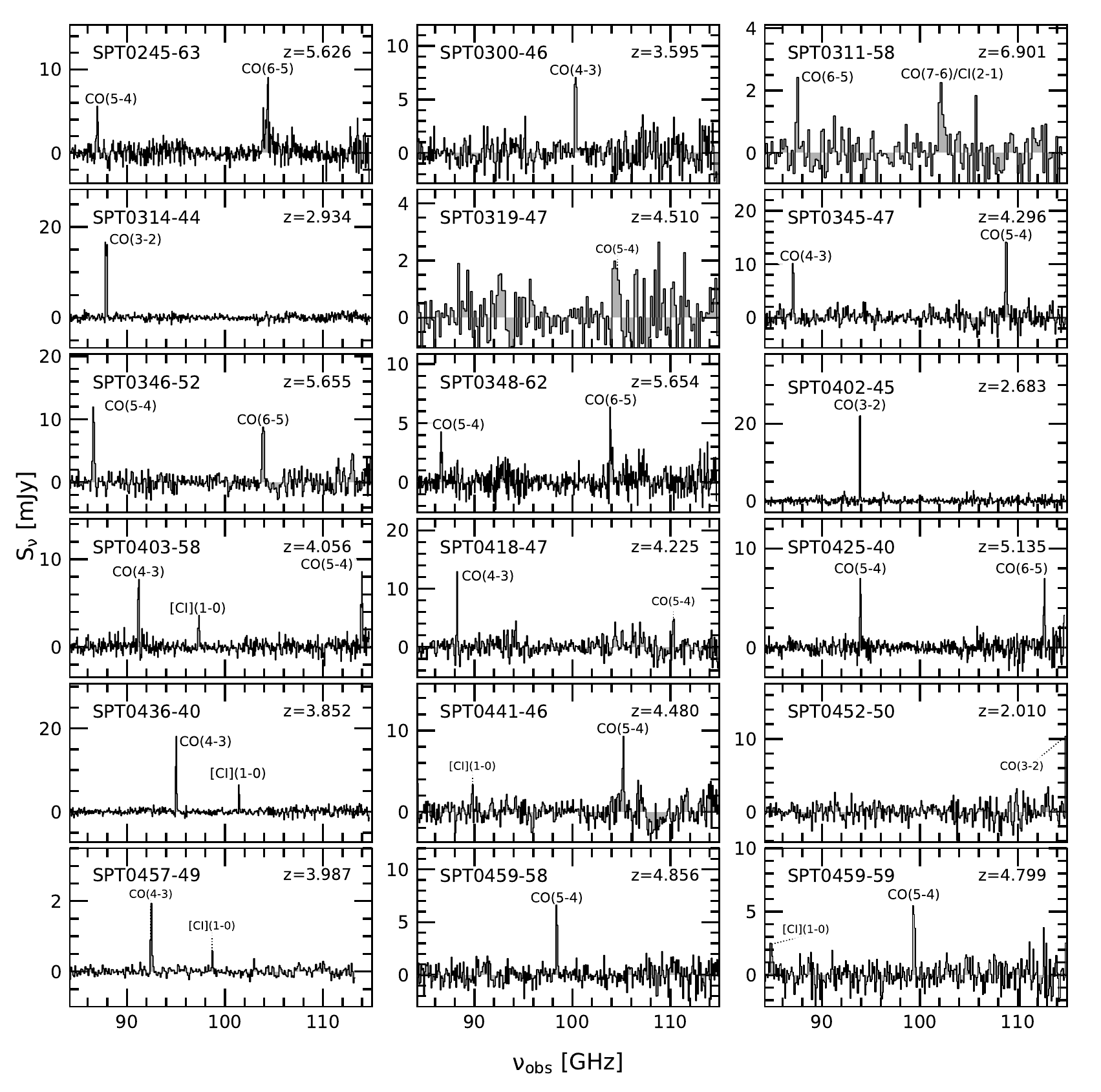}
	\caption{(Continued) The $3\1{mm}$ ALMA spectra (spanning $84.2$$-$$114.9\1{GHz}$) of all SPT DSFGs. Marginal detections ($>$$3\sigma$) are designated by the smaller font and dashed line.  Though no spectroscopic lines were detected in SPT0457-49 initially, two lines were detected when the source was reobserved, as shown above.  }
	\label{fig:3mmthumbs_SPT0245-SPT0459}
\end{figure*}

\begin{figure*}[h] 
	\includegraphics[width=\textwidth]{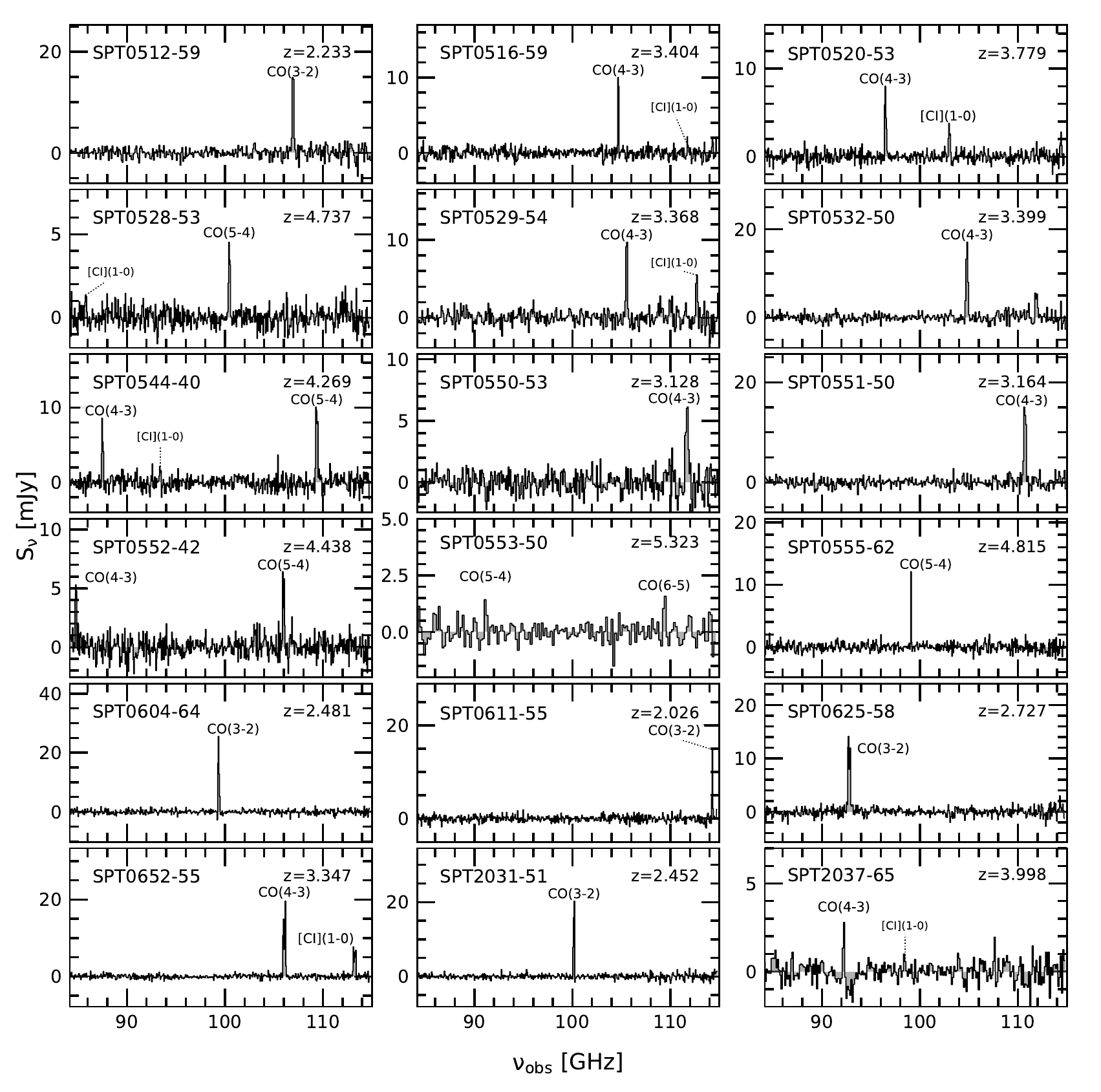}
	\caption{(Continued) The $3\1{mm}$ ALMA spectra (spanning $84.2$$-$$114.9\1{GHz}$) of all SPT DSFGs. Marginal detections ($>$$3\sigma$) are designated by the smaller font and dashed line.    }
	\label{fig:3mmthumbs_SPT0512-SPT2037}
\end{figure*}

\begin{figure*}[h] 
	\includegraphics[width=\textwidth]{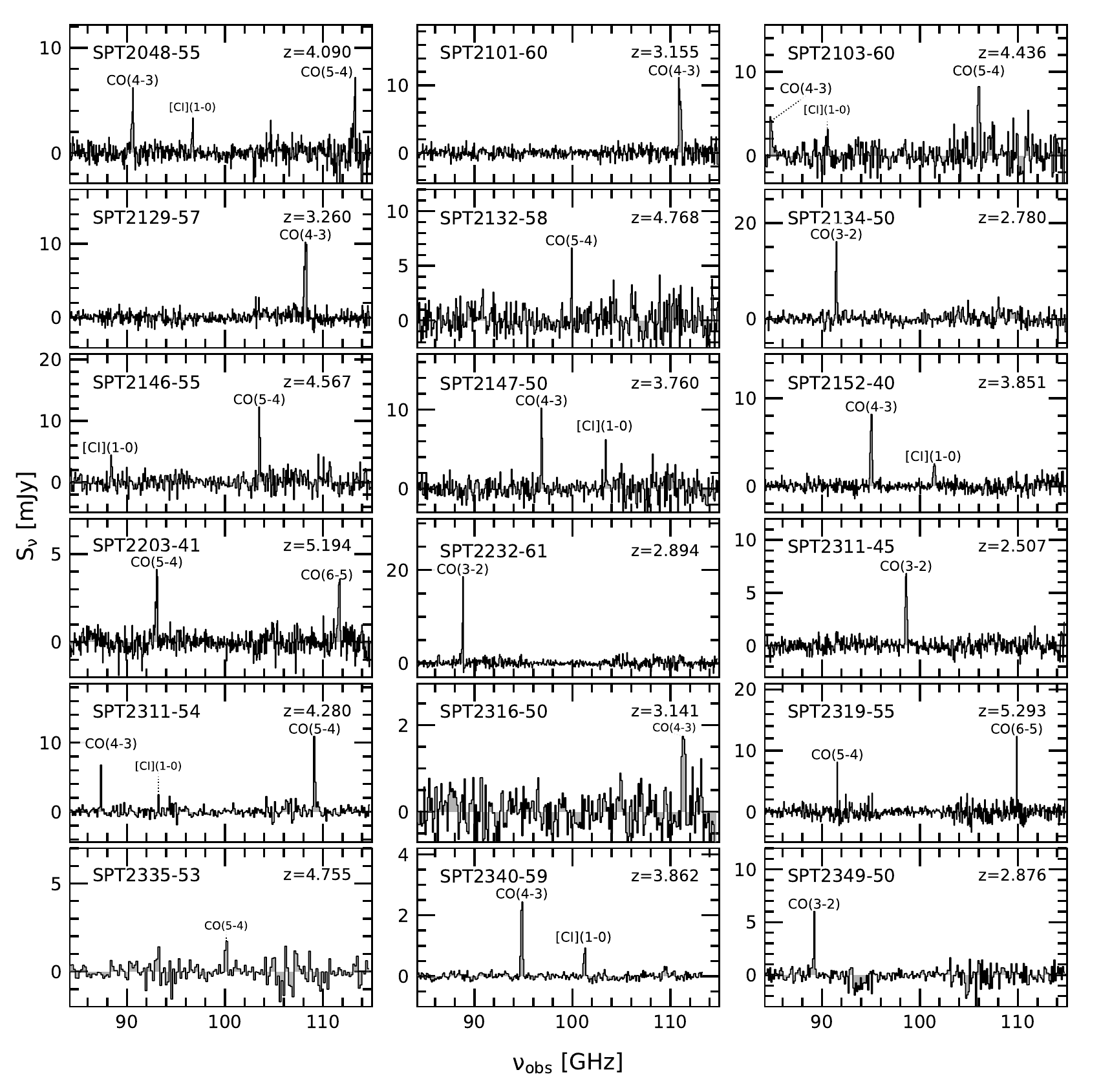}
	\caption{(Continued) The $3\1{mm}$ ALMA spectra (spanning $84.2-114.9\1{GHz}$) of all SPT DSFGs. Marginal detections ($>$$3\sigma$) are designated by the smaller font and dashed line.  A tentative line in SPT2340-59 was identified in \citet{strandet16} at $94.79\1{GHz}$.  When SPT2340-59 was reobserved in Cycle 7, two lines were detected, shown above.  }
	\label{fig:3mmthumbs_SPT2048-SPT2349-50}
\end{figure*}

\begin{figure*}[h] 
	\includegraphics[width=\textwidth]{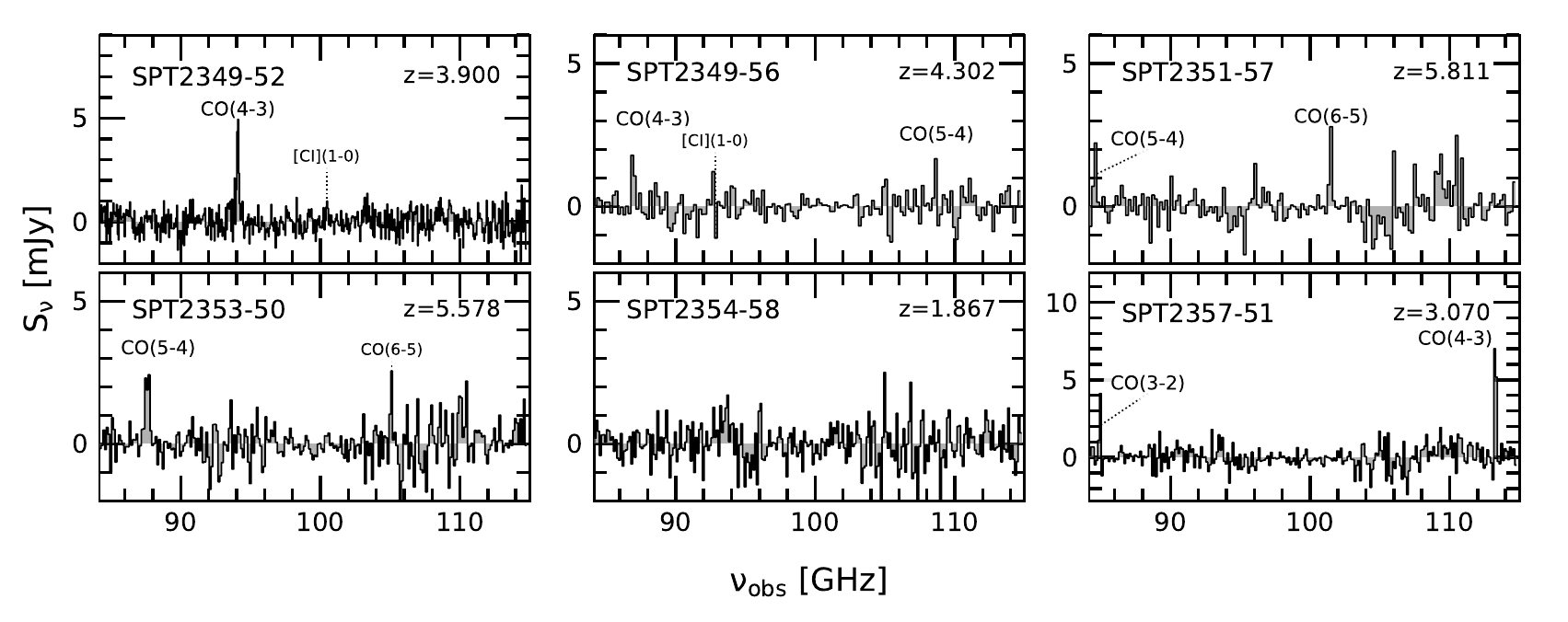}
	\caption{(Continued) The $3\1{mm}$ ALMA spectra (spanning $84.2-114.9\1{GHz}$) of all SPT DSFGs. Marginal detections ($>$$3\sigma$) are designated by the smaller font and dashed line.  No spectroscopic lines were detected in SPT2354-58 because it falls into a spectroscopic redshift desert, depicted in Fig.~\ref{fig:almab3}.  }
	\label{fig:3mmthumbs_SPT2349-52-SPT2357}
\end{figure*}

%========================================================
% Section C All follow up spectroscopy
\section{Ancillary Spectroscopic Observations}
\label{ap:specfollowup}
%========================================================
\renewcommand{\theHtable}{C.\arabic{table}}%<---!!!!---
\renewcommand{\theHfigure}{C.\arabic{figure}}%<---!!!!---
\setcounter{table}{0}
\setcounter{figure}{0}
\renewcommand\thetable{\thesection.\arabic{table}}  
\renewcommand\thefigure{\thesection.\arabic{figure}}  
%\phantomsection
In this appendix, we show the supplementary observations that resolve redshift ambiguities in our ALMA observations.  Ancillary spectroscopic observations were performed with targeted line scans using APEX/FLASH, ALMA and ATCA. 

%========================================================
% Section C.1 Follow-ups with APEX/FLASH
\subsection*{APEX/FLASH Confirmations} \label{sec:apexflash}
\begin{figure*}[h]
\begin{tabular}{cc}
  \includegraphics[width=65mm]{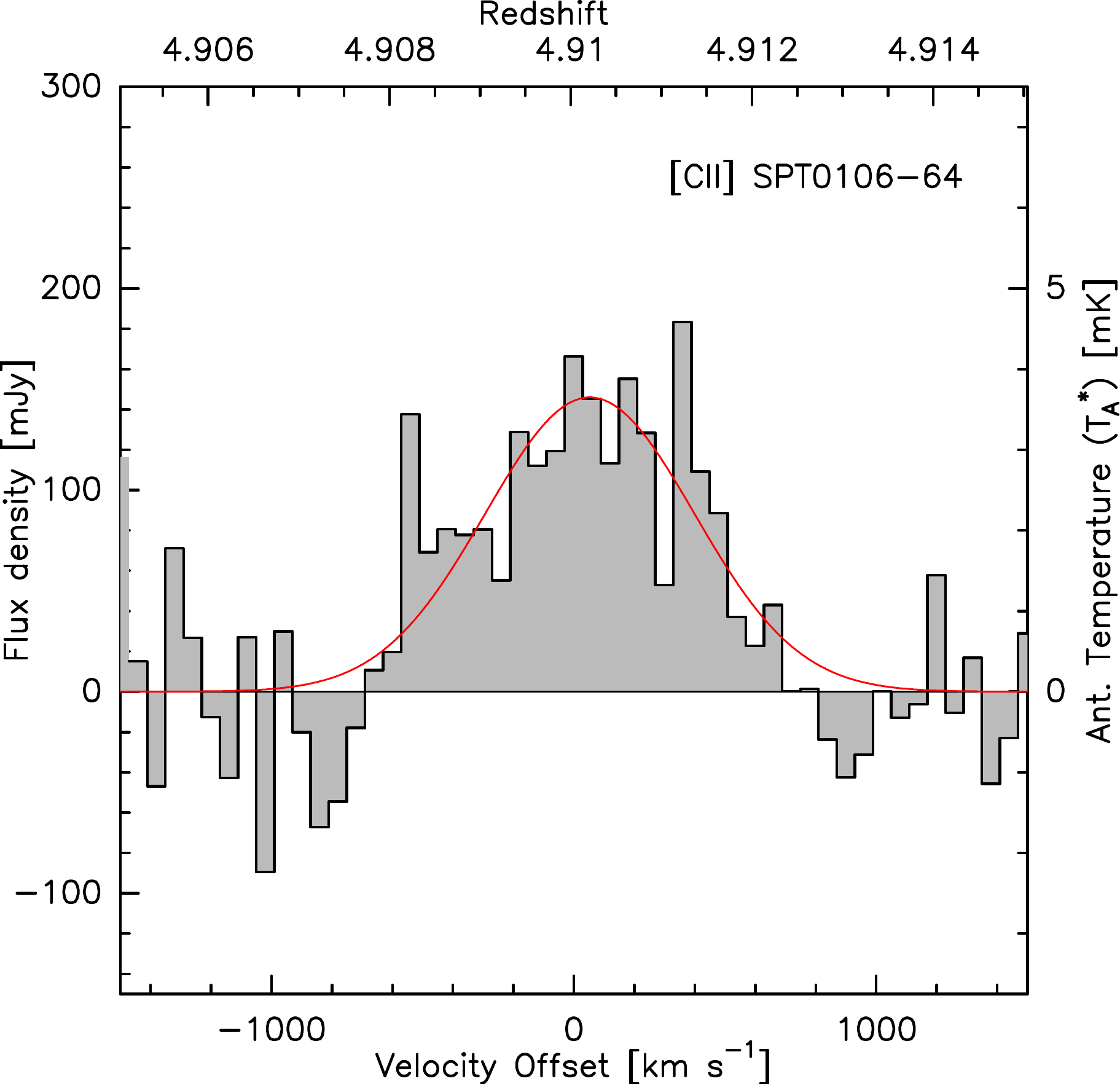} &   \includegraphics[width=65mm]{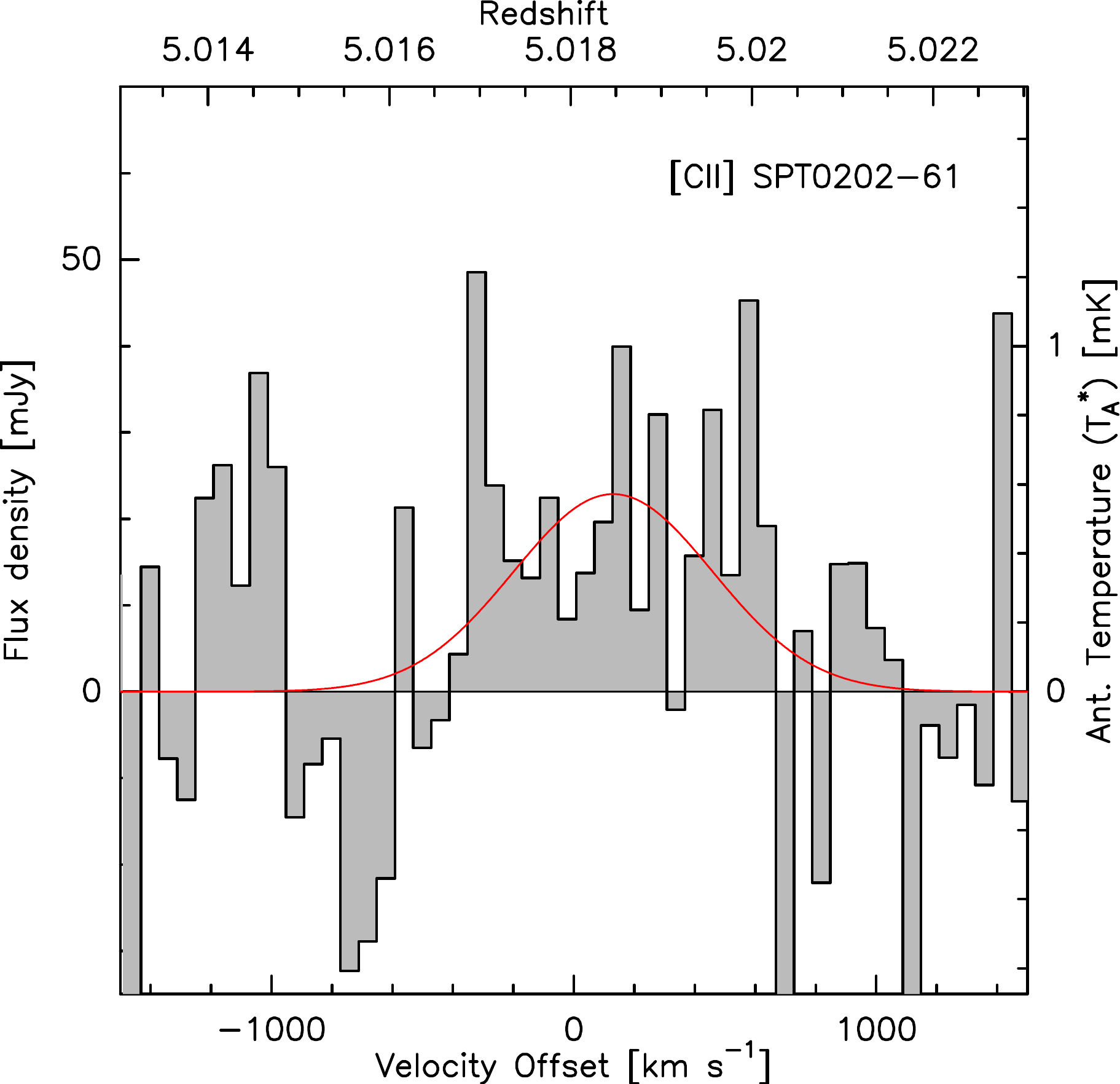} \\
(a) SPT0106-64 & (b) SPT0202-61 \\[6pt]
 \includegraphics[width=65mm]{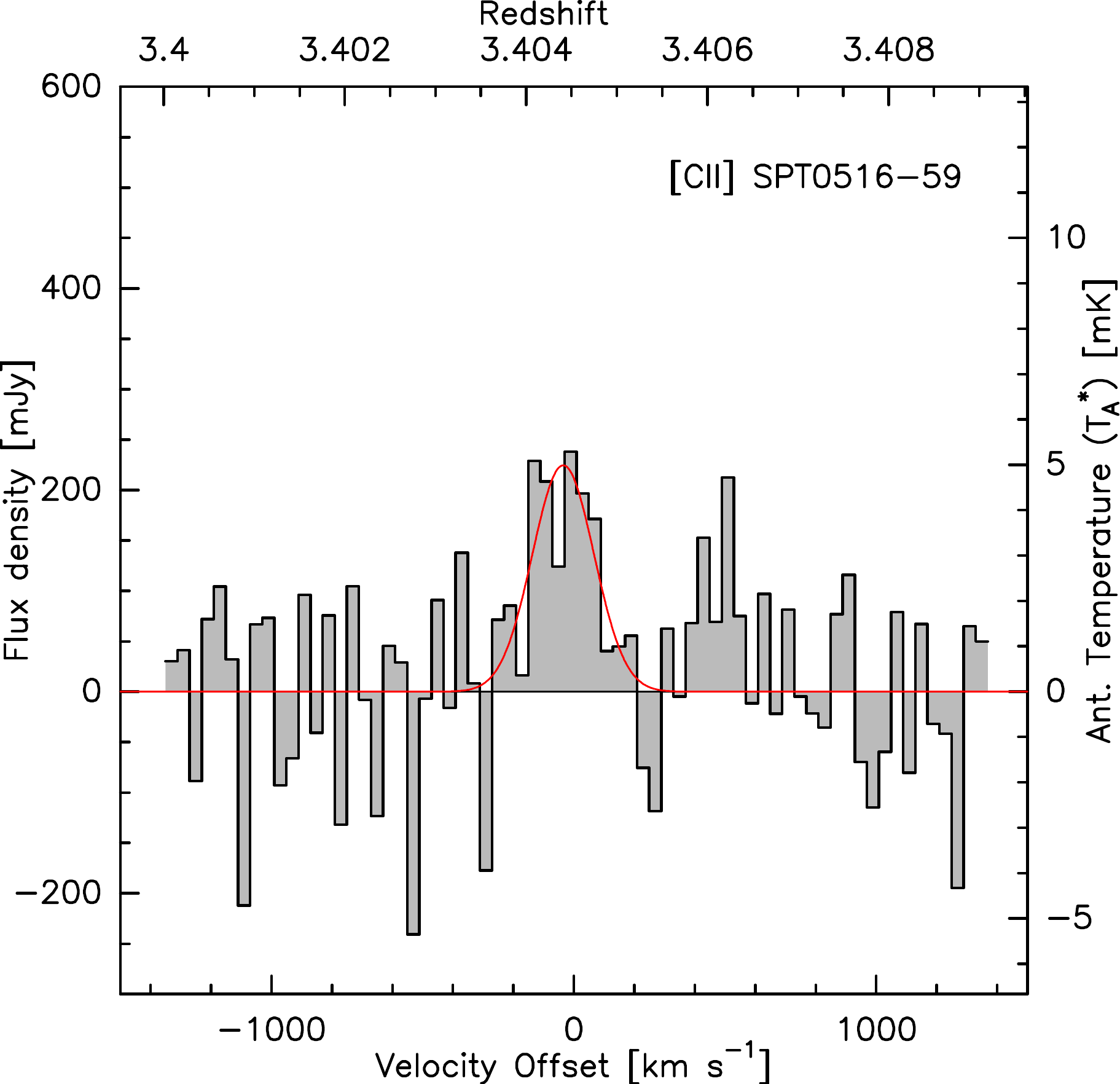} &   \includegraphics[width=65mm]{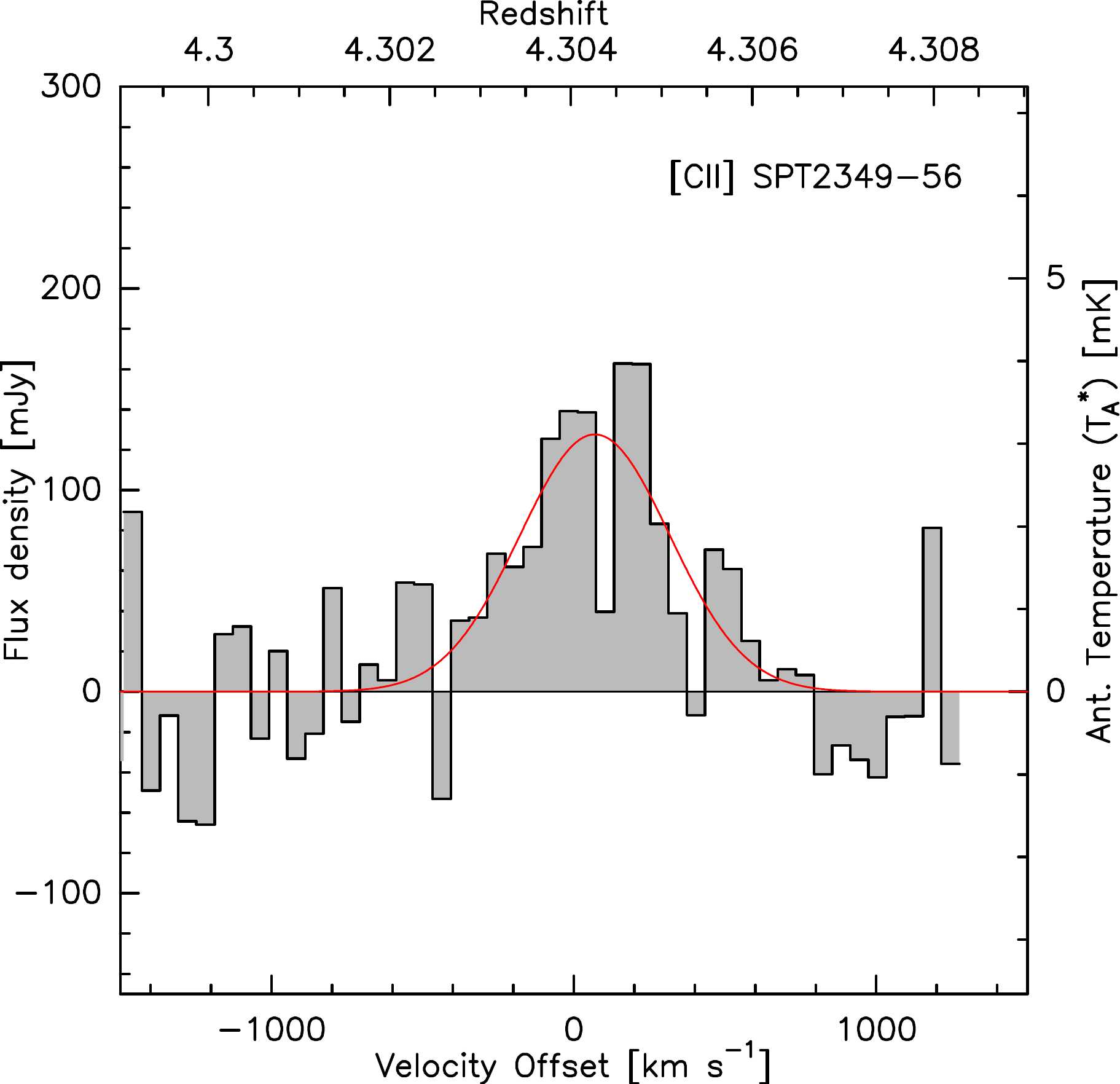} \\
(c) SPT0516-59 & (d) SPT2349-56 \\[6pt]
\end{tabular}
\caption{APEX/FLASH [CII] redshift confirmations \label{fig:apexflash}}
\end{figure*}

%========================================================
Because of its brightness, the [CII] line was used to confirm eight sources at $z$$\gtrapprox$$3$.  Because four of the spectra have already been published in \citet{weiss13} and \citet{strandet16}, we present only the four new [CII] observations.
The observations were carried out using APEX/FLASH~\citep{klein14} in the $345~\1{GHz}$ and $460~\1{GHz}$ transmission window.  The data were obtained as part of project IDs M-095.F-9528-2015 and M-097.F-9519-2016 using Max Planck Society observing time in the period March -- August 2015 and April -- September 2016.  All observations were performed in ``good" weather conditions with an average precipitable water vapor $<$$1.5\1{mm}$, yielding typical system temperatures of $\mathrm{T_{sys}}$$=$$240\1{K}$.  A full detail of the observation scheme and data processing procedure is detailed in~\citet{gullberg15}.  Fig.~\ref{fig:apexflash} shows the four new [CII] spectra used in this work.  

%---------------------------------------------
% FIGURE C1 APEX/FLASH [CII] z-confirmations
%---------------------------------------------

%========================================================
% Section C.2 ALMA
\subsection*{ALMA Targeted Line Scans} \label{sec:almazconfirm}
\begin{figure*}[h!]
	\centering
	\includegraphics[width=\textwidth]{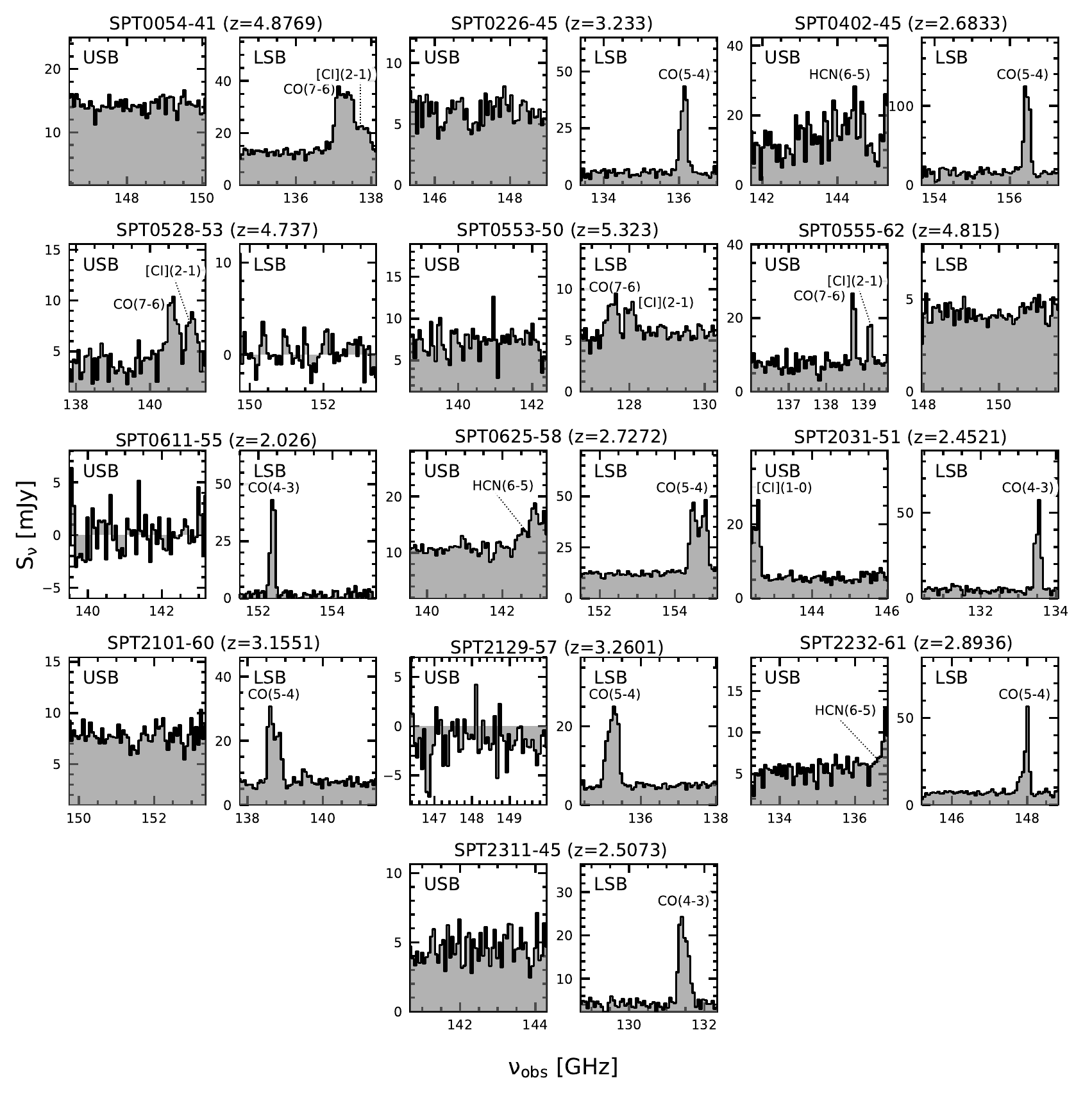}
	\caption{ALMA $2\1{mm}$ spectra for sources with redshifts based on a single $3\1{mm}$ emission line. For each source we show the USB and LSB spectra in the left and right panel, respectively. Each sideband has a total bandwidth of $3.75\1{GHz}$.}
	\label{fig:almab4_cy6}
\end{figure*}

\begin{figure*}[h!]
	\centering
	\includegraphics[width=\textwidth]{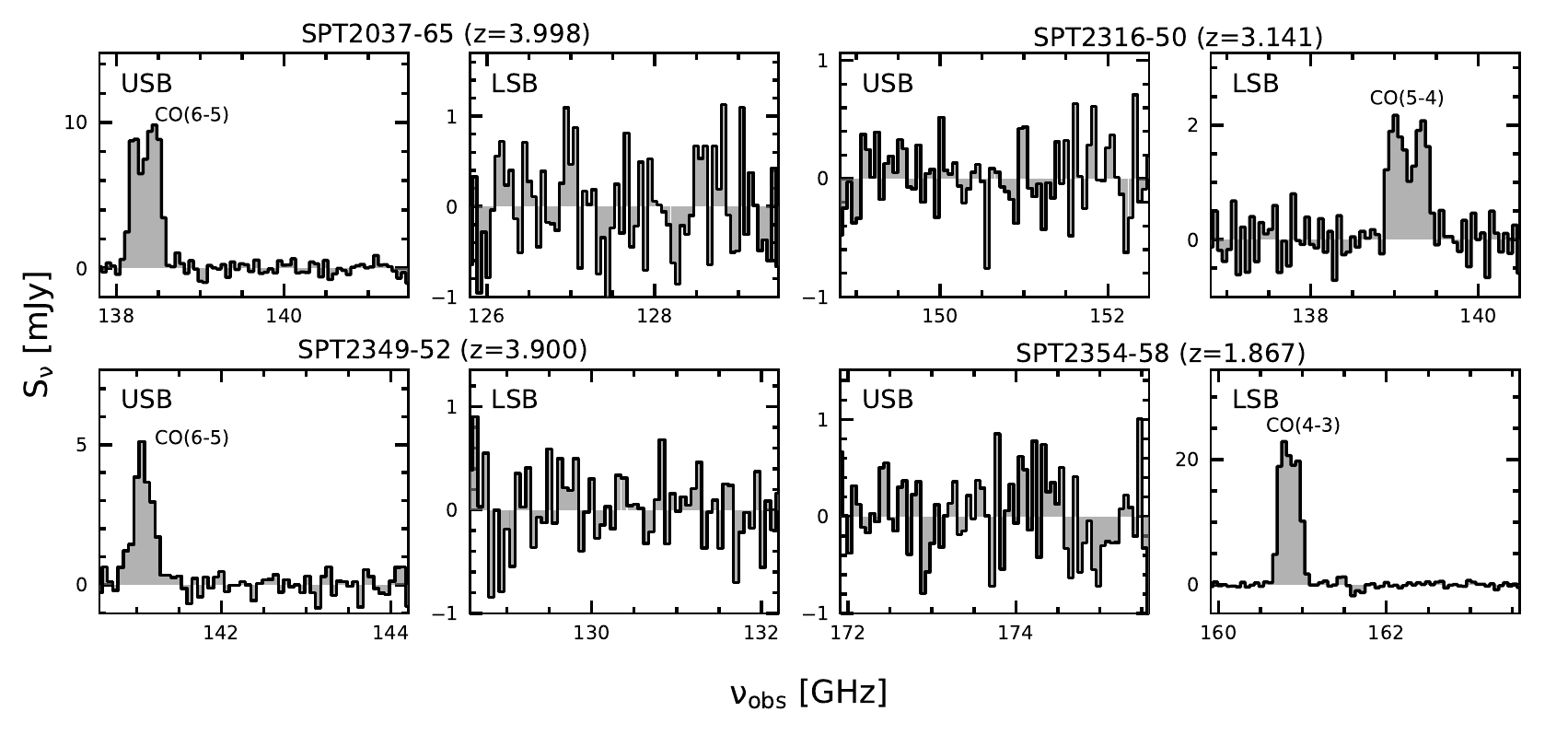}
	\caption{ALMA spectra for sources with redshifts based on a single $3\1{mm}$ emission line.  Because of the low redshift of SPT2354-58 ($z=1.867$), it was observed at $1.6\1{mm}$.  The other sources were observed at $2\1{mm}$.  For each source we show the USB and LSB spectra in the left and right panel, respectively.  Each sideband has a total bandwidth of $3.75\1{GHz}$.}
	\label{fig:almab4_cy7}
\end{figure*}

%========================================================
% ALMA B4 (2mm) followups
A total of 17 sources in the ALMA $3\1{mm}$ spectral scans that contained a single CO line were confirmed through targeted line scans.  Though a photometric redshift could be used to further identify which CO transition was likely detected, the large error bars associated with photometric redshifts allow for some ambiguity.  Using the existing CO line detections and the photometric redshift as a guide, we are able to specifically target CO lines that would be visible for degenerate redshift solution.  The sidebands were configured such that an observation would yield at least one CO line detection for each source at a likely redshift option.  The first 13 sources were observed between December 2018 - January 2019 (project ID: 2018.1.01254.S), using ALMA Band 4 ($125$$-$$163\1{GHz}$).  The line scans are shown in Fig.~\ref{fig:almab4_cy6}.  An additional four sources were confirmed between December 2019 -- January 2020 (project ID: 2019.1.00486.S), using ALMA Bands 4 and 5 ($163$$-$$211\1{GHz}$).  The line scans are shown in Fig.~\ref{fig:almab4_cy7}.  

For both sets of observations, the flux density and bandpass calibrations were based on observations of J0519-4546, J0538-4405, J1924-2914, J2056-4714, J2258-2758, and J2357-5311.  The phase calibration was determined using nearby quasars.  Between 42-48 antennas were used to complete the observations, resulting in synthesized beam sizes between $3.1$$-$$0.8\arcsec$.  The observing time for each science block ranged from $5$$-$$13\1{minutes}$ on-source, excluding overheads, with average single-sideband system temperatures of $\mathrm{T_{sys}}$$=$$56$$-$$86\1{K}$. The data were processed using the \texttt{CASA} package. The images were created using natural weighting and the subsequent spectra were created with a channel width of $16\1{MHz}$.  The average noise per channel in the resulting spectra is $0.4-0.6\1{mJy \, beam^{-1}}$. 

%---------------------------------------------
% FIGURE C2 ALMA B4 z-confirm
%---------------------------------------------

%---------------------------------------------
% FIGURE C3 ALMA B4/B5 z-confirm
%---------------------------------------------

%========================================================
% Section C.3 ATCA 
\subsection*{ATCA Targeted Line Scans} \label{sec:atcazconfirm}
\begin{figure}[h!] 
	\includegraphics[width=\columnwidth]{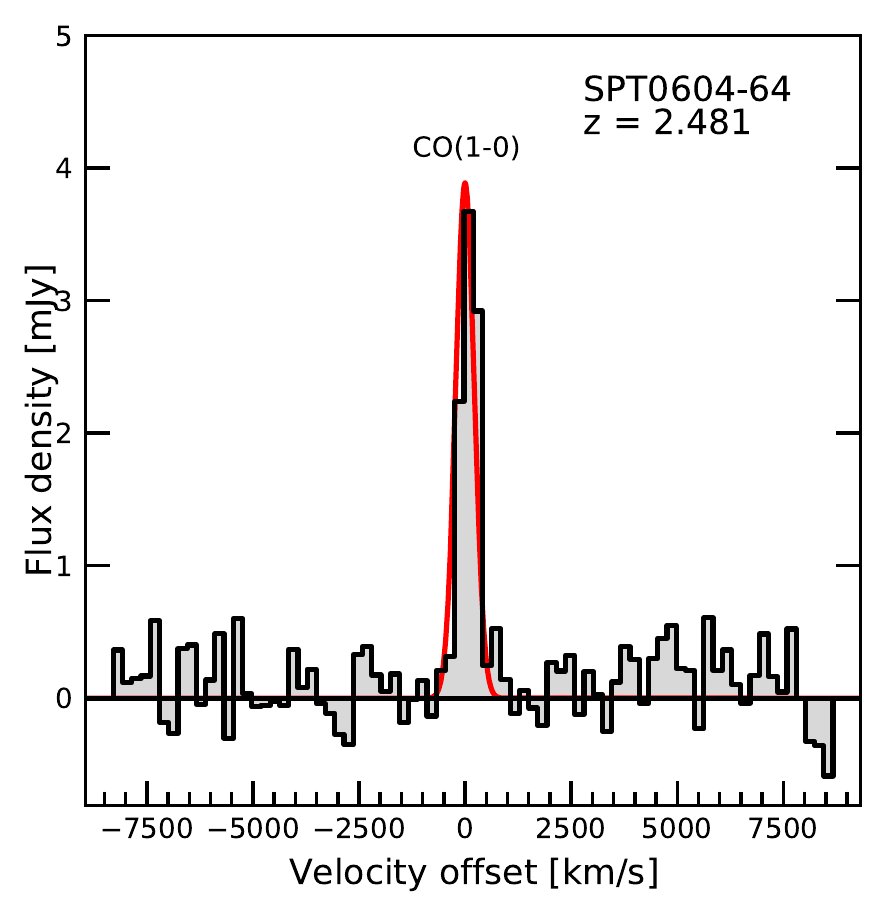}
	\caption{ATCA spectrum of CO(2-1) for SPT0604-64.  The data were fitted with a Gaussian (\textit{red}) and confirm the redshift at $z$$=$$2.481$. }
	\label{fig:0604atca}
\end{figure}

%========================================================
A subset of the SPT DSFGs were also observed with the Australia Telescope Compact Array (ATCA) as part of a survey to observe the CO(1-0) and CO(2-1) line emission ($\nu_{\mathrm{rest}} = 115.2712$ and $230.5380\1{GHz}$, respectively).  This survey was conducted as part of project IDs C2744 and C2818 and most of the observations were published in \citet{aravena16}.  However, observations for one source, SPT0604-64, were obtained after publication and are shown in Fig.~\ref{fig:0604atca}.  We summarize the observations briefly here, but more detail can be found in \citet{aravena16}.

In order to obtain these observations, we used the Compact Array Broadband Backend (CABB) configured in the wide bandwidth mode~\citep{wilson11}.  This leads to a total bandwidth of $2\1{GHz}$ per correlator window and a spectral resolution of $1\1{MHz}$ per channel ($\sim6 - 10\1{km\,s^{-1}}$ per channel for the relevant frequency range).  The $7\1{mm}$ receivers were tuned to the frequency range $30$$-$$50\1{GHz}$, which covers the redshift ranges $1.38$$-$$2.84$ for CO(1-0).  The H214 array configuration at these observing frequencies leads to typical beam sizes of $5$$-$$6\arcsec$.  We expect the flux calibration to be accurate to within $15\%$, based on the comparison of the Uranus and 1934-638 fluxes. The software packages \texttt{Miriad}~\citep{sault95} and \texttt{CASA} were used for editing, calibration and imaging.

%---------------------------------------------
% FIGURE C4 ATCA z-confirm
%---------------------------------------------

%========================================================
% Section C.4 Still needing confirmation
\subsection*{Sources Awaiting Spectroscopic Confirmation} \label{ap:1liners}
\begin{figure}[h] 
	\includegraphics[width=\columnwidth]{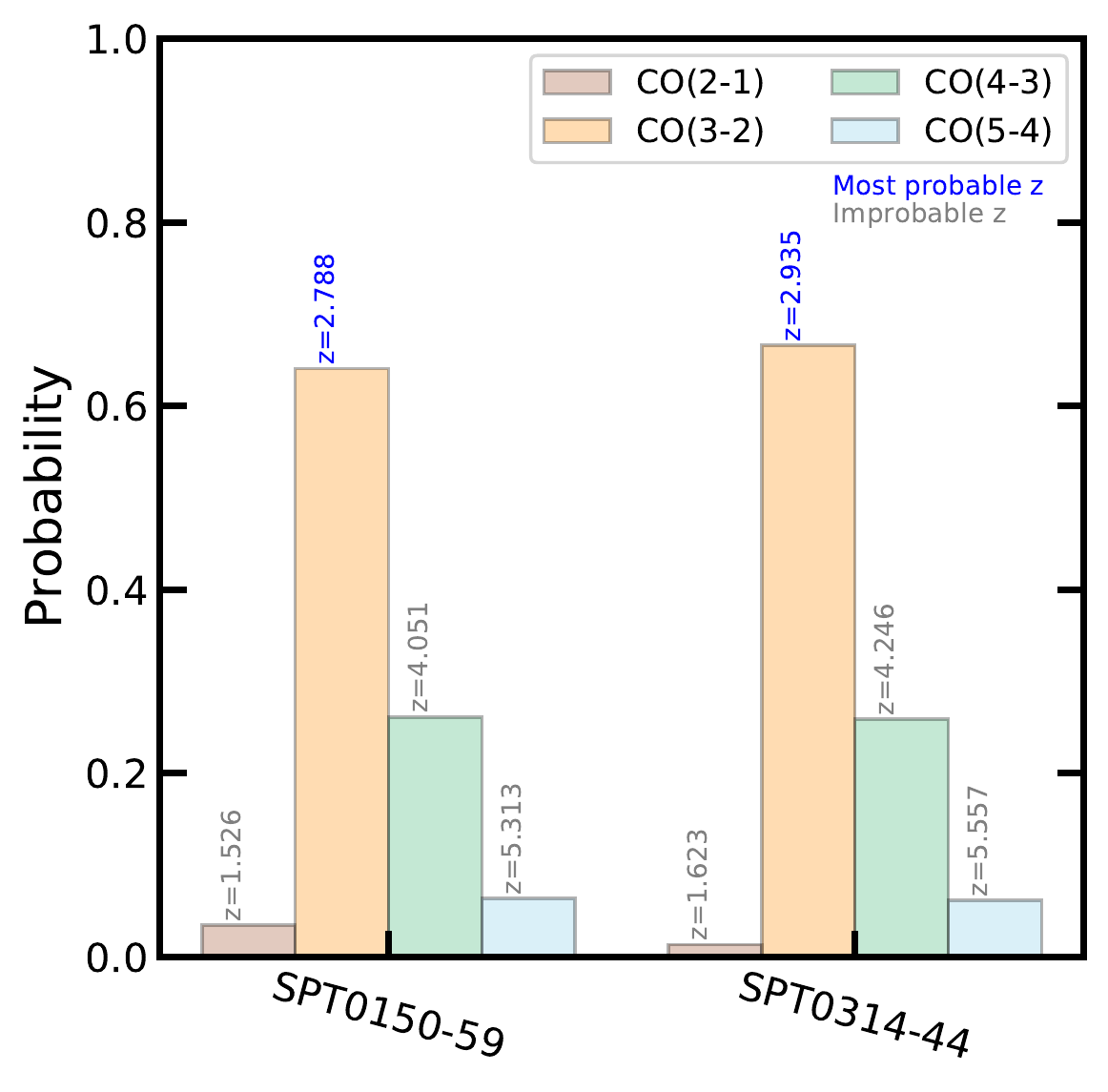}
	\caption{Histograms showing the probability of the redshift options for each source with a single line detection in the ALMA 3 mm spectrum based on the dust temperature distribution shown in Fig.~\ref{fig:td_dist}. The sources are sorted by most probable redshift. The bars represent the probability of each line identification, and the most probable redshift values are denoted in blue. }
	\label{fig:1linerprobs}
\end{figure}

%========================================================
As mentioned in the main text, we have detected a single CO line for two sources (SPT0150-59 and SPT0314-44).  However, these sources are still awaiting ancillary spectroscopic observations to confirm the redshifts.  Using the dust temperature distribution given in Fig.~\ref{fig:td_dist} as a prior, we assign a likelihood of the source being at each possible redshift solution.  These probabilities are displayed in Fig.~\ref{fig:1linerprobs} for the two yet-unconfirmed sources.  For both sources, the most probable identification is CO(3-2).  It should be noted that if the line was the higher-J option, CO(4-3), an additional CO(5-4) line would have been observed in the $3\1{mm}$ window.  

%========================================================
% Section D Photometry
\section{Photometry}
\label{ap:fml}
%========================================================
\renewcommand{\theHtable}{D.\arabic{table}}%<---!!!!---
\renewcommand{\theHfigure}{D.\arabic{figure}}%<---!!!!---
\setcounter{table}{0}
\setcounter{figure}{0}
\renewcommand\thetable{\thesection.\arabic{table}}  
\renewcommand\thefigure{\thesection.\arabic{figure}} 

In this Appendix, we show Tab.~\ref{tab:fml} which contains the values obtained from the photometric observations described in Sec.~\ref{sec:photobs}.  

% this is to prevent photometry table from going up and getting broken apart by all the figures
\clearpage

%---------------------------------------------
% TABLE D1 All flux densities
%---------------------------------------------
\startlongtable
\begin{deluxetable*}{ccccccccccc} 
\tabletypesize{\footnotesize}
\tablecaption{Photometry for all SPT-selected sources \label{tab:fml}}
\tablewidth{0pc}
\tabletypesize{\scriptsize}
\tablehead{
\colhead{Source} & \colhead{$\mathrm{z_{spec}}$} & \colhead{$S_{3000\1{\mu m}}$} & $S_{2000\1{\mu m}}$ & $S_{1400\1{\mu m}}$ & $S_{870\1{\mu m}}$ & $S_{500\1{\mu m}}$& $S_{350\1{\mu m}}$ & $S_{250\1{\mu m}}$ & $S_{160\1{\mu m}}$ & $S_{100\1{\mu m}}$ \\
 & &\colhead{[mJy]} & \colhead{[mJy]} & \colhead{[mJy]} & \colhead{[mJy]} & \colhead{[mJy]} & \colhead{[mJy]} & \colhead{[mJy]} & \colhead{[mJy]} & \colhead{[mJy]} }
\startdata
SPT0002-52 & 2.351 & $0.47\pm0.03$ & $3.1\pm0.9$ & $11\pm6$ & $50\pm4$ & $202\pm10$ & $284\pm9$ & $333\pm10$ & $234\pm21$ & $94\pm5$    \\ 
SPT0020-51 & 4.123 &$1.25\pm0.05$ & $7\pm1$     & $20\pm5$ & $71\pm4$ & $144\pm9$  & $121\pm7$ & $72\pm8$   & $<26$       & $<8$        \\ 
SPT0027-50 & 3.444 & $1.41\pm0.05$ & $9\pm1$     & $36\pm5$ & $138\pm8$& $316\pm8$  & $326\pm7$ & $233\pm7$  & $86\pm13$  & $16\pm2$    \\ 
SPT0054-41 & 4.877 & $1.80\pm0.06$ & $10\pm2$    & $33\pm5$ & $98\pm7$ & $174\pm10$ & $126\pm9$ & $79\pm9$   &            &             \\ 
SPT0103-45 & 3.090 & $1.46\pm0.23$ & $8\pm2$     & $31\pm5$ & $125\pm6$& $232\pm8$  & $213\pm7$ & $133\pm11$ &  $<47$     &    $<13$     \\ 
SPT0106-64 & 4.910 & $2.15\pm0.07$ & $13\pm1$    & $43\pm5$ & $145\pm8$& $237\pm9$  & $256\pm10$& $152\pm8$  & $64\pm14$   &    $<12$     \\
SPT0109-47 & 3.614 & $1.11\pm0.04$ & $7\pm2$     & $16\pm5$ & $109\pm8$& $214\pm8$  & $219\pm9$ & $166\pm9$  & $57\pm10$   &    $<8$     \\
SPT0112-55 & 3.443 & $0.18\pm0.03$ & $4\pm1$     & $17\pm5$ & $36\pm4$ & $37\pm7$   & $38\pm6$  & $<19$      &            &             \\ 
SPT0113-46 & 4.233 & $1.28\pm0.20$ & $9\pm2$     & $22\pm5$ & $79\pm8$ & $89\pm6$   & $56\pm6$  & $25\pm6$   & $<23$       & $<7$        \\ 
SPT0125-47 & 2.515 & $1.88\pm0.29$ & $8\pm1$     & $32\pm5$ & $144\pm9$& $507\pm10$ & $656\pm11$& $778\pm13$ & $423\pm48$ & $116\pm6$   \\ 
SPT0125-50 & 3.957 & $1.51\pm0.24$ & $7\pm1$     & $29\pm5$ & $109\pm9$& $162\pm7$  & $181\pm7$ & $156\pm8$  & $66\pm14$  & $13\pm3$    \\ 
SPT0136-63 & 4.299 & $1.12\pm0.05$ & $7\pm1$     & $22\pm4$ & $69\pm4$ & $122\pm6$  & $81\pm7$  & $42\pm5$   &            &             \\ 
SPT0147-64 & 4.803 & $1.87\pm0.06$ & $11\pm1$    & $28\pm5$ & $83\pm5$ & $122\pm7$  & $99\pm7$  & $49\pm6$   &            &             \\ 
SPT0150-59 & 2.788 & $1.04\pm0.04$ & $7\pm1$     & $23\pm5$ & $63\pm3$ & $170\pm9$  & $208\pm7$ & $174\pm8$  &            &             \\ 
SPT0155-62 & 4.349 & $4.12\pm0.07$ & $24\pm2$    & $67\pm5$ & $174\pm7$ & $200\pm7$ & $135\pm7$ & $58\pm4$   & $<25$       & $<11$        \\ 
SPT0202-61 & 5.018 & $2.73\pm0.06$ & $14\pm1$    & $41\pm4$ & $109\pm7$ & $146\pm7$ & $150\pm8$ & $128\pm8$  & $44\pm11$  & $10\pm3$    \\ 
SPT0226-45 & 3.233 & $0.59\pm0.03$ & $6\pm1$     & $20\pm6$ & $66\pm5$ & $172\pm15$ & $184\pm9$ & $185\pm10$ &            &             \\ 
SPT0243-49 & 5.702 & $3.16\pm0.48$ & $11\pm1$    & $29\pm5$ & $85\pm5$ & $58\pm7$   & $44\pm7$  & $25\pm6$   & $<32$      & $<10$        \\ 
SPT0245-63 & 5.626 & $1.31\pm0.05$ & $6\pm1$     & $16\pm4$ & $61\pm5$ & $59\pm7$   & $52\pm6$  & $50\pm6$   & $49\pm8$   & $31\pm3$    \\ 
SPT0300-46 & 3.595 & $1.01\pm0.16$ & $6\pm1$     & $17\pm5$ & $57\pm5$ & $153\pm7$  & $130\pm6$ & $85\pm8$   & $<38$      & $<10$        \\ 
SPT0311-58 & 6.901 & $1.28\pm0.05$ & $7\pm1$     & $18\pm4$ & $37\pm7$ & $52\pm8$   & $38\pm6$  & $29\pm8$   &            &             \\ 
SPT0314-44 & 2.935 & $1.53\pm0.06$ & $12\pm2$    & $38\pm5$ & $104\pm6$& $390\pm12$ & $443\pm11$& $337\pm11$ & $117\pm15$ &  $<12$       \\ 
SPT0319-47 & 4.510 & $1.2\pm0.2$   & $7\pm1$     & $21\pm5$ & $67\pm6$ & $103\pm7$  & $102\pm6$ & $69\pm7$   & $33\pm8$   & $<7$        \\ 
SPT0345-47 & 4.296 & $1.48\pm0.24$ & $6\pm1$     & $23\pm5$ & $89\pm6$ & $215\pm8$  & $275\pm7$ & $233\pm6$  & $84\pm12$  & $13\pm2$    \\ 
SPT0346-52 & 5.655 & $2.82\pm0.43$ & $12\pm1$    & $39\pm5$ & $131\pm8$& $204\pm8$  & $181\pm7$ & $122\pm7$  & $33\pm9$   & $<6$        \\ 
SPT0348-62 & 5.654 & $0.81\pm0.04$ & $4\pm1$     & $17\pm6$ & $52\pm4$ & $55\pm7$   & $45\pm6$  & $29\pm6$   & $<26$       & $<6$        \\ 
SPT0402-45 & 2.683 & $1.5\pm0.06$  & $13\pm1$    & $47\pm5$ & $200\pm12$& $555\pm12$& $758\pm13$& $796\pm16$ &            &             \\ 
SPT0403-58 & 4.056 & $0.71\pm0.03$ & $5\pm1$     & $18\pm5$ & $40\pm5$ & $87\pm8$   & $82\pm8$  & $56\pm7$   & $<32$      & $<10$        \\ 
SPT0418-47 & 4.225 & $0.79\pm0.13$ & $9\pm1$     & $32\pm5$ & $108\pm11$ & $175\pm7$& $166\pm6$ & $114\pm6$  & $45\pm8$   & $<7$        \\ 
SPT0425-40 & 5.135 & $0.97\pm0.04$ & $8\pm1$     & $24\pm5$ & $60\pm6$ & $117\pm8$  & $125\pm6$ & $71\pm6$   &            &             \\ 
SPT0436-40 & 3.852 & $0.8\pm0.04$  & $7\pm1$     & $23\pm6$ & $75\pm6$ & $118\pm9$  & $112\pm7$ & $71\pm7$   &            &             \\ 
SPT0441-46 & 4.480 & $1.26\pm0.2$  & $7\pm1$     & $25\pm5$ & $80\pm9$ & $106\pm7$  & $98\pm6$  & $60\pm7$   & $<27$      & $<7$       \\ 
SPT0452-50 & 2.011 & $0.67\pm0.11$ & $5\pm1$     & $18\pm4$ & $43\pm4$ & $94\pm7$   & $81\pm6$  & $54\pm5$   & $<29$      & $<7$        \\ 
SPT0457-49 & 3.988 & $0.28\pm0.07$ & $2\pm1$     & $7\pm4$ & $26\pm3$  & $71\pm6$   & $62\pm6$  & $40\pm4$   & $<26$       & $<7$        \\  
SPT0459-58 & 4.856 & $0.96\pm0.16$ & $6\pm1$     & $16\pm3$ & $53\pm6$ & $80\pm7$   & $65\pm6$  & $44\pm6$   & $<28$       & $<7$        \\ 
SPT0459-59 & 4.799 & $1.19\pm0.19$ & $7\pm1$     & $22\pm3$ & $61\pm5$ & $75\pm8$   & $67\pm7$  & $54\pm8$   & $<28$       & $<11$       \\ 
SPT0512-59 & 2.233 & $0.98\pm0.16$ & $5\pm1$     & $20\pm3$ & $75\pm6$ & $257\pm8$  & $369\pm7$ & $305\pm7$  & $139\pm18$ & $33\pm4$    \\ 
SPT0516-59 & 3.404 & $0.38\pm0.03$ & $3\pm1$     & $14\pm3$ & $33\pm3$ & $89\pm7$   & $113\pm6$ & $89\pm6$   & $52\pm11$  & $20\pm3$    \\ 
SPT0520-53 & 3.779 & $0.9\pm0.03$  & $5\pm1$     & $14\pm4$ & $56\pm5$ & $97\pm7$   & $90\pm6$  & $56\pm6$   & $<22$       & $<8$        \\ 
SPT0528-53 & 4.737 & $0.55\pm0.03$ & $3\pm1$     & $11\pm6$ & $27\pm3$ & $41\pm6$   & $39\pm6$  & $27\pm8$   &            &             \\ 
SPT0529-54 & 3.368 & $1.51\pm0.23$ & $9\pm1$     & $34\pm3$ & $118\pm7$& $174\pm10$ & $141\pm10$& $88\pm7$   & $<68$      & $<27$        \\ 
SPT0532-50 & 3.399 & $3.04\pm0.47$ & $11\pm1$    & $38\pm4$ & $118\pm8$& $290\pm8$  & $298\pm8$ & $216\pm7$  & $69\pm13$  & $<8$        \\ 
SPT0538-50 & 2.786 &               &$9\pm1$      & $30\pm3$ & $125\pm5$& $360\pm9$  & $426\pm9$ & $344\pm8$  & $142\pm16$ & $31\pm2$    \\ 
SPT0544-40 & 4.269 & $1.45\pm0.05$ & $6\pm2$     & $10\pm5$ & $87\pm7$ & $132\pm8$  & $121\pm6$ & $68\pm6$   & $<39$      & $<10$        \\  
SPT0550-53 & 3.128 & $0.61\pm0.12$ & $4\pm1$     & $18\pm3$ & $53\pm6$ & $97\pm11$  & $89\pm10$ & $69\pm10$  & $28\pm9$   & $8\pm2$     \\ 
SPT0551-48 & 2.583 &               & $11\pm1$    & $23\pm5$ & $139\pm4$& $420\pm12$& $600\pm20$& $633\pm12$  &            &             \\ 
SPT0551-50 & 3.164 & $1.04\pm0.17$ & $5\pm1$     & $24\pm3$ & $74\pm6$ & $197\pm8$  & $190\pm7$ & $149\pm7$  & $63\pm14$  & $13\pm3$    \\ 
SPT0552-42 & 4.438 & $1.14\pm0.05$ & $7\pm1$     & $21\pm5$ & $48\pm4$ & $74\pm8$   & $48\pm6$  & $37\pm6$   &            &             \\ 
SPT0553-50 & 5.323 & $1.08\pm0.05$ & $3\pm1$     & $10\pm7$ & $43\pm4$ & $67\pm9$   & $47\pm6$  & $27\pm7$   & $<39$      & $<8$        \\ 
SPT0555-62 & 4.815 & $0.94\pm0.04$ & $9\pm1$     & $23\pm5$ & $56\pm3$ & $87\pm7$   & $81\pm5$  & $79\pm6$   & $56\pm17$  & $23\pm4$    \\ 
SPT0604-64 & 2.481 & $1.33\pm0.03$ & $11\pm1$    & $46\pm6$ & $150\pm8$& $440\pm12$ & $620\pm13$& $509\pm13$ & $207\pm25$ & $51\pm4$    \\ 
SPT0611-55 & 2.026 & $0.17\pm0.03$ & $0.7\pm0.4$ & $3\pm2$  & $65\pm7$ & $250\pm10$ & $316\pm10$& $261\pm11$ &            &             \\ 
SPT0625-58 & 2.727 & $1.62\pm0.06$ & $9\pm1$     & $34\pm5$ & $125\pm6$& $321\pm9$  & $394\pm9$ & $324\pm7$  & $150\pm18$ & $26\pm5$    \\ 
SPT0652-55 & 3.347 & $1.86\pm0.06$ & $15\pm1$    & $42\pm6$ & $172\pm7$& $325\pm8$  & $297\pm8$ & $186\pm8$  & $<37$      & $<12$        \\ 
SPT2031-51 & 2.452 & $0.79\pm0.03$ & $4\pm1$     & $19\pm5$ & $65\pm3$ & $225\pm7$  & $246\pm7$ & $227\pm8$  & $91\pm13$  & $22\pm4$    \\ 
SPT2037-65 & 3.998 & $8.82\pm0.08$ & $20\pm1$    & $42\pm4$ & $131\pm6$& $237\pm12$ & $199\pm9$ & $129\pm10$ &            &             \\ 
SPT2048-55 & 4.090 & $1.89\pm0.06$ & $7\pm1$     & $17\pm4$ & $54\pm4$ & $80\pm9$   & $91\pm6$  & $48\pm9$   & $<24$       & $<7$        \\  
SPT2101-60 & 3.155 & $0.88\pm0.04$ & $6\pm1$     & $20\pm4$ & $62\pm6$ & $186\pm7$  & $235\pm8$ & $170\pm8$  &            &             \\ 
SPT2103-60 & 4.436 & $0.99\pm0.16$ & $9\pm1$     & $28\pm5$ & $78\pm6$ & $111\pm7$  & $84\pm5$  & $48\pm5$   & $<23$       & $<7$        \\  
SPT2129-57 & 3.260 & $0.86\pm0.04$ & $5\pm1$     & $22\pm4$ & $87\pm6$ & $148\pm8$  & $184\pm9$ & $155\pm6$  & $92\pm16$  & $34\pm3$    \\ 
SPT2132-58 & 4.768 & $1.42\pm0.23$ & $6\pm1$     & $29\pm5$ & $58\pm5$ & $80\pm7$   & $75\pm7$  & $57\pm11$  & $<37$      & $<12$        \\  
SPT2134-50 & 2.780 & $1.13\pm0.18$ & $6\pm1$     & $21\pm5$ & $101\pm7$& $269\pm9$  & $332\pm9$ & $350\pm9$  & $196\pm22$ & $49\pm3$    \\ 
SPT2146-55 & 4.567 & $1.18\pm0.19$ & $5\pm1$     & $17\pm4$ & $55\pm4$ & $83\pm9$   & $69\pm12$ & $65\pm12$  & $<29$      & $<8$        \\ 
SPT2147-50 & 3.760 & $0.76\pm0.12$ & $6\pm1$     & $21\pm4$ & $61\pm5$ & $121\pm8$  & $115\pm7$ & $72\pm7$   & $<28$       & $9\pm2$     \\ 
SPT2152-40 & 3.851 & $1.29\pm0.05$ & $8\pm2$     & $18\pm5$ & $97\pm7$ & $113\pm7$  & $113\pm6$ & $85\pm7$   & $37\pm11$  & $<12$       \\ 
SPT2203-41 & 5.194 & $0.96\pm0.04$ & $7\pm1$     & $30\pm5$ & $63\pm6$ & $79\pm7$   & $48\pm5$  & $38\pm4$   &            &             \\ 
SPT2232-61 & 2.894 & $0.73\pm0.04$ & $7\pm1$     & $22\pm5$ & $60\pm5$ & $210\pm8$  & $215\pm8$ & $168\pm9$  &            &             \\ 
SPT2311-45 & 2.507 & $0.78\pm0.04$ & $5\pm1$     & $15\pm12$& $60\pm4$ & $155\pm8$  & $198\pm8$ & $164\pm6$  & $90\pm12$  & $12\pm4$    \\ 
SPT2311-54 & 4.280 & $0.62\pm0.04$ & $5\pm1$     & $18\pm3$ & $44\pm3$ & $95\pm7$   & $106\pm7$ & $85\pm10$  & $<32$      & $12\pm3$    \\ 
SPT2316-50 & 3.141 & $0.21\pm0.03$ & $1.9\pm1.3$ & $9\pm7$  & $31\pm4$ & $66\pm11$  & $60\pm11$ & $38\pm11$  &            &             \\ 
SPT2319-55 & 5.293 & $0.76\pm0.04$ & $5\pm1$     & $15\pm3$ & $38\pm3$ & $49\pm7$   & $44\pm6$  & $33\pm6$   & $<25$       & $<8$        \\  
SPT2332-53 & 2.726 &               & $4.3\pm0.5$ & $25\pm2$ & $168\pm6$& $304\pm5$  & $564\pm16$& $585\pm37$ & $233\pm37$ & $57\pm7$    \\ 
SPT2335-53 & 4.756 & $0.33\pm0.03$ & $4\pm1$     & $12\pm3$ & $36\pm5$ & $79\pm10$  & $65\pm8$  & $43\pm9$   &            &             \\ 
SPT2340-59 & 3.862 & $0.26\pm0.04$ & $4\pm1$     & $13\pm4$ & $34\pm4$ & $71\pm9$   & $66\pm7$  & $42\pm9$   & $<29$      & $<1$        \\ 
SPT2349-50 & 2.876 & $0.5\pm0.04$  & $4\pm1$     & $11\pm6$ & $43\pm3$ & $128\pm8$  & $136\pm7$ & $129\pm9$  & $<77$      & $<38$       \\  
SPT2349-52 & 3.900 & $0.16\pm0.02$ & $4\pm1$     & $12\pm5$ & $37\pm5$ & $73\pm10$  & $62\pm8$  & $45\pm9$   &            &             \\ 
SPT2349-56 & 4.302 & $0.25\pm0.03$ & $6\pm1$     & $19\pm3$ & $57\pm8$ & $85\pm6$   & $72\pm6$  & $37\pm6$   & $<33$      & $<12$        \\ 
SPT2351-57 & 5.811 & $0.75\pm0.04$ & $7\pm1$     & $16\pm3$ & $35\pm3$ & $74\pm6$   & $56\pm6$  & $44\pm5$   & $<44$      & $<10$        \\  
SPT2353-50 & 5.578 & $0.88\pm0.04$ & $5\pm1$     & $16\pm3$ & $41\pm4$ & $56\pm7$   & $52\pm6$  & $30\pm7$   & $<41$      & $<12$        \\  
SPT2354-58 & 1.867 & $0.6\pm0.04$  & $3\pm1$     & $14\pm3$ & $66\pm5$ & $278\pm8$  & $469\pm9$ & $614\pm11$ & $532\pm59$ & $239\pm11$  \\ 
SPT2357-51 & 3.070 & $0.41\pm0.03$ & $5\pm1$     & $18\pm3$ & $53\pm5$ & $123\pm8$  & $112\pm6$ & $71\pm5$   & $<34$      & $<8$        \\ 
\enddata
\tablecomments{Non-detections are shown as $3\sigma$ upper limits.  The uncertainties do not include absolute calibration errors.  Additionally, the $2\1{mm}$ and $1.4\1{mm}$ SPT flux densities are deboosted and were presented in \citet{everett20}.  This table is available in machine-readable format at: \href{https://github.com/spt-smg/publicdata}{https://github.com/spt-smg/publicdata}}
\end{deluxetable*}

\clearpage

%========================================================
% Section E All 3mm spectroscopy
\section{All ALMA $3\1{mm}$ Spectroscopic Line Observations}
\label{ap:3mm}
%========================================================
\setcounter{table}{0}
\setcounter{figure}{0}
\renewcommand{\theHtable}{E.\arabic{table}}%<---!!!!---
\renewcommand{\theHfigure}{E.\arabic{figure}}%<---!!!!---
\renewcommand\thetable{\thesection.\arabic{table}}  
\renewcommand\thefigure{\thesection.\arabic{figure}} 

A summary of all spectroscopic data obtained for the final SPT-selected DSFG sample is given below, in Tab.~\ref{tab:lines}.  Blind $3\1{mm}$ scans were conducted for all 81 sources across ALMA Cycles 0, 1, 3 and 4.  The Cycle 0 data was published in \citet{weiss13} and Cycle 1 data was published in \citet{strandet16}.  The remaining Cycles are presented for the first time in this work.  While the spectra are given in Figs.~\ref{fig:3mmthumbs_SPT0002-SPT0243}-\ref{fig:3mmthumbs_SPT2349-52-SPT2357}, the lines detected can be found in the ``Lines from $3\1{mm}$ scans" column.  Any ancillary spectroscopic data obtained is presented in the ``New lines $\&$ comments" column.  All of the combined observational efforts have yielded secure redshifts for the complete flux-limited sample of 81 sources from the $2500\1{deg^2}$ SPT survey.  Of these, 79 sources had multiple spectroscopic lines, detected either solely from the $3\1{mm}$ window or with ancillary spectroscopic observations.  Only two of the redshifts are based on a single line, but with a robust line identification based on our analysis of the dust temperature distribution.  All together, this is the largest and most complete collection of spectroscopic redshifts for high-redshift DSFGs obtained so far at mm wavelengths.

%---------------------------------------------
% TABLE E1 Redshifts
%---------------------------------------------
\begin{longrotatetable}
\begin{deluxetable*}{lccccll}
\tabletypesize{\scriptsize}
\movetabledown=20mm
\tablecaption{Redshifts and line identifications \label{tab:lines}}
\tablehead{
\colhead{Source} & \colhead{Cycle} &\colhead{$z$} & \colhead{Figure }&\colhead{Typical RMS} & \colhead{Lines from 3\,mm scans} &  \colhead{New lines \& comments} \\
&  & & \colhead{number} &\colhead{[mJy/beam]} &  &   
}
\startdata
SPT2354-58 & 1  & $1.867(1)$  & \ref{fig:3mmthumbs_SPT2349-52-SPT2357}, \ref{fig:almab4_cy7} & 0.91  & None$^{a}$        & $\mathrm{OH^+}^{a}$, CO(4-3) from ALMA             \\[0.25ex]
SPT0452-50 & 0  & $2.0105(8)$ & \ref{fig:3mmthumbs_SPT0245-SPT0459} & 1.53  & CO(3-2)$^{b}$     & $\mathrm{CO(1-0)}^{c}$ from ATCA                              \\[0.25ex]
SPT0611-55 & 3  & $2.026(1)$ & \ref{fig:3mmthumbs_SPT0512-SPT2037}, \ref{fig:almab4_cy6} & 0.65  & CO(3-2)           & CO(4-3) from ALMA                                             \\[0.25ex]
SPT0512-59 & 0  & $2.2334(1)$ & \ref{fig:3mmthumbs_SPT0512-SPT2037} & 1.22  & CO(3-2)$^{b}$     & $\mathrm{CO(6-5)}^{a}$ from ALMA; $\cii^{d}$ from SPIRE FTS   \\[0.25ex]
SPT0002-52 & 1  & $2.351(1)$  & \ref{fig:3mmthumbs_SPT0002-SPT0243} & 0.89  & CO(3-2)$^{a}$     & CO(5-4)$^{a}$ from APEX                                       \\[0.25ex]
SPT2031-51 & 4  & $2.4521(1)$ & \ref{fig:3mmthumbs_SPT0512-SPT2037}, \ref{fig:almab4_cy6} & 0.58  & CO(3-2)           & CO(4-3) from ALMA                                             \\[0.25ex]
SPT0604-64 & 3  & $2.48071(5)$ & \ref{fig:3mmthumbs_SPT0512-SPT2037}, \ref{fig:0604atca} & 0.67  & CO(3-2)           & CO(1-0) with ATCA                                    \\[0.25ex]
SPT2311-45 & 4  & $2.5073(1)$ & \ref{fig:3mmthumbs_SPT2048-SPT2349-50}, \ref{fig:almab4_cy6} & 0.59  & CO(3-2)           & CO(4-3) from ALMA                                             \\[0.25ex]
SPT0125-47 & 0  & $2.5149(1)$ & \ref{fig:3mmthumbs_SPT0002-SPT0243} & 1.47  & CO(3-2)$^{b}$     & $\ci(1-0)^{c}$ from ATCA                                      \\[0.25ex]
SPT0551-48 & --  & $2.5833(2)$ & \ref{fig:3mmthumbs_SPT0512-SPT2037} & --      & CO(7-6)$^{a}$ CO(8-7)$^{a}$ $\ci(2-1)^{a}$    & $3\1{mm}$ lines from Z-Spec; $\co{1}{0}^{e}$ from ATCA; No ALMA data \\[0.25ex]
SPT0402-45 & 4  & $2.6833(2)$ & \ref{fig:3mmthumbs_SPT0245-SPT0459}, \ref{fig:almab4_cy6} & 0.61  & CO(3-2)           & CO(5-4) from ALMA                                            \\[0.25ex]
SPT2332-53 & --  & $2.7256(2)$ & \ref{fig:3mmthumbs_SPT2048-SPT2349-50} & --  & CO(7-6)$^{b}$ $\mathrm{C_{IV}}1549\angstrom$$^{b}$ $\mathrm{Ly\alpha}$$^{b}$ & $3\1{mm}$ lines from Z-Spec; $\mathrm{CO}(1-0)^{e}$ from ATCA \\[0.25ex]
SPT0625-58 & 3  & $2.7272(2)$ & \ref{fig:3mmthumbs_SPT0512-SPT2037}, \ref{fig:almab4_cy6} & 0.65  & CO(3-2)           & CO(5-4) from ALMA                                            \\[0.25ex]
SPT2134-50 & 0  & $2.78(2)$  &  \ref{fig:3mmthumbs_SPT2048-SPT2349-50} & 1.67  & CO(3-2)$^{b}$     & CO(7-6)$^{b}$ CO(8-7)$^{b}$ from Z-Spec and SMA              \\[0.25ex]
SPT0538-50 & --  & $2.7855(1)$ & \ref{fig:3mmthumbs_SPT0512-SPT2037} & --      & CO(7-6)$^{b}$ CO(8-7)$^{b}$ $\mathrm{Si_{IV}} 1400 \angstrom^{b}$ & $3\1{mm}$ lines from Z-Spec; $\mathrm{CO}(1-0)^{e}$ and $\co{3}{2}^{f}$ from ATCA \\[0.25ex]
\textbf{SPT0150-59} & 3 & $2.7882(2)$ & \ref{fig:3mmthumbs_SPT0002-SPT0243} & 0.70 & CO(3-2)           & $\mathrm{z_{phot}} = 3.4 \pm 0.7$                             \\[0.25ex]
SPT2349-50 & 1  & $2.8759(3)$  & \ref{fig:3mmthumbs_SPT2048-SPT2349-50} & 0.94  & CO(3-2)$^{a}$     & CO(7-6) from APEX/SEPIA$^{a}$                                      \\[0.25ex]
SPT2232-61 & 4  & $2.8936(2)$ & \ref{fig:3mmthumbs_SPT2048-SPT2349-50}, \ref{fig:almab4_cy6} & 0.65  & CO(3-2)           & CO(5-4) from ALMA                                            \\[0.25ex]
\textbf{SPT0314-44} & 4 & $2.9345(1)$ & \ref{fig:3mmthumbs_SPT0245-SPT0459} & 0.56  & CO(3-2)           & $\mathrm{z_{phot}} = 2.9 \pm 0.4$                             \\[0.25ex]
SPT2357-51 & 1  & $3.07(4)$   & \ref{fig:3mmthumbs_SPT2349-52-SPT2357} & 0.94  & CO(3-2)$^a$ CO(4-3)$^a$   & Lyman-$\alpha^a$ and $\mathrm{OII}_{3727 \angstrom}^a$ from VLT/X-shooter \\[0.25ex]
SPT0103-45 & 0  & $3.0901(4)$  & \ref{fig:3mmthumbs_SPT0002-SPT0243} & 1.52  & CO(3-2)$^b$ CO(4-3)$^b$   &                                                               \\[0.25ex]
SPT0550-53 & 0  & $3.1276(7)$  & \ref{fig:3mmthumbs_SPT0512-SPT2037} & 1.20  & CO(4-3)$^b$       & $\mathrm{CO}(8-7)^{a}$ and $\mathrm{H_2O}(2_{02}-1_{11})^{a}$ from ALMA; $\cii^{d}$ from APEX \\[0.25ex]
SPT2316-50 & 4  & $3.1413(8)$ & \ref{fig:3mmthumbs_SPT2048-SPT2349-50}, \ref{fig:almab4_cy7} & 0.60  & CO(4-3)           & CO(5-4) from ALMA                                            \\[0.25ex]
SPT2101-60 & 3  & $3.1551(3)$ & \ref{fig:3mmthumbs_SPT2048-SPT2349-50}, \ref{fig:almab4_cy6} & 0.75  & CO(4-3)           & CO(5-4) from ALMA                                            \\[0.25ex]
SPT0551-50 & 3  & $3.1642(3)$ & \ref{fig:3mmthumbs_SPT0512-SPT2037} & 1.19  & CO(4-3)$^b$       & $\cii^{d}$ and $\mathrm{CO}(8-7)^{a}$ from APEX               \\[0.25ex]
SPT0226-45 & 4  & $3.233(1)$ & \ref{fig:3mmthumbs_SPT0002-SPT0243}, \ref{fig:almab4_cy6}  & 0.58  & CO(4-3)           & CO(5-4) from ALMA                                            \\[0.25ex]
SPT2129-57 & 4  & $3.2601(3)$ & \ref{fig:3mmthumbs_SPT2048-SPT2349-50}, \ref{fig:almab4_cy6} & 0.66  & CO(4-3)           & CO(5-4) from ALMA                                            \\[0.25ex]
SPT0652-55 & 4  & $3.3466(3)$ & \ref{fig:3mmthumbs_SPT0512-SPT2037} & 0.50  & CO(4-3) [CI](1-0) &                                                               \\[0.25ex]
SPT0529-54 & 0  & $3.3684(8)$ & \ref{fig:3mmthumbs_SPT0512-SPT2037} & 1.19  & CO(4-3)$^b$ [CI](1-0)$b$ $^{13}$CO(4-3)$^b$ &                                     \\[0.25ex]
SPT0532-50 & 0  & $3.3986(4)$ & \ref{fig:3mmthumbs_SPT0512-SPT2037} & 1.18  & CO(4-3)$^b$ [CI](1-0)$b$ $^{13}$CO(4-3)$^b$ &                                     \\[0.25ex]
SPT0516-59 & 3  & $3.4039(4)$  & \ref{fig:3mmthumbs_SPT0512-SPT2037}, \ref{fig:apexflash} & 0.66  & CO(4-3)           & $\cii$ from APEX/FLASH                                        \\[0.25ex]
SPT0112-55 & 3, 7   & $3.443(1)$& \ref{fig:3mmthumbs_SPT0002-SPT0243} & 0.67, 0.29  & CO(4-3) [CI](1-0) &                                                               \\[0.25ex]
SPT0027-50 & 3  & $3.4436(3)$ & \ref{fig:3mmthumbs_SPT0002-SPT0243} & 0.65  & CO(4-3) [CI](1-0) &                                                               \\[0.25ex]
SPT0300-46 & 0  & $3.5948(4)$ & \ref{fig:3mmthumbs_SPT0245-SPT0459} & 1.65  & CO(4-3)$^b$       & CO(10-9)$^{a}$ from ALMA; $\cii^{d}$ from APEX                \\[0.25ex]
SPT0109-47 & 3  & $3.6139(5)$ & \ref{fig:3mmthumbs_SPT0002-SPT0243} & 0.66  & CO(4-3) [CI](1-0) &                                                               \\[0.25ex]
SPT2147-50 & 0  & $3.7604(2)$ & \ref{fig:3mmthumbs_SPT2048-SPT2349-50} & 1.68  & CO(4-3) [CI](1-0) &                                                               \\[0.25ex]
SPT0520-53 & 4  & $3.7785(6)$ & \ref{fig:3mmthumbs_SPT0512-SPT2037} & 0.58  & CO(4-3) [CI](1-0) &                                                               \\[0.25ex]
SPT2152-40 & 4  & $3.8507(6)$ & \ref{fig:3mmthumbs_SPT2048-SPT2349-50} & 0.52  & CO(4-3) [CI](1-0) &                                                               \\[0.25ex]
SPT0436-40 & 4  & $3.8519(4)$ & \ref{fig:3mmthumbs_SPT0245-SPT0459} & 0.61  & CO(4-3) [CI](1-0) &                                                               \\[0.25ex]
SPT2340-59 & 1, 7   & $3.862(3)$ & \ref{fig:3mmthumbs_SPT2048-SPT2349-50} & 0.91, 0.20 & CO(4-3) [CI](1-0) &                                                               \\[0.25ex]
SPT2349-52 & 4  & $3.9000(5)$ & \ref{fig:3mmthumbs_SPT2349-52-SPT2357}, \ref{fig:almab4_cy7} & 0.57  & CO(4-3)           & CO(6-5) from ALMA                                             \\[0.25ex]
SPT0125-50 & 0  & $3.957(9)$  & \ref{fig:3mmthumbs_SPT0002-SPT0243} & 1.50  & CO(4-3)$^b$ [CI](1-0)$^b$ & $\mathrm{CO}(10-9)^{a}$ and $H_{2}O^{g}$ abs line from ALMA \\[0.25ex]
SPT0457-49 & 0, 7   & $3.988(2)$& \ref{fig:3mmthumbs_SPT0245-SPT0459} & 1.59, 0.23  & CO(4-3) [CI](1-0) &                                                               \\[0.25ex]
SPT2037-65 & 3  & $3.9977(8)$ & \ref{fig:3mmthumbs_SPT0512-SPT2037}, \ref{fig:almab4_cy7} & 0.76  & CO(4-3)           & CO(6-5) from ALMA                                             \\[0.25ex]
SPT0403-58 & 4  & $4.0564(1)$ & \ref{fig:3mmthumbs_SPT0245-SPT0459} & 0.60  & CO(4-3) CO(5-4) [CI](1-0) &                                                       \\[0.25ex]
SPT2048-55 & 3  & $4.0898(2)$ & \ref{fig:3mmthumbs_SPT2048-SPT2349-50} & 0.75  & CO(4-3) CO(5-4) [CI](1-0) &                                                       \\[0.25ex]
SPT0020-51 & 3  & $4.1227(7)$ & \ref{fig:3mmthumbs_SPT0002-SPT0243} & 0.65  & CO(4-3) CO(5-4) [CI](1-0) &                                                       \\[0.25ex]
SPT0418-47 & 0  & $4.2246(4)$ & \ref{fig:3mmthumbs_SPT0245-SPT0459} & 1.69  & CO(4-3) CO(5-4)           &                                                       \\[0.25ex]
SPT0113-46 & 0  & $4.2334(3)$ & \ref{fig:3mmthumbs_SPT0002-SPT0243} & 1.49  & CO(5-4) CO(4-3) [CI](1-0) &                                                       \\[0.25ex]
SPT0544-40 & 3  & $4.2692(5)$ & \ref{fig:3mmthumbs_SPT0512-SPT2037} & 0.81  & CO(4-3) CO(5-4) [CI](1-0) &                                                       \\[0.25ex]
SPT2311-54 & 1  & $4.2796(3)$ & \ref{fig:3mmthumbs_SPT2048-SPT2349-50} & 0.93  & CO(4-3)$^a$ CO(5-4)$^a$ [CI](1-0)$^a$ &                                           \\[0.25ex]
SPT0345-47 & 0  & $4.2958(6)$ & \ref{fig:3mmthumbs_SPT0245-SPT0459} & 1.58  & CO(4-3) CO(5-4)   &                                                               \\[0.25ex]
SPT0136-63 & 3  & $4.2991(2)$ & \ref{fig:3mmthumbs_SPT0002-SPT0243} & 0.71  & CO(4-3) CO(5-4) [CI](1-0) &                                                       \\[0.25ex]
SPT2349-56 & 1  & $4.302(2)$  & \ref{fig:3mmthumbs_SPT2349-52-SPT2357}, \ref{fig:apexflash} & 0.89  & CO(4-3)$^a$       & $\cii$ from APEX/FLASH                                        \\[0.25ex]
SPT0155-62 & 3  & $4.3492(6)$ & \ref{fig:3mmthumbs_SPT0002-SPT0243} & 0.71  & CO(4-3) CO(5-4) [CI](1-0) &                                                       \\[0.25ex]
SPT2103-60 & 0  & $4.4359(2)$ & \ref{fig:3mmthumbs_SPT2048-SPT2349-50} & 1.76  & CO(5-4) CO(4-3)   &                                                               \\[0.25ex]
SPT0552-42 & 3  & $4.4376(9)$ & \ref{fig:3mmthumbs_SPT0512-SPT2037} & 0.80  & CO(4-3) CO(5-4)   &                                                               \\[0.25ex]
SPT0441-46 & 0  & $4.4803(3)$ & \ref{fig:3mmthumbs_SPT0245-SPT0459} & 1.54  & CO(5-4)$^b$ [CI](1-0)$^b$ & $\cii^{b}$ with APEX; $\mathrm{CO}(11-10)^{a}$, $\mathrm{H_2O}(2_{20} - 2_{11})^{a}$ $\&$ $\mathrm{NH}_3^{}$ from ALMA \\[0.25ex]
SPT0319-47 & 0  & $4.5187(1)$  & \ref{fig:3mmthumbs_SPT0245-SPT0459} & 1.61  & CO(5-4)$^b$       & $\mathrm{CO}(12-11)^{a}$ from ALMA; $\cii^{a}$ from APEX     \\[0.25ex]
SPT2146-55 & 0  & $4.567(2)$ & \ref{fig:3mmthumbs_SPT2048-SPT2349-50} & 1.70  & CO(5-4)$^b$ [CI](1-0)$^b$ &                                                       \\[0.25ex]
SPT0528-53 & 4  & $4.737(4)$  & \ref{fig:3mmthumbs_SPT0512-SPT2037}, \ref{fig:almab4_cy6} & 0.58  & CO(5-4)           & CO(7-6) from ALMA                                            \\[0.25ex]
SPT2335-53 & 1  & $4.7555(9)$ & \ref{fig:3mmthumbs_SPT2048-SPT2349-50} & 0.88  & CO(5-4)$^a$       & [CII]$^{a}$ from APEX                                          \\[0.25ex]
SPT2132-58 & 0  & $4.7678(5)$ & \ref{fig:3mmthumbs_SPT2048-SPT2349-50} & 1.72  & CO(5-4)$^b$       & $\mathrm{CO}(12-11)^{b}$ and $\nii^{a}$ from ALMA            \\[0.25ex]
SPT0459-59 & 0  & $4.7989(2)$ & \ref{fig:3mmthumbs_SPT0245-SPT0459} & 1.23  & CO(5-4)$^b$ [CI](1-0)$^b$ &                                                       \\[0.25ex]
SPT0147-64 & 3  & $4.8031(3)$ & \ref{fig:3mmthumbs_SPT0002-SPT0243} & 0.70  & CO(5-4) [CI](1-0) &                                                               \\[0.25ex]
SPT0555-62 & 3  & $4.815(3)$  & \ref{fig:3mmthumbs_SPT0512-SPT2037}, \ref{fig:almab4_cy6} & 1.20  & CO(5-4)           & CO(7-6) $\&$ $\cii(2-1)$ from ALMA                           \\[0.25ex]
SPT0459-58 & 0  & $4.8562(3)$ & \ref{fig:3mmthumbs_SPT0245-SPT0459} & 1.22  & CO(5-4)$^b$       & $\mathrm{CO}(11-10)^{a}$ from ALMA                           \\[0.25ex]
SPT0054-41 & 4  & $4.8769(2)$ & \ref{fig:3mmthumbs_SPT0002-SPT0243}, \ref{fig:almab4_cy6} & 0.56  & CO(5-4)           & CO(7-6) $\&$ $\cii(2-1)$ from ALMA                           \\[0.25ex]
SPT0106-64 & 3  & $4.9104(2)$ & \ref{fig:3mmthumbs_SPT0002-SPT0243}, \ref{fig:apexflash} & 0.70  & CO(5-4)           & $\cii$ from APEX/FLASH                                       \\[0.25ex]
SPT0202-61 & 3  & $5.0182(2)$ & \ref{fig:3mmthumbs_SPT0002-SPT0243}, \ref{fig:apexflash} & 0.71  & CO(5-4) CO(6-5)   & $\cii$ from APEX/FLASH                                       \\[0.25ex]
SPT0425-40 & 4  & $5.1353(1)$ & \ref{fig:3mmthumbs_SPT0245-SPT0459} & 0.60  & CO(5-4) CO(6-5)   &                                                               \\[0.25ex]
SPT2203-41 & 4  & $5.1937(1)$ & \ref{fig:3mmthumbs_SPT2048-SPT2349-50} & 0.50  & CO(5-4) CO(6-5)   &                                                               \\[0.25ex]
SPT2319-55 & 1  & $5.2927(6)$ & \ref{fig:3mmthumbs_SPT2048-SPT2349-50} & 0.89  & CO(5-4)$^a$ CO(6-5)$^a$ & CO(14-13)$^j$                                           \\[0.25ex]
SPT0553-50 & 3  & $5.3201(1)$ & \ref{fig:3mmthumbs_SPT0512-SPT2037}, \ref{fig:almab4_cy6} & 0.78  & CO(5-4) CO(6-5)   & CO(7-6) $\&$ $\cii(2-1)$ from ALMA                           \\[0.25ex]
SPT2353-50 & 1  & $5.5781(2)$ & \ref{fig:3mmthumbs_SPT2349-52-SPT2357} & 0.95  & CO(5-4)$^a$       & $\cii^a$ from APEX                                           \\[0.25ex]
SPT0245-63 & 3  & $5.6256(2)$ & \ref{fig:3mmthumbs_SPT0245-SPT0459} & 0.89  & CO(5-4) CO(6-5)   &                                                               \\[0.25ex]
SPT0348-62 & 3  & $5.6541(1)$ & \ref{fig:3mmthumbs_SPT0245-SPT0459} & 0.88  & CO(5-4) CO(6-5)   &                                                               \\[0.25ex]
SPT0346-52 & 0  & $5.6554(1)$ & \ref{fig:3mmthumbs_SPT0245-SPT0459} & 1.53  & CO(5-4)$^b$ CO(6-5)$^b$ H$_{2}$O$^b$ H$_{2}$O$^{+,b}$    &                        \\[0.25ex]
SPT0243-49 & 0  & $5.7022(2)$ & \ref{fig:3mmthumbs_SPT0002-SPT0243} & 1.77  & CO(5-4)$^b$ CO(6-5)$^b$   &                                                       \\[0.25ex]
SPT2351-57 & 1  & $5.8114(2)$ & \ref{fig:3mmthumbs_SPT2349-52-SPT2357} & 0.90  & CO(6-5)$^a$ CO(5-4)$^a$   &                                                       \\[0.25ex]
SPT0311-58 & 3  & $6.9011(4)$ & \ref{fig:3mmthumbs_SPT0245-SPT0459} & 0.88  & CO(6-5)$^{h,i}$ CO(7-6)$^{h,i}$ $\ci(2-1)^{h,i}$   & $\cii^h$ from APEX and ALMA; CO(3-2)$^h$ from ATCA \\[0.25ex]
\enddata
\tablecomments{
The parenthesis at the end of the redshift represents the uncertainty on the last digit presented.
The column named `Cycle' refers to the ALMA Cycle in which the $3\1{mm}$ scan was obtained.  
The column named `Typical RMS' gives the typical RMS found in each ALMA $3\1{mm}$ spectrum, given a channel width of $62.5\1{MHz}$.  
The $3\1{mm}$ spectra for three sources (SPT0551-48, SPT0538-50 and SPT2332-53) were obtained using Z-Spec.  
The values for spectroscopic redshift differ slightly from previous works because the data was binned to a common binning and was fitted through MCMC.
The spectroscopic redshifts obtained, however, are consistent within their error bars to previously published data.
Comments in the right column indicates any follow-up spectroscopy or photometric redshift.  
}
\tablenotetext{a}{Published by \citet{strandet16}}
\tablenotetext{b}{Published by \citet{weiss13}}
\tablenotetext{c}{Published by \citet{aravena16}}
\tablenotetext{d}{Published by \citet{gullberg15}}
\tablenotetext{e}{Published by \citet{aravena13}}
\tablenotetext{f}{Published by \citet{spilker15}}
\tablenotetext{g}{Published by \citet{spilker14}}
\tablenotetext{h}{Published by \citet{strandet17}}
\tablenotetext{i}{Published by \citet{marrone18}}
\tablenotetext{j}{Obtained as part of ALMA project ID: 2017.1.01340.S}
\end{deluxetable*}
\end{longrotatetable}

%========================================================
% Section F All individual source properties
\section{Individual source properties}
\label{ap:srcpropstable}
%========================================================
\renewcommand{\theHtable}{F.\arabic{table}}%<---!!!!---
\renewcommand{\theHfigure}{F.\arabic{figure}}%<---!!!!---
\setcounter{table}{0}
\setcounter{figure}{0}
\renewcommand\thetable{\thesection.\arabic{table}}  
\renewcommand\thefigure{\thesection.\arabic{figure}} 
The properties for the individual SPT sources are derived using a modified blackbody model, which is described in detail in Sec.~\ref{sec:sed}.  The resulting fits are given in Sec.~\ref{sec:sedresults}.  The values for dust temperature (\tdust), FIR luminosity ($\mathrm{L_{FIR}}$), dust mass ($\mathrm{M_{dust}}$) and star formation rate (SFR) are calculated as described in Sec.~\ref{sec:indv_src_props} and given in Tab.~\ref{tab:srcprops}.  If available, the magnification from the lens models described in \citet{spilker16} are used to give the intrinsic values.  However, for the 42 sources without lens models, the median magnification of all sources with lens models ($\big< \mu_{870\1{\mu m}} \big> = 5.5$) is adopted.  

\startlongtable
\begin{deluxetable*}{lllllclll}
\tabletypesize{\footnotesize}
\tablecaption{Individual source properties \label{tab:srcprops}}
\tablewidth{0pc}
\tablecolumns{9}
\tablehead{
\colhead{Source} & \colhead{$\mathrm{z_{spec}}$} & \colhead{$\mathrm{z_{phot}}$} & \colhead{$\mathrm{T_{dust}}$} & \colhead{$\lambda_{\mathrm{peak}}$} & \colhead{$\mu$} & \colhead{\lfir} & \colhead{$\mathrm{M_{dust}}$} & \colhead{SFR} %     
\\
& & & [K] & $\micron$ & & $\times 10^{13}\1{L_\odot}$ & $\times 10^{9}\1{M_\odot}$ & $\times 10^{3}\1{M_\odot \, yr^{-1}}$ 
} 
\startdata
SPT0002-52 &    2.351 & 2.3(0.3) & 59(16)   & 86(6)     & $5.5^{\star}$     & 0.8(0.2)   & $0.96^{+0.38}_{-1.08}$ & 2.6(1.7) \\ 
SPT0020-51 &    4.123 & 4.2(0.7) & 50(15)   & 92(9)     & 5.48(0.45)        & 0.9(0.2)   & $1.93^{+0.41}_{-0.52}$ & 2.4(0.6) \\ 
SPT0027-50 &    3.444 & 3.7(0.5) & 52(7)    & 88(4)     & 5.49(0.43)        & 1.5(0.2)   & $2.04^{+0.28}_{-0.28}$ & 3.2(0.5) \\ 
SPT0054-41 &    4.877 & 4.3(0.7) & 58(11)   & 89(8)     & $5.5^{\star}$     & 1.2(0.2)   & $1.74^{+0.75}_{-2.14}$ & 4.3(2.8) \\ 
SPT0103-45 &    3.090 & 4.1(0.5) & 33(4)    & 114(9)    & 5.34(0.49)        & 0.7(0.1)   & $2.85^{+0.56}_{-0.69}$ & 2.5(0.7) \\ 
SPT0106-64 &    4.910 & 4.5(0.4) & 66(10)   & 76(6)     & $5.5^{\star}$     & 1.8(0.2)   & $1.94^{+0.80}_{-2.41}$ & 7.0(4.5) \\ 
SPT0109-47 &    3.614 & 3.8(0.5) & 58(9)    & 86(6)     & 12.25(17.51)      & 0.5(0.7)   & $0.33^{+1.16}_{-0.64}$ & 1.3(1.9) \\ 
SPT0112-55 &    3.443 & 6.2(1.0) & 56(19)   & 108(4)    & $5.5^{\star}$     & 0.19(0.05) & $0.20^{+0.09}_{-0.21}$ & 0.5(0.3) \\ 
SPT0113-46 &    4.233 & 5.6(0.9) & 31(4)    & 113(14)   & 23.86(0.51)       & 0.10(0.02) & $0.48^{+0.15}_{-0.28}$ & 0.3(0.1) \\ 
SPT0125-47 &    2.515 & 2.6(0.3) & 66(23)   & 83(5)     & 5.47(0.76)        & 2.4(0.6)   & $4.38^{+0.68}_{-0.94}$ & 4.9(0.8) \\ 
SPT0125-50 &    3.957 & 4.5(0.4) & 68(12)   & 74(5)     & 14.17(1.05)       & 0.48(0.06) & $0.86^{+0.13}_{-0.15}$ & 1.1(0.1) \\ 
SPT0136-63 &    4.299 & 4.5(0.8) & 43(8)    & 99(9)     & $5.5^{\star}$     & 0.7(0.1)   & $1.34^{+0.59}_{-1.65}$ & 1.6(1.1) \\ 
SPT0147-64 &    4.803 & 5.3(0.6) & 48(12)   & 91(10)    & $5.5^{\star}$     & 1.0(0.2)   & $2.27^{+1.00}_{-3.25}$ & 2.6(1.7) \\ 
SPT0150-59 &    2.788 & 3.7(0.6) & 37(6)    & 95(7)     & $5.5^{\star}$     & 0.6(0.1)   & $2.06^{+0.84}_{-2.42}$ & 2.1(1.3) \\ 
SPT0155-62 &    4.349 & 6.2(0.6) & 30(2)    & 112(9)    & $5.5^{\star}$     & 1.1(0.2)   & $6.20^{+2.91}_{-7.14}$ & 2.8(1.9) \\ 
SPT0202-61 &    5.018 & 5.6(0.5) & 67(20)   & 66(5)     & 8.28(1.04)        & 1.0(0.2)   & $2.31^{+0.39}_{-0.52}$ & 2.7(0.4) \\ 
SPT0226-45 &    3.233 & 3.9(0.6) & 64(12)   & 88(7)     & $5.5^{\star}$     & 0.8(0.1)   & $0.79^{+0.32}_{-0.90}$ & 2.7(1.8) \\ 
SPT0243-49 &    5.702 & 6.8(0.7) & 34(7)    & 103(14)   & 5.09(0.46)        & 0.7(0.2)   & $6.53^{+2.23}_{-4.71}$ & 2.4(0.9) \\ 
SPT0245-63 &    5.626 & 5.4(0.8) & 60(17)   & 56(2)     & 1.00(0.00)        & 3.7(0.5)   & $9.41^{+0.92}_{-1.03}$ & 18.1(1.0) \\ 
SPT0300-46 &    3.595 & 3.9(0.7) & 44(8)    & 96(10)    & 3.53(0.82)        & 1.0(0.3)   & $2.77^{+0.76}_{-1.22}$ & 3.2(1.1) \\ 
SPT0311-58 &    6.901 & 6.4(1.0) & 57(22)   & 71(11)    & 2.0(0.2)$^a,c$    & 1.4(0.3)   & $0.88^{+0.44}_{-1.42}$ & 2.8(1.9) \\ 
SPT0314-44 &    2.934 & 2.8(0.5) & 47(8)    & 93(5)     & $5.5^{\star}$     & 1.5(0.2)   & $2.82^{+1.05}_{-4.02}$ & 3.2(2.1) \\ 
SPT0319-47 &    4.510 & 4.8(0.6) & 59(20)   & 84(9)     & 2.88(0.28)        & 1.6(0.3)   & $3.06^{+0.63}_{-0.83}$ & 5.2(1.1) \\ 
SPT0345-47 &    4.296 & 3.3(0.4) & 71(15)   & 64(4)     & 7.95(0.48)        & 1.1(0.1)   & $1.44^{+0.18}_{-0.21}$ & 3.2(0.3) \\ 
SPT0346-52 &    5.655 & 4.6(0.5) & 79(15)   & 68(6)     & 5.57(0.12)        & 1.9(0.2)   & $3.21^{+0.54}_{-0.65}$ & 6.4(0.7) \\ 
SPT0348-62 &    5.654 & 5.2(0.9) & 63(16)   & 89(2)     & 1.18(0.01)$^c$    & 2.2(0.3)   & $2.95^{+0.33}_{-0.39}$ & 9.6(0.7) \\ 
SPT0402-45 &    2.683 & 2.8(0.3) & 60(11)   & 89(5)     & $5.5^{\star}$     & 2.5(0.4)   & $2.50^{+1.02}_{-3.03}$ & 7.9(5.0) \\ 
SPT0403-58 &    4.056 & 4.3(0.7) & 55(20)   & 88(9)     & 1.66(0.19)        & 1.8(0.4)   & $2.67^{+0.58}_{-0.73}$ & 5.8(1.5) \\ 
SPT0418-47 &    4.225 & 4.2(0.5) & 58(11)   & 91(2)     & 32.70(2.66)       & 0.18(0.03) & $0.11^{+0.01}_{-0.02}$ & 0.6(0.1) \\ 
SPT0425-40 &    5.135 & 3.7(1.1) & 69(11)   & 72(8)     & $5.5^{\star}$     & 0.8(0.1)   & $0.58^{+0.25}_{-0.70}$ & 3.7(2.4) \\ 
SPT0436-40 &    3.852 & 6.1(1.3) & 51(8)    & 101(9)    & $5.5^{\star}$     & 0.7(0.1)   & $0.94^{+0.38}_{-1.10}$ & 2.1(1.4) \\ 
SPT0441-46 &    4.480 & 5.0(0.6) & 51(17)   & 93(10)    & 12.73(0.96)       & 0.34(0.06) & $0.70^{+0.16}_{-0.23}$ & 1.1(0.3) \\ 
SPT0452-50 &    2.010 & 4.4(0.7) & 21(2)    & 154(14)   & 1.71(0.10)        & 0.20(0.06) & $7.11^{+1.35}_{-1.98}$ & 1.1(0.4) \\ 
SPT0457-49 &    3.987 & 3.6(1.0) & 43(9)    & 88(11)    & $5.5^{\star, c}$  & 0.26(0.05) & $0.30^{+0.14}_{-0.40}$ & 1.2(0.8) \\ 
SPT0459-58 &    4.856 & 4.6(0.8) & 56(16)   & 87(10)    & 4.97(0.57)        & 0.7(0.1)   & $1.28^{+0.32}_{-0.45}$ & 2.4(0.7) \\ 
SPT0459-59 &    4.799 & 5.3(0.7) & 49(16)   & 93(11)    & 3.64(0.39)        & 1.0(0.2)   & $2.14^{+0.54}_{-0.82}$ & 3.2(1.0) \\ 
SPT0512-59 &    2.233 & 2.6(0.4) & 34(6)    & 101(6)    & $5.5^{\star}$     & 0.6(0.1)   & $2.53^{+1.00}_{-2.87}$ & 1.4(0.9) \\ 
SPT0516-59 &    3.404 & 3.2(0.6) & 63(13)   & 84(7)     & $5.5^{\star}$     & 0.5(0.1)   & $0.52^{+0.22}_{-0.55}$ & 1.6(1.0) \\ 
SPT0520-53 &    3.779 & 4.2(0.7) & 43(7)    & 98(9)     & $5.5^{\star}$     & 0.5(0.1)   & $1.30^{+0.55}_{-1.39}$ & 1.6(1.1) \\ 
SPT0528-53 &    4.737 & 4.3(0.9) & 58(22)   & 87(12)    & $5.5^{\star}$     & 0.3(0.1)   & $0.66^{+0.31}_{-0.89}$ & 1.1(0.8) \\ 
SPT0529-54 &    3.368 & 4.7(0.5) & 32(3)    & 122(10)   & 13.23(0.85)       & 0.25(0.04) & $1.15^{+0.22}_{-0.36}$ & 0.8(0.3) \\ 
SPT0532-50 &    3.399 & 4.0(0.5) & 44(7)    & 88(6)     & 10.04(0.57)       & 0.75(0.09) & $3.56^{+0.51}_{-0.56}$ & 1.7(0.2) \\ 
SPT0538-50 &    2.785 & 3.0(0.4) & 42(6)    & 92(7)     & 20.10(2.98)       & 0.34(0.07) & $1.05^{+0.21}_{-0.29}$ & 0.7(0.2) \\ 
SPT0544-40 &    4.269 & 4.0(0.6) & 53(18)   & 85(9)     & $5.5^{\star}$     & 0.9(0.2)   & $2.17^{+0.93}_{-2.81}$ & 2.2(1.5) \\ 
SPT0550-53 &    3.128 & 4.1(0.6) & 35(8)    & 112(12)   & $5.5^{\star, b}$  & 0.3(0.1)   & $1.13^{+0.52}_{-1.28}$ & 0.9(0.6) \\ 
SPT0551-48 &    2.583 & 2.8(0.3) & 62(21)   & 85(9)     & $5.5^{\star}$     & 2.0(0.5)   & $2.25^{+0.89}_{-3.17}$ & 6.1(4.0) \\ 
SPT0551-50 &    3.164 & 3.5(0.5) & 44(8)    & 93(6)     & $5.5^{\star, b}$  & 0.8(0.1)   & $1.91^{+0.81}_{-2.25}$ & 1.7(1.1) \\ 
SPT0552-42 &    4.438 & 4.9(0.9) & 36(5)    & 102(11)   & $5.5^{\star}$     & 0.4(0.1)   & $1.33^{+0.60}_{-1.88}$ & 1.5(1.0) \\ 
SPT0553-50 &    5.323 & 4.2(0.9) & 59(18)   & 84(13)    & $5.5^{\star, c}$  & 0.6(0.1)   & $1.47^{+0.72}_{-2.17}$ & 1.8(1.2) \\ 
SPT0555-62 &    4.815 & 4.8(0.7) & 62(15)   & 73(7)     & $5.5^{\star}$     & 0.7(0.1)   & $0.72^{+0.30}_{-0.92}$ & 2.9(1.9) \\ 
SPT0604-64 &    2.481 & 3.0(0.4) & 42(8)    & 98(5)     & $5.5^{\star}$     & 1.4(0.3)   & $2.82^{+1.10}_{-3.38}$ & 3.0(1.9) \\ 
SPT0611-55 &    2.026 & 2.5(0.4) & 53(9)    & 104(2)    & $5.5^{\star}$     & 0.5(0.1)   & $0.43^{+0.17}_{-0.52}$ & 0.8(0.5) \\ 
SPT0625-58 &    2.727 & 3.4(0.4) & 39(6)    & 93(5)     & $5.5^{\star}$     & 1.1(0.2)   & $2.96^{+1.16}_{-3.32}$ & 2.4(1.5) \\ 
SPT0652-55 &    3.347 & 4.2(0.5) & 40(4)    & 108(7)    & $5.5^{\star}$     & 1.3(0.2)   & $2.86^{+1.23}_{-3.29}$ & 4.0(2.6) \\ 
SPT2031-51 &    2.452 & 2.9(0.4) & 37(6)    & 100(6)    & 3.89(0.16)        & 0.8(0.2)   & $2.55^{+0.31}_{-0.37}$ & 1.8(0.3) \\ 
SPT2037-65 &    3.998 & 5.1(0.6) & 35(4)    & 89(3)     & $5.5^{\star}$     & 1.3(0.2)   & $17.23^{+6.67}_{-19.52}$ & 3.8(2.4) \\ 
SPT2048-55 &    4.090 & 5.1(0.7) & 33(31)   & 85(8)     & 6.25(0.70)        & 0.4(0.2)   & $3.86^{+0.70}_{-0.85}$ & 1.0(0.2) \\ 
SPT2101-60 &    3.155 & 3.1(0.6) & 52(10)   & 89(9)     & $5.5^{\star}$     & 0.9(0.1)   & $1.47^{+0.63}_{-1.64}$ & 2.7(1.8) \\ 
SPT2103-60 &    4.436 & 5.1(0.7) & 44(9)    & 102(3)    & 27.84(1.76)       & 0.14(0.02) & $0.20^{+0.02}_{-0.03}$ & 0.4(0.1) \\ 
SPT2129-57 &    3.260 & 4.1(0.4) & 56(13)   & 87(6)     & $5.5^{\star}$     & 0.8(0.1)   & $1.17^{+0.49}_{-1.30}$ & 2.5(1.6) \\ 
SPT2132-58 &    4.768 & 5.2(0.6) & 51(19)   & 90(11)    & 5.72(0.54)        & 0.7(0.1)   & $1.50^{+0.35}_{-0.57}$ & 2.1(0.6) \\ 
SPT2134-50 &    2.780 & 3.0(0.3) & 61(20)   & 83(6)     & 21.00(2.42)       & 0.33(0.07) & $0.59^{+0.09}_{-0.11}$ & 0.8(0.1) \\ 
SPT2146-55 &    4.567 & 4.6(0.8) & 47(19)   & 88(11)    & 6.65(0.41)        & 0.5(0.1)   & $1.43^{+0.32}_{-0.55}$ & 1.7(0.4) \\ 
SPT2147-50 &    3.760 & 4.2(0.7) & 48(9)    & 93(8)     & 6.55(0.42)        & 0.54(0.07) & $0.93^{+0.15}_{-0.22}$ & 1.3(0.3) \\ 
SPT2152-40 &    3.851 & 4.9(0.5) & 47(11)   & 93(9)     & $5.5^{\star}$     & 0.7(0.1)   & $1.79^{+0.80}_{-2.27}$ & 2.2(1.4) \\ 
SPT2203-41 &    5.194 & 5.3(0.9) & 53(10)   & 95(2)     & $5.5^{\star}$     & 0.6(0.1)   & $0.67^{+0.27}_{-0.72}$ & 2.2(1.4) \\ 
SPT2232-61 &    2.894 & 3.1(0.5) & 45(8)    & 100(6)    & $5.5^{\star}$     & 0.7(0.1)   & $1.32^{+0.50}_{-1.61}$ & 2.3(1.5) \\ 
SPT2311-45 &    2.507 & 3.2(0.5) & 35(6)    & 97(6)     & $5.5^{\star}$     & 0.4(0.1)   & $1.51^{+0.61}_{-1.64}$ & 1.0(0.6) \\ 
SPT2311-54 &    4.280 & 4.3(0.6) & 61(12)   & 76(7)     & 1.95(0.09)        & 1.7(0.2)   & $1.48^{+0.22}_{-0.31}$ & 5.9(0.9) \\ 
SPT2316-50 &    3.141 & 3.6(1.0) & 48(18)   & 106(5)    & $5.5^{\star}$     & 0.2(0.1)   & $0.30^{+0.12}_{-0.36}$ & 0.6(0.4) \\ 
SPT2319-55 &    5.293 & 5.8(0.9) & 60(17)   & 86(10)    & 7.89(1.85)        & 0.3(0.1)   & $0.46^{+0.14}_{-0.26}$ & 1.2(0.4) \\ 
SPT2332-53 &    2.726 & 2.1(0.3) & 64(12)   & 77(6)     & $5.5^{\star}$     & 1.7(0.3)   & $1.57^{+0.67}_{-1.99}$ & 3.8(2.4) \\ 
SPT2335-53 &    4.755 & 4.3(1.0) & 61(16)   & 81(2)     & $5.5^{\star, c}$  & 0.3(0.1)   & $0.17^{+0.06}_{-0.19}$ & 2.0(1.3) \\ 
SPT2340-59 &    3.862 & 4.3(0.9) & 55(14)   & 93(3)     & 3.37(0.31)        & 0.5(0.1)   & $0.48^{+0.07}_{-0.08}$ & 2.0(0.3) \\ 
SPT2349-50 &    2.876 & 3.7(0.6) & 49(15)   & 94(7)     & 2.15(0.09)        & 1.3(0.3)   & $2.49^{+0.39}_{-0.47}$ & 3.9(0.7) \\ 
SPT2349-52 &    3.900 & 4.1(1.0) & 63(17)   & 84(3)     & $5.5^{\star}$     & 0.19(0.04) & $0.12^{+0.05}_{-0.14}$ & 1.2(0.8) \\ 
SPT2349-56 &    4.302 & 4.6(0.7) & 59(11)   & 84(2)     & 1.00(0.00)$^c$    & 1.7(0.3)   & $0.99^{+0.10}_{-0.11}$ & 6.3(0.5) \\ 
SPT2351-57 &    5.811 & 4.6(1.1) & 70(15)   & 70(7)     & $5.5^{\star, b}$  & 0.5(0.1)   & $0.40^{+0.17}_{-0.49}$ & 2.9(1.9) \\ 
SPT2353-50 &    5.578 & 5.7(0.9) & 66(16)   & 81(10)    & $5.5^{\star, b}$  & 0.5(0.1)   & $0.67^{+0.31}_{-0.80}$ & 2.1(1.4) \\ 
SPT2354-58 &    1.867 & 1.9(0.3) & 61(17)   & 87(4)     & 6.29(0.36)        & 0.9(0.2)   & $1.39^{+0.16}_{-0.18}$ & 2.8(0.3) \\ 
SPT2357-51 &    3.070 & 3.8(0.6) & 43(6)    & 110(3)    & 2.85(0.12)        & 0.7(0.1)   & $1.29^{+0.14}_{-0.16}$ & 2.4(0.2) \\ 
\enddata
\tablecomments{Estimates of dust temperatures and FIR luminosities are derived from the modified blackbody fits described in Sec.~\ref{sec:sed}.  Using these parameters, the dust masses and star formation rates are then derived using Eq.~\ref{eq:mdust} and Eq.~\ref{eq:sfr}, respectively.  While not explictly given in the table, the total IR luminosity can be obtained using the SFR, as shown in Eq.~\ref{eq:sfr}.  The FIR luminosities, star formation rates and dust masses were corrected for gravitational amplification ($\mu$) according to Table 4 of~\citet{spilker16} where such models exist.  Where multiple source components exist, we use a flux weighted average to estimate $\mu$.  Though the intrinsic values are presented above, the apparent values can be obtained by multiplying by the magnification factor.  This table is available in machine-readable format at \href{https://github.com/spt-smg/publicdata}{https://github.com/spt-smg/publicdata}.}  
\tablenotetext{\star}{The sources that have not been modeled have parameters which have been corrected by the median gravitational amplification $\langle \mu_{870\1{\mu m}} \rangle = 5.5$~\citep{spilker16}.}
\tablenotetext{a}{Published by \citet{marrone18}}
\tablenotetext{b}{Presented in \citet{spilker16} as a cluster lens}
\tablenotetext{c}{DSFGs with observed multiplicities, as discussed in Sec.~\ref{dis:multiplicity}}.
\end{deluxetable*}

\bibliography{redshifts}

\begin{thebibliography}{}
\expandafter\ifx\csname natexlab\endcsname\relax\def\natexlab#1{#1}\fi

\bibitem[{{Abazajian} {et~al.}(2019){Abazajian}, {Addison}, {Adshead}, {Ahmed},
  {Allen}, {Alonso}, {Alvarez}, {Anderson}, {Arnold}, {Baccigalupi}, {Bailey},
  {Barkats}, {Barron}, {Barry}, {Bartlett}, {Basu Thakur}, {Battaglia},
  {Baxter}, {Bean}, {Bebek}, {Bender}, {Benson}, {Berger}, {Bhimani},
  {Bischoff}, {Bleem}, {Bocquet}, {Boddy}, {Bonato}, {Bond}, {Borrill},
  {Bouchet}, {Brown}, {Bryan}, {Burkhart}, {Buza}, {Byrum}, {Calabrese},
  {Calafut}, {Caldwell}, {Carlstrom}, {Carron}, {Cecil}, {Challinor}, {Chang},
  {Chinone}, {Cho}, {Cooray}, {Crawford}, {Crites}, {Cukierman}, {Cyr-Racine},
  {de Haan}, {de Zotti}, {Delabrouille}, {Demarteau}, {Devlin}, {Di Valentino},
  {Dobbs}, {Duff}, {Duivenvoorden}, {Dvorkin}, {Edwards}, {Eimer}, {Errard},
  {Essinger-Hileman}, {Fabbian}, {Feng}, {Ferraro}, {Filippini}, {Flauger},
  {Flaugher}, {Fraisse}, {Frolov}, {Galitzki}, {Galli}, {Ganga}, {Gerbino},
  {Gilchriese}, {Gluscevic}, {Green}, {Grin}, {Grohs}, {Gualtieri}, {Guarino},
  {Gudmundsson}, {Habib}, {Haller}, {Halpern}, {Halverson}, {Hanany},
  {Harrington}, {Hasegawa}, {Hasselfield}, {Hazumi}, {Heitmann}, {Henderson},
  {Henning}, {Hill}, {Hlozek}, {Holder}, {Holzapfel}, {Hubmayr},
  {Huffenberger}, {Huffer}, {Hui}, {Irwin}, {Johnson}, {Johnstone}, {Jones},
  {Karkare}, {Katayama}, {Kerby}, {Kernovsky}, {Keskitalo}, {Kisner}, {Knox},
  {Kosowsky}, {Kovac}, {Kovetz}, {Kuhlmann}, {Kuo}, {Kurita}, {Kusaka},
  {Lahteenmaki}, {Lawrence}, {Lee}, {Lewis}, {Li}, {Linder}, {Loverde},
  {Lowitz}, {Madhavacheril}, {Mantz}, {Matsuda}, {Mauskopf}, {McMahon},
  {McQuinn}, {Meerburg}, {Melin}, {Meyers}, {Millea}, {Mohr}, {Moncelsi},
  {Mroczkowski}, {Mukherjee}, {M{\"u}nchmeyer}, {Nagai}, {Nagy}, {Namikawa},
  {Nati}, {Natoli}, {Negrello}, {Newburgh}, {Niemack}, {Nishino}, {Nordby},
  {Novosad}, {O'Connor}, {Obied}, {Padin}, {Pandey}, {Partridge}, {Pierpaoli},
  {Pogosian}, {Pryke}, {Puglisi}, {Racine}, {Raghunathan}, {Rahlin},
  {Rajagopalan}, {Raveri}, {Reichanadter}, {Reichardt}, {Remazeilles}, {Rocha},
  {Roe}, {Roy}, {Ruhl}, {Salatino}, {Saliwanchik}, {Schaan}, {Schillaci},
  {Schmittfull}, {Scott}, {Sehgal}, {Shandera}, {Sheehy}, {Sherwin},
  {Shirokoff}, {Simon}, {Slosar}, {Somerville}, {Spergel}, {Staggs}, {Stark},
  {Stompor}, {Story}, {Stoughton}, {Suzuki}, {Tajima}, {Teply}, {Thompson},
  {Timbie}, {Tomasi}, {Treu}, {Tristram}, {Tucker}, {Umilt{\`a}}, {van
  Engelen}, {Vieira}, {Vieregg}, {Vogelsberger}, {Wang}, {Watson}, {White},
  {Whitehorn}, {Wollack}, {Kimmy Wu}, {Xu}, {Yasini}, {Yeck}, {Yoon}, {Young},
  \& {Zonca}}]{abazajian19}
{Abazajian}, K., {Addison}, G., {Adshead}, P., {et~al.} 2019, arXiv e-prints,
  arXiv:1907.04473

\bibitem[{{Aravena} {et~al.}(2013){Aravena}, {Murphy}, {Aguirre}, {Ashby},
  {Benson}, {Bothwell}, {Brodwin}, {Carlstrom}, {Chapman}, {Crawford}, {de
  Breuck}, {Fassnacht}, {Gonzalez}, {Greve}, {Gullberg}, {Hezaveh}, {Holder},
  {Holzapfel}, {Keisler}, {Malkan}, {Marrone}, {McIntyre}, {Reichardt},
  {Sharon}, {Spilker}, {Stalder}, {Stark}, {Vieira}, \& {Wei{\ss}}}]{aravena13}
{Aravena}, M., {Murphy}, E.~J., {Aguirre}, J.~E., {et~al.} 2013, \mnras, 433,
  498

\bibitem[{{Aravena} {et~al.}(2016){Aravena}, {Spilker}, {Bethermin},
  {Bothwell}, {Chapman}, {de Breuck}, {Furstenau}, {G{\'o}nzalez-L{\'o}pez},
  {Greve}, {Litke}, {Ma}, {Malkan}, {Marrone}, {Murphy}, {Stark}, {Strandet},
  {Vieira}, {Weiss}, {Welikala}, {Wong}, \& {Collier}}]{aravena16}
{Aravena}, M., {Spilker}, J.~S., {Bethermin}, M., {et~al.} 2016, \mnras, 457,
  4406

\bibitem[{{Asboth} {et~al.}(2016){Asboth}, {Conley}, {Sayers}, {B{\'e}thermin},
  {Chapman}, {Clements}, {Cooray}, {Dannerbauer}, {Farrah}, {Glenn}, {Golwala},
  {Halpern}, {Ibar}, {Ivison}, {Maloney}, {Marques-Chaves}, {Martinez-Navajas},
  {Oliver}, {P{\'e}rez-Fournon}, {Riechers}, {Rowan-Robinson}, {Scott},
  {Siegel}, {Vieira}, {Viero}, {Wang}, {Wardlow}, \& {Wheeler}}]{asboth16}
{Asboth}, V., {Conley}, A., {Sayers}, J., {et~al.} 2016, \mnras, 462, 1989

\bibitem[{{Balm}(2012)}]{balm12}
{Balm}, P. 2012, in Astronomical Society of the Pacific Conference Series, Vol.
  461, Astronomical Data Analysis Software and Systems XXI, ed. P.~{Ballester},
  D.~{Egret}, \& N.~P.~F. {Lorente}, 733

\bibitem[{{Baugh} {et~al.}(2005){Baugh}, {Lacey}, {Frenk}, {Granato}, {Silva},
  {Bressan}, {Benson}, \& {Cole}}]{baugh05}
{Baugh}, C.~M., {Lacey}, C.~G., {Frenk}, C.~S., {et~al.} 2005, \mnras, 356,
  1191

\bibitem[{{Benson} {et~al.}(2014){Benson}, {Ade}, {Ahmed}, {Allen}, {Arnold},
  {Austermann}, {Bender}, {Bleem}, {Carlstrom}, {Chang}, {Cho}, {Cliche},
  {Crawford}, {Cukierman}, {de Haan}, {Dobbs}, {Dutcher}, {Everett}, {Gilbert},
  {Halverson}, {Hanson}, {Harrington}, {Hattori}, {Henning}, {Hilton},
  {Holder}, {Holzapfel}, {Irwin}, {Keisler}, {Knox}, {Kubik}, {Kuo}, {Lee},
  {Leitch}, {Li}, {McDonald}, {Meyer}, {Montgomery}, {Myers}, {Natoli},
  {Nguyen}, {Novosad}, {Padin}, {Pan}, {Pearson}, {Reichardt}, {Ruhl},
  {Saliwanchik}, {Simard}, {Smecher}, {Sayre}, {Shirokoff}, {Stark}, {Story},
  {Suzuki}, {Thompson}, {Tucker}, {Vanderlinde}, {Vieira}, {Vikhlinin}, {Wang},
  {Yefremenko}, \& {Yoon}}]{benson14}
{Benson}, B.~A., {Ade}, P.~A.~R., {Ahmed}, Z., {et~al.} 2014, in Society of
  Photo-Optical Instrumentation Engineers (SPIE) Conference Series, Vol. 9153,
  \procspie, 91531P

\bibitem[{{B{\'e}thermin} {et~al.}(2015{\natexlab{a}}){B{\'e}thermin}, {De
  Breuck}, {Sargent}, \& {Daddi}}]{bethermin15b}
{B{\'e}thermin}, M., {De Breuck}, C., {Sargent}, M., \& {Daddi}, E.
  2015{\natexlab{a}}, \aap, 576, L9

\bibitem[{{B{\'e}thermin} {et~al.}(2012{\natexlab{a}}){B{\'e}thermin}, {Daddi},
  {Magdis}, {Sargent}, {Hezaveh}, {Elbaz}, {Le Borgne}, {Mullaney}, {Pannella},
  {Buat}, {Charmandaris}, {Lagache}, \& {Scott}}]{bethermin12b}
{B{\'e}thermin}, M., {Daddi}, E., {Magdis}, G., {et~al.} 2012{\natexlab{a}},
  \apjl, 757, L23

\bibitem[{{B{\'e}thermin} {et~al.}(2012{\natexlab{b}}){B{\'e}thermin}, {Le
  Floc'h}, {Ilbert}, {Conley}, {Lagache}, {Amblard}, {Arumugam}, {Aussel},
  {Berta}, {Bock}, {Boselli}, {Buat}, {Casey}, {Castro-Rodr{\'{\i}}guez},
  {Cava}, {Clements}, {Cooray}, {Dowell}, {Eales}, {Farrah}, {Franceschini},
  {Glenn}, {Griffin}, {Hatziminaoglou}, {Heinis}, {Ibar}, {Ivison},
  {Kartaltepe}, {Levenson}, {Magdis}, {Marchetti}, {Marsden}, {Nguyen},
  {O'Halloran}, {Oliver}, {Omont}, {Page}, {Panuzzo}, {Papageorgiou},
  {Pearson}, {P{\'e}rez-Fournon}, {Pohlen}, {Rigopoulou}, {Roseboom},
  {Rowan-Robinson}, {Salvato}, {Schulz}, {Scott}, {Seymour}, {Shupe}, {Smith},
  {Symeonidis}, {Trichas}, {Tugwell}, {Vaccari}, {Valtchanov}, {Vieira},
  {Viero}, {Wang}, {Xu}, \& {Zemcov}}]{bethermin12a}
{B{\'e}thermin}, M., {Le Floc'h}, E., {Ilbert}, O., {et~al.}
  2012{\natexlab{b}}, \aap, 542, A58

\bibitem[{{B{\'e}thermin} {et~al.}(2015{\natexlab{b}}){B{\'e}thermin}, {Daddi},
  {Magdis}, {Lagos}, {Sargent}, {Albrecht}, {Aussel}, {Bertoldi}, {Buat},
  {Galametz}, {Heinis}, {Ilbert}, {Karim}, {Koekemoer}, {Lacey}, {Le Floc'h},
  {Navarrete}, {Pannella}, {Schreiber}, {Smol{\v c}i{\'c}}, {Symeonidis}, \&
  {Viero}}]{bethermin15a}
{B{\'e}thermin}, M., {Daddi}, E., {Magdis}, G., {et~al.} 2015{\natexlab{b}},
  \aap, 573, A113

\bibitem[{{B{\'e}thermin} {et~al.}(2017){B{\'e}thermin}, {Wu}, {Lagache},
  {Davidzon}, {Ponthieu}, {Cousin}, {Wang}, {Dor{\'e}}, {Daddi}, \&
  {Lapi}}]{bethermin17}
{B{\'e}thermin}, M., {Wu}, H.-Y., {Lagache}, G., {et~al.} 2017, \aap, 607, A89

\bibitem[{{B{\'e}thermin} {et~al.}(2018){B{\'e}thermin}, {Greve}, {De Breuck},
  {Vieira}, {Aravena}, {Chapman}, {Chen}, {Dong}, {Hayward}, {Hezaveh},
  {Marrone}, {Narayanan}, {Phadke}, {Reuter}, {Spilker}, {Stark}, {Strandet},
  \& {Wei{\ss}}}]{bethermin18}
{B{\'e}thermin}, M., {Greve}, T.~R., {De Breuck}, C., {et~al.} 2018, \aap, 620,
  A115

\bibitem[{{Blain}(1996)}]{blain96}
{Blain}, A.~W. 1996, \mnras, 283, 1340

\bibitem[{{Blain} {et~al.}(2003){Blain}, {Barnard}, \& {Chapman}}]{blain03}
{Blain}, A.~W., {Barnard}, V.~E., \& {Chapman}, S.~C. 2003, \mnras, 338, 733

\bibitem[{{Blain} {et~al.}(2004){Blain}, {Chapman}, {Smail}, \&
  {Ivison}}]{blain04}
{Blain}, A.~W., {Chapman}, S.~C., {Smail}, I., \& {Ivison}, R. 2004, \apj, 611,
  725

\bibitem[{{Blain} \& {Longair}(1993)}]{blain93}
{Blain}, A.~W., \& {Longair}, M.~S. 1993, \mnras, 264, 509

\bibitem[{{Blain} {et~al.}(2002){Blain}, {Smail}, {Ivison}, {Kneib}, \&
  {Frayer}}]{blain02}
{Blain}, A.~W., {Smail}, I., {Ivison}, R.~J., {Kneib}, J.-P., \& {Frayer},
  D.~T. 2002, \physrep, 369, 111

\bibitem[{{Bothwell} {et~al.}(2013){Bothwell}, {Aguirre}, {Chapman}, {Marrone},
  {Vieira}, {Ashby}, {Aravena}, {Benson}, {Bock}, {Bradford}, {Brodwin},
  {Carlstrom}, {Crawford}, {de Breuck}, {Downes}, {Fassnacht}, {Gonzalez},
  {Greve}, {Gullberg}, {Hezaveh}, {Holder}, {Holzapfel}, {Ibar}, {Ivison},
  {Kamenetzky}, {Keisler}, {Lupu}, {Ma}, {Malkan}, {McIntyre}, {Murphy},
  {Nguyen}, {Reichardt}, {Rosenman}, {Spilker}, {Stalder}, {Stark}, {Strandet},
  {Vernet}, {Wei{\ss}}, \& {Welikala}}]{bothwell13b}
{Bothwell}, M.~S., {Aguirre}, J.~E., {Chapman}, S.~C., {et~al.} 2013, \apj,
  779, 67

\bibitem[{{Bothwell} {et~al.}(2017){Bothwell}, {Aguirre}, {Aravena},
  {Bethermin}, {Bisbas}, {Chapman}, {De Breuck}, {Gonzalez}, {Greve},
  {Hezaveh}, {Ma}, {Malkan}, {Marrone}, {Murphy}, {Spilker}, {Strandet},
  {Vieira}, \& {Wei{\ss}}}]{bothwell17}
{Bothwell}, M.~S., {Aguirre}, J.~E., {Aravena}, M., {et~al.} 2017, \mnras, 466,
  2825

\bibitem[{{Brisbin} {et~al.}(2017){Brisbin}, {Miettinen, Oskari}, {Aravena,
  Manuel}, {Smolci\'{}c, Vernesa}, {Delvecchio, Ivan}, {Jiang, Chunyan},
  {Magnelli, Benjamin}, {Albrecht, Marcus}, {Arancibia, Alejandra Mu\~noz},
  {Aussel, Herv\'e}, {Baran, Nikola}, {Bertoldi, Frank}, {B\'ethermin,
  Matthieu}, {Capak, Peter}, {Casey, Caitlin M.}, {Civano, Francesca},
  {Hayward, Christopher C.}, {Ilbert, Olivier}, {Karim, Alexander}, {Le Fevre,
  Olivier}, {Marchesi, Stefano}, {McCracken, Henry Joy}, {Navarrete, Felipe},
  {Novak, Mladen}, {Riechers, Dominik}, {Padilla, Nelson}, {Salvato, Mara},
  {Scott, Kimberly}, {Schinnerer, Eva}, {Sheth, Kartik}, \& {Tasca,
  Lidia}}]{brisbin17}
{Brisbin}, D., {Miettinen, Oskari}, {Aravena, Manuel}, {et~al.} 2017, A\&A,
  608, A15

\bibitem[{{Bussmann} {et~al.}(2013){Bussmann}, {P{\'e}rez-Fournon}, {Amber},
  {Calanog}, {Gurwell}, {Dannerbauer}, {De Bernardis}, {Fu}, {Harris}, {Krips},
  {Lapi}, {Maiolino}, {Omont}, {Riechers}, {Wardlow}, {Baker}, {Birkinshaw},
  {Bock}, {Bourne}, {Clements}, {Cooray}, {De Zotti}, {Dunne}, {Dye}, {Eales},
  {Farrah}, {Gavazzi}, {Gonz{\'a}lez Nuevo}, {Hopwood}, {Ibar}, {Ivison},
  {Laporte}, {Maddox}, {Mart{\'{\i}}nez-Navajas}, {Michalowski}, {Negrello},
  {Oliver}, {Roseboom}, {Scott}, {Serjeant}, {Smith}, {Smith}, {Streblyanska},
  {Valiante}, {van der Werf}, {Verma}, {Vieira}, {Wang}, \&
  {Wilner}}]{bussmann13}
{Bussmann}, R.~S., {P{\'e}rez-Fournon}, I., {Amber}, S., {et~al.} 2013, \apj,
  779, 25

\bibitem[{{Carlstrom} {et~al.}(2011){Carlstrom}, {Ade}, {Aird}, {Benson},
  {Bleem}, {Busetti}, {Chang}, {Chauvin}, {Cho}, {Crawford}, {Crites}, {Dobbs},
  {Halverson}, {Heimsath}, {Holzapfel}, {Hrubes}, {Joy}, {Keisler}, {Lanting},
  {Lee}, {Leitch}, {Leong}, {Lu}, {Lueker}, {Luong-van}, {McMahon}, {Mehl},
  {Meyer}, {Mohr}, {Montroy}, {Padin}, {Plagge}, {Pryke}, {Ruhl}, {Schaffer},
  {Schwan}, {Shirokoff}, {Spieler}, {Staniszewski}, {Stark}, {Tucker},
  {Vanderlinde}, {Vieira}, \& {Williamson}}]{carlstrom11}
{Carlstrom}, J.~E., {Ade}, P.~A.~R., {Aird}, K.~A., {et~al.} 2011, \pasp, 123,
  568

\bibitem[{{Casey}(2012)}]{casey12}
{Casey}, C.~M. 2012, \mnras, 425, 3094

\bibitem[{{Casey}(2016)}]{casey16}
---. 2016, \apj, 824, 36

\bibitem[{{Casey}(2020)}]{casey20}
---. 2020, arXiv e-prints, arXiv:2007.11012

\bibitem[{{Casey} {et~al.}(2014){Casey}, {Narayanan}, \& {Cooray}}]{casey14}
{Casey}, C.~M., {Narayanan}, D., \& {Cooray}, A. 2014, \physrep, 541, 45

\bibitem[{{Chapman} {et~al.}(2003){Chapman}, {Blain}, {Ivison}, \&
  {Smail}}]{chapman03}
{Chapman}, S.~C., {Blain}, A.~W., {Ivison}, R.~J., \& {Smail}, I.~R. 2003,
  \nat, 422, 695

\bibitem[{{Chapman} {et~al.}(2005){Chapman}, {Blain}, {Smail}, \&
  {Ivison}}]{chapman05}
{Chapman}, S.~C., {Blain}, A.~W., {Smail}, I., \& {Ivison}, R.~J. 2005, \apj,
  622, 772

\bibitem[{Chapman {et~al.}(2015)Chapman, Bertoldi, Smail, Steidel, Blain,
  Geach, Gurwell, Ivison, Petitpas, \& Reddy}]{chapman15}
Chapman, S.~C., Bertoldi, F., Smail, I., {et~al.} 2015, Monthly Notices of the
  Royal Astronomical Society, 453, 951

\bibitem[{{Chary} \& {Elbaz}(2001)}]{chary01}
{Chary}, R., \& {Elbaz}, D. 2001, \apj, 556, 562

\bibitem[{{Conley} {et~al.}(2011){Conley}, {Cooray}, {Vieira}, {Gonz{\'a}lez
  Solares}, {Kim}, {Aguirre}, {Amblard}, {Auld}, {Baker}, {Beelen}, {Blain},
  {Blundell}, {Bock}, {Bradford}, {Bridge}, {Brisbin}, {Burgarella},
  {Carpenter}, {Chanial}, {Chapin}, {Christopher}, {Clements}, {Cox},
  {Djorgovski}, {Dowell}, {Eales}, {Earle}, {Ellsworth-Bowers}, {Farrah},
  {Franceschini}, {Frayer}, {Fu}, {Gavazzi}, {Glenn}, {Griffin}, {Gurwell},
  {Halpern}, {Ibar}, {Ivison}, {Jarvis}, {Kamenetzky}, {Krips}, {Levenson},
  {Lupu}, {Mahabal}, {Maloney}, {Maraston}, {Marchetti}, {Marsden},
  {Matsuhara}, {Mortier}, {Murphy}, {Naylor}, {Neri}, {Nguyen}, {Oliver},
  {Omont}, {Page}, {Papageorgiou}, {Pearson}, {P{\'e}rez-Fournon}, {Pohlen},
  {Rangwala}, {Rawlings}, {Raymond}, {Riechers}, {Rodighiero}, {Roseboom},
  {Rowan-Robinson}, {Schulz}, {Scott}, {Scott}, {Serra}, {Seymour}, {Shupe},
  {Smith}, {Symeonidis}, {Tugwell}, {Vaccari}, {Valiante}, {Valtchanov},
  {Verma}, {Viero}, {Vigroux}, {Wang}, {Wiebe}, {Wright}, {Xu}, {Zeimann},
  {Zemcov}, \& {Zmuidzinas}}]{conley11}
{Conley}, A., {Cooray}, A., {Vieira}, J.~D., {et~al.} 2011, \apjl, 732, L35+

\bibitem[{{Cooke} {et~al.}(2018){Cooke}, {Smail}, {Swinbank}, {Stach}, {An},
  {Gullberg}, {Almaini}, {Simpson}, {Wardlow}, {Blain}, {Chapman}, {Chen},
  {Conselice}, {Coppin}, {Farrah}, {Maltby}, {Micha{\l}owski}, {Scott},
  {Simpson}, {Thomson}, \& {van der Werf}}]{cooke18}
{Cooke}, E.~A., {Smail}, I., {Swinbank}, A.~M., {et~al.} 2018, \apj, 861, 100

\bibitem[{{Coppin} {et~al.}(2009){Coppin}, {Smail}, {Alexander}, {Weiss},
  {Walter}, {Swinbank}, {Greve}, {Kovacs}, {De Breuck}, {Dickinson}, {Ibar},
  {Ivison}, {Reddy}, {Spinrad}, {Stern}, {Brandt}, {Chapman}, {Dannerbauer},
  {van Dokkum}, {Dunlop}, {Frayer}, {Gawiser}, {Geach}, {Huynh}, {Knudsen},
  {Koekemoer}, {Lehmer}, {Menten}, {Papovich}, {Rix}, {Schinnerer}, {Wardlow},
  \& {van der Werf}}]{coppin09}
{Coppin}, K.~E.~K., {Smail}, I., {Alexander}, D.~M., {et~al.} 2009, \mnras,
  395, 1905

\bibitem[{{Cortzen} {et~al.}(2020){Cortzen}, {Magdis}, {Valentino}, {Daddi},
  {Liu}, {Rigopoulou}, {Sargent}, {Riechers}, {Cormier}, {Hodge}, {Walter},
  {Elbaz}, {B{\'e}thermin}, {Greve}, {Kokorev}, \& {Toft}}]{cortzen20}
{Cortzen}, I., {Magdis}, G.~E., {Valentino}, F., {et~al.} 2020, \aap, 634, L14

\bibitem[{{Cunningham} {et~al.}(2020){Cunningham}, {Chapman}, {Aravena}, {De
  Breuck}, {B{\'e}thermin}, {Chen}, {Dong}, {Gonzalez}, {Greve}, {Litke}, {Ma},
  {Malkan}, {Marrone}, {Miller}, {Phadke}, {Reuter}, {Rotermund}, {Spilker},
  {Stark}, {Strandet}, {Vieira}, \& {Wei{\ss}}}]{cunningham20}
{Cunningham}, D.~J.~M., {Chapman}, S.~C., {Aravena}, M., {et~al.} 2020, \mnras,
  494, 4090

\bibitem[{da~Cunha {et~al.}(2013)da~Cunha, Groves, Walter, Decarli, Weiss,
  Bertoldi, Carilli, Daddi, Elbaz, Ivison, \& et~al.}]{dacunha13}
da~Cunha, E., Groves, B., Walter, F., {et~al.} 2013, The Astrophysical Journal,
  766, 13

\bibitem[{{da Cunha} {et~al.}(2015){da Cunha}, {Walter}, {Smail}, {Swinbank},
  {Simpson}, {Decarli}, {Hodge}, {Weiss}, {van der Werf}, {Bertoldi},
  {Chapman}, {Cox}, {Danielson}, {Dannerbauer}, {Greve}, {Ivison}, {Karim}, \&
  {Thomson}}]{dacunha15}
{da Cunha}, E., {Walter}, F., {Smail}, I.~R., {et~al.} 2015, \apj, 806, 110

\bibitem[{{Danielson} {et~al.}(2017){Danielson}, {Swinbank}, {Smail},
  {Simpson}, {Casey}, {Chapman}, {da Cunha}, {Hodge}, {Walter}, \&
  {Wardlow}}]{danielson17}
{Danielson}, A.~L.~R., {Swinbank}, A.~M., {Smail}, I., {et~al.} 2017, \apj,
  840, 78

\bibitem[{{Dong} {et~al.}(2019){Dong}, {Spilker}, {Gonzalez}, {Apostolovski},
  {Aravena}, {B{\'e}thermin}, {Chapman}, {Chen}, {Hayward}, {Hezaveh}, {Litke},
  {Ma}, {Marrone}, {Morningstar}, {Phadke}, {Reuter}, {Sreevani}, {Stark},
  {Vieira}, \& {Wei{\ss}}}]{dong19}
{Dong}, C., {Spilker}, J.~S., {Gonzalez}, A.~H., {et~al.} 2019, \apj, 873, 50

\bibitem[{{Dowell} {et~al.}(2014){Dowell}, {Conley}, {Glenn}, {Arumugam},
  {Asboth}, {Aussel}, {Bertoldi}, {B{\'e}thermin}, {Bock}, {Boselli}, {Bridge},
  {Buat}, {Burgarella}, {Cabrera-Lavers}, {Casey}, {Chapman}, {Clements},
  {Conversi}, {Cooray}, {Dannerbauer}, {De Bernardis}, {Ellsworth-Bowers},
  {Farrah}, {Franceschini}, {Griffin}, {Gurwell}, {Halpern}, {Hatziminaoglou},
  {Heinis}, {Ibar}, {Ivison}, {Laporte}, {Marchetti},
  {Mart{\'{\i}}nez-Navajas}, {Marsden}, {Morrison}, {Nguyen}, {O'Halloran},
  {Oliver}, {Omont}, {Page}, {Papageorgiou}, {Pearson}, {Petitpas},
  {P{\'e}rez-Fournon}, {Pohlen}, {Riechers}, {Rigopoulou}, {Roseboom},
  {Rowan-Robinson}, {Sayers}, {Schulz}, {Scott}, {Seymour}, {Shupe}, {Smith},
  {Streblyanska}, {Symeonidis}, {Vaccari}, {Valtchanov}, {Vieira}, {Viero},
  {Wang}, {Wardlow}, {Xu}, \& {Zemcov}}]{dowell14}
{Dowell}, C.~D., {Conley}, A., {Glenn}, J., {et~al.} 2014, \apj, 780, 75

\bibitem[{{Dudzevi{\v{c}}i{\={u}}t{\.{e}}}
  {et~al.}(2020){Dudzevi{\v{c}}i{\={u}}t{\.{e}}}, {Smail}, {Swinbank}, {Stach},
  {Almaini}, {da Cunha}, {An}, {Arumugam}, {Birkin}, {Blain}, {Chapman},
  {Chen}, {Conselice}, {Coppin}, {Dunlop}, {Farrah}, {Geach}, {Gullberg},
  {Hartley}, {Hodge}, {Ivison}, {Maltby}, {Scott}, {Simpson}, {Simpson},
  {Thomson}, {Walter}, {Wardlow}, {Weiss}, \& {van der Werf}}]{dudzeviciute20}
{Dudzevi{\v{c}}i{\={u}}t{\.{e}}}, U., {Smail}, I., {Swinbank}, A.~M., {et~al.}
  2020, \mnras, 494, 3828

\bibitem[{{Dunne} {et~al.}(2003){Dunne}, {Eales}, \& {Edmunds}}]{dunne03}
{Dunne}, L., {Eales}, S.~A., \& {Edmunds}, M.~G. 2003, \mnras, 341, 589

\bibitem[{{Eales} {et~al.}(2010){Eales}, {Dunne}, {Clements}, {Cooray}, {de
  Zotti}, {Dye}, {Ivison}, {Jarvis}, {Lagache}, {Maddox}, {Negrello},
  {Serjeant}, {Thompson}, {Kampen}, {Amblard}, {Andreani}, {Baes}, {Beelen},
  {Bendo}, {Benford}, {Bertoldi}, {Bock}, {Bonfield}, {Boselli}, {Bridge},
  {Buat}, {Burgarella}, {Carlberg}, {Cava}, {Chanial}, {Charlot},
  {Christopher}, {Coles}, {Cortese}, {Dariush}, {da Cunha}, {Dalton}, {Danese},
  {Dannerbauer}, {Driver}, {Dunlop}, {Fan}, {Farrah}, {Frayer}, {Frenk},
  {Geach}, {Gardner}, {Gomez}, {Gonz{\'a}lez-Nuevo}, {Gonz{\'a}lez-Solares},
  {Griffin}, {Hardcastle}, {Hatziminaoglou}, {Herranz}, {Hughes}, {Ibar},
  {Jeong}, {Lacey}, {Lapi}, {Lawrence}, {Lee}, {Leeuw}, {Liske},
  {L{\'o}pez-Caniego}, {M{\"u}ller}, {Nandra}, {Panuzzo}, {Papageorgiou},
  {Patanchon}, {Peacock}, {Pearson}, {Phillipps}, {Pohlen}, {Popescu},
  {Rawlings}, {Rigby}, {Rigopoulou}, {Robotham}, {Rodighiero}, {Sansom},
  {Schulz}, {Scott}, {Smith}, {Sibthorpe}, {Smail}, {Stevens}, {Sutherland},
  {Takeuchi}, {Tedds}, {Temi}, {Tuffs}, {Trichas}, {Vaccari}, {Valtchanov},
  {van der Werf}, {Verma}, {Vieria}, {Vlahakis}, \& {White}}]{eales10}
{Eales}, S., {Dunne}, L., {Clements}, D., {et~al.} 2010, \pasp, 122, 499

\bibitem[{{Engel} {et~al.}(2010){Engel}, {Tacconi}, {Davies}, {Neri}, {Smail},
  {Chapman}, {Genzel}, {Cox}, {Greve}, {Ivison}, {Blain}, {Bertoldi}, \&
  {Omont}}]{engel10}
{Engel}, H., {Tacconi}, L.~J., {Davies}, R.~I., {et~al.} 2010, \apj, 724, 233

\bibitem[{{Everett} {et~al.}(2020){Everett}, {Zhang}, {Crawford}, {Vieira},
  {Aravena}, {Archipley}, {Austermann}, {Benson}, {Bleem}, {Carlstrom},
  {Chang}, {Chapman}, {Crites}, {de Haan}, {Dobbs}, {George}, {Halverson},
  {Harrington}, {Holder}, {Holzapfel}, {Hrubes}, {Knox}, {Lee}, {Luong-Van},
  {Mangian}, {Marrone}, {McMahon}, {Meyer}, {Mocanu}, {Mohr}, {Natoli},
  {Padin}, {Pryke}, {Reichardt}, {Reuter}, {Ruhl}, {Sayre}, {Schaffer},
  {Shirokoff}, {Spilker}, {Stalder}, {Staniszewski}, {Stark}, {Story},
  {Switzer}, {Vanderlinde}, {Weiss}, \& {Williamson}}]{everett20}
{Everett}, W.~B., {Zhang}, L., {Crawford}, T.~M., {et~al.} 2020, arXiv
  e-prints, arXiv:2003.03431

\bibitem[{{Foreman-Mackey} {et~al.}(2013){Foreman-Mackey}, {Hogg}, {Lang}, \&
  {Goodman}}]{foremanmackey13}
{Foreman-Mackey}, D., {Hogg}, D.~W., {Lang}, D., \& {Goodman}, J. 2013, \pasp,
  125, 306

\bibitem[{{Fu} {et~al.}(2013){Fu}, {Cooray}, {Feruglio}, {Ivison}, {Riechers},
  {Gurwell}, {Bussmann}, {Harris}, {Altieri}, {Aussel}, {Baker}, {Bock},
  {Boylan-Kolchin}, {Bridge}, {Calanog}, {Casey}, {Cava}, {Chapman},
  {Clements}, {Conley}, {Cox}, {Farrah}, {Frayer}, {Hopwood}, {Jia}, {Magdis},
  {Marsden}, {Mart{\'{\i}}nez-Navajas}, {Negrello}, {Neri}, {Oliver}, {Omont},
  {Page}, {P{\'e}rez-Fournon}, {Schulz}, {Scott}, {Smith}, {Vaccari},
  {Valtchanov}, {Vieira}, {Viero}, {Wang}, {Wardlow}, \& {Zemcov}}]{fu13}
{Fu}, H., {Cooray}, A., {Feruglio}, C., {et~al.} 2013, \nat, 498, 338

\bibitem[{Fudamoto {et~al.}(2017)Fudamoto, Ivison, Oteo, Krips, Zhang, Weiss,
  Dannerbauer, Omont, Chapman, Christensen, Arumugam, Bertoldi, Bremer,
  Clements, Dunne, Eales, Greenslade, Maddox, Martinez-Navajas, Michalowski,
  Pérez-Fournon, Riechers, Simpson, Stalder, Valiante, \& van~der
  Werf}]{fudamoto17}
Fudamoto, Y., Ivison, R.~J., Oteo, I., {et~al.} 2017, Monthly Notices of the
  Royal Astronomical Society, 472, 2028

\bibitem[{{Genzel} {et~al.}(2015){Genzel}, {Tacconi}, {Lutz}, {Saintonge},
  {Berta}, {Magnelli}, {Combes}, {Garc{\'\i}a-Burillo}, {Neri}, {Bolatto},
  {Contini}, {Lilly}, {Boissier}, {Boone}, {Bouch{\'e}}, {Bournaud}, {Burkert},
  {Carollo}, {Colina}, {Cooper}, {Cox}, {Feruglio}, {F{\"o}rster Schreiber},
  {Freundlich}, {Gracia-Carpio}, {Juneau}, {Kovac}, {Lippa}, {Naab}, {Salome},
  {Renzini}, {Sternberg}, {Walter}, {Weiner}, {Weiss}, \& {Wuyts}}]{genzel15}
{Genzel}, R., {Tacconi}, L.~J., {Lutz}, D., {et~al.} 2015, \apj, 800, 20

\bibitem[{{Gladders} {et~al.}(2013){Gladders}, {Rigby}, {Sharon}, {Wuyts},
  {Abramson}, {Dahle}, {Persson}, {Monson}, {Kelson}, {Benford}, {Murphy},
  {Bayliss}, {Finkelstein}, {Koester}, {Bans}, {Baxter}, \&
  {Helsby}}]{gladders12}
{Gladders}, M.~D., {Rigby}, J.~R., {Sharon}, K., {et~al.} 2013, \apj, 764, 177

\bibitem[{{Greve} {et~al.}(2008){Greve}, {Pope}, {Scott}, {Ivison}, {Borys},
  {Conselice}, \& {Bertoldi}}]{greve08}
{Greve}, T.~R., {Pope}, A., {Scott}, D., {et~al.} 2008, \mnras, 389, 1489

\bibitem[{{Greve} {et~al.}(2012){Greve}, {Vieira}, {Wei{\ss}}, {Aguirre},
  {Aird}, {Ashby}, {Benson}, {Bleem}, {Bradford}, {Brodwin}, {Carlstrom},
  {Chang}, {Chapman}, {Crawford}, {de Breuck}, {de Haan}, {Dobbs}, {Downes},
  {Fassnacht}, {Fazio}, {George}, {Gladders}, {Gonzalez}, {Halverson},
  {Hezaveh}, {High}, {Holder}, {Holzapfel}, {Hoover}, {Hrubes}, {Johnson},
  {Keisler}, {Knox}, {Lee}, {Leitch}, {Lueker}, {Luong-Van}, {Malkan},
  {Marrone}, {McIntyre}, {McMahon}, {Mehl}, {Menten}, {Meyer}, {Montroy},
  {Murphy}, {Natoli}, {Padin}, {Plagge}, {Pryke}, {Reichardt}, {Rest},
  {Rosenman}, {Ruel}, {Ruhl}, {Schaffer}, {Sharon}, {Shaw}, {Shirokoff},
  {Stalder}, {Stanford}, {Staniszewski}, {Stark}, {Story}, {Vanderlinde},
  {Walsh}, {Welikala}, \& {Williamson}}]{greve12}
{Greve}, T.~R., {Vieira}, J.~D., {Wei{\ss}}, A., {et~al.} 2012, \apj, 756, 101

\bibitem[{{Griffin} {et~al.}(2010){Griffin}, {Abergel}, {Abreu}, {Ade},
  {Andr{\'e}}, {Augueres}, {Babbedge}, {Bae}, {Baillie}, {Baluteau}, {Barlow},
  {Bendo}, {Benielli}, {Bock}, {Bonhomme}, {Brisbin}, {Brockley-Blatt},
  {Caldwell}, {Cara}, {Castro-Rodriguez}, {Cerulli}, {Chanial}, {Chen},
  {Clark}, {Clements}, {Clerc}, {Coker}, {Communal}, {Conversi}, {Cox},
  {Crumb}, {Cunningham}, {Daly}, {Davis}, {de Antoni}, {Delderfield}, {Devin},
  {di Giorgio}, {Didschuns}, {Dohlen}, {Donati}, {Dowell}, {Dowell}, {Duband},
  {Dumaye}, {Emery}, {Ferlet}, {Ferrand}, {Fontignie}, {Fox}, {Franceschini},
  {Frerking}, {Fulton}, {Garcia}, {Gastaud}, {Gear}, {Glenn}, {Goizel},
  {Griffin}, {Grundy}, {Guest}, {Guillemet}, {Hargrave}, {Harwit}, {Hastings},
  {Hatziminaoglou}, {Herman}, {Hinde}, {Hristov}, {Huang}, {Imhof}, {Isaak},
  {Israelsson}, {Ivison}, {Jennings}, {Kiernan}, {King}, {Lange}, {Latter},
  {Laurent}, {Laurent}, {Leeks}, {Lellouch}, {Levenson}, {Li}, {Li},
  {Lilienthal}, {Lim}, {Liu}, {Lu}, {Madden}, {Mainetti}, {Marliani}, {McKay},
  {Mercier}, {Molinari}, {Morris}, {Moseley}, {Mulder}, {Mur}, {Naylor},
  {Nguyen}, {O'Halloran}, {Oliver}, {Olofsson}, {Olofsson}, {Orfei}, {Page},
  {Pain}, {Panuzzo}, {Papageorgiou}, {Parks}, {Parr-Burman}, {Pearce},
  {Pearson}, {P{\'e}rez-Fournon}, {Pinsard}, {Pisano}, {Podosek}, {Pohlen},
  {Polehampton}, {Pouliquen}, {Rigopoulou}, {Rizzo}, {Roseboom}, {Roussel},
  {Rowan-Robinson}, {Rownd}, {Saraceno}, {Sauvage}, {Savage}, {Savini},
  {Sawyer}, {Scharmberg}, {Schmitt}, {Schneider}, {Schulz}, {Schwartz},
  {Shafer}, {Shupe}, {Sibthorpe}, {Sidher}, {Smith}, {Smith}, {Smith},
  {Spencer}, {Stobie}, {Sudiwala}, {Sukhatme}, {Surace}, {Stevens}, {Swinyard},
  {Trichas}, {Tourette}, {Triou}, {Tseng}, {Tucker}, {Turner}, {Vaccari},
  {Valtchanov}, {Vigroux}, {Virique}, {Voellmer}, {Walker}, {Ward}, {Waskett},
  {Weilert}, {Wesson}, {White}, {Whitehouse}, {Wilson}, {Winter}, {Woodcraft},
  {Wright}, {Xu}, {Zavagno}, {Zemcov}, {Zhang}, \& {Zonca}}]{griffin10}
{Griffin}, M.~J., {Abergel}, A., {Abreu}, A., {et~al.} 2010, \aap, 518, L3

\bibitem[{{Gullberg} {et~al.}(2015){Gullberg}, {De Breuck}, {Vieira},
  {Wei{\ss}}, {Aguirre}, {Aravena}, {B{\'e}thermin}, {Bradford}, {Bothwell},
  {Carlstrom}, {Chapman}, {Fassnacht}, {Gonzalez}, {Greve}, {Hezaveh},
  {Holzapfel}, {Husband}, {Ma}, {Malkan}, {Marrone}, {Menten}, {Murphy},
  {Reichardt}, {Spilker}, {Stark}, {Strandet}, \& {Welikala}}]{gullberg15}
{Gullberg}, B., {De Breuck}, C., {Vieira}, J.~D., {et~al.} 2015, \mnras, 449,
  2883

\bibitem[{{Harris} {et~al.}(2012){Harris}, {Baker}, {Frayer}, {Smail},
  {Swinbank}, {Riechers}, {van der Werf}, {Auld}, {Baes}, {Bussmann},
  {Buttiglione}, {Cava}, {Clements}, {Cooray}, {Dannerbauer}, {Dariush},
  {DeZotti}, {Dunne}, {Dye}, {Eales}, {Fritz}, {Gonzalez-Nuevo}, {Hopwood},
  {Ibar}, {Ivison}, {Jarvis}, {Maddox}, {Negrello}, {Rigby}, {Smith}, {Temi},
  \& {Wardlow}}]{harris12}
{Harris}, A.~I., {Baker}, A.~J., {Frayer}, D.~T., {et~al.} 2012, ArXiv
  e-prints, arXiv:1204.4706

\bibitem[{{Hayward} {et~al.}(2011){Hayward}, {Kere{\v s}}, {Jonsson},
  {Narayanan}, {Cox}, \& {Hernquist}}]{hayward11}
{Hayward}, C.~C., {Kere{\v s}}, D., {Jonsson}, P., {et~al.} 2011, \apj, 743,
  159

\bibitem[{{Hayward} {et~al.}(2013){Hayward}, {Narayanan}, {Kere{\v s}},
  {Jonsson}, {Hopkins}, {Cox}, \& {Hernquist}}]{hayward13}
{Hayward}, C.~C., {Narayanan}, D., {Kere{\v s}}, D., {et~al.} 2013, \mnras,
  428, 2529

\bibitem[{{Helou} {et~al.}(1988){Helou}, {Khan}, {Malek}, \&
  {Boehmer}}]{helou88}
{Helou}, G., {Khan}, I.~R., {Malek}, L., \& {Boehmer}, L. 1988, \apjs, 68, 151

\bibitem[{{Heyminck} {et~al.}(2006){Heyminck}, {Kasemann}, {G{\"u}sten}, {de
  Lange}, \& {Graf}}]{heyminck06}
{Heyminck}, S., {Kasemann}, C., {G{\"u}sten}, R., {de Lange}, G., \& {Graf},
  U.~U. 2006, \aap, 454, L21

\bibitem[{{Hezaveh} \& {Holder}(2011)}]{hezaveh11}
{Hezaveh}, Y.~D., \& {Holder}, G.~P. 2011, \apj, 734, 52

\bibitem[{{Hezaveh} {et~al.}(2012){Hezaveh}, {Marrone}, \&
  {Holder}}]{hezaveh12a}
{Hezaveh}, Y.~D., {Marrone}, D.~P., \& {Holder}, G.~P. 2012, \apj, 761, 20

\bibitem[{{Hezaveh} {et~al.}(2013){Hezaveh}, {Marrone}, {Fassnacht}, {Spilker},
  {Vieira}, {Aguirre}, {Aird}, {Aravena}, {Ashby}, {Bayliss}, {Benson},
  {Bleem}, {Bothwell}, {Brodwin}, {Carlstrom}, {Chang}, {Chapman}, {Crawford},
  {Crites}, {De Breuck}, {de Haan}, {Dobbs}, {Fomalont}, {George}, {Gladders},
  {Gonzalez}, {Greve}, {Halverson}, {High}, {Holder}, {Holzapfel}, {Hoover},
  {Hrubes}, {Husband}, {Hunter}, {Keisler}, {Lee}, {Leitch}, {Lueker},
  {Luong-Van}, {Malkan}, {McIntyre}, {McMahon}, {Mehl}, {Menten}, {Meyer},
  {Mocanu}, {Murphy}, {Natoli}, {Padin}, {Plagge}, {Reichardt}, {Rest}, {Ruel},
  {Ruhl}, {Sharon}, {Schaffer}, {Shaw}, {Shirokoff}, {Stalder}, {Staniszewski},
  {Stark}, {Story}, {Vanderlinde}, {Wei{\ss}}, {Welikala}, \&
  {Williamson}}]{hezaveh13}
{Hezaveh}, Y.~D., {Marrone}, D.~P., {Fassnacht}, C.~D., {et~al.} 2013, \apj,
  767, 132

\bibitem[{{Hill} {et~al.}(2020){Hill}, {Chapman}, {Scott}, {Apostolovski},
  {Aravena}, {Bethermin}, {Bradford}, {de Breuck}, {Canning}, {Dong},
  {Gonzalez}, {Greve}, {Hayward}, {Hezaveh}, {Litke}, {Malkan}, {Marrone},
  {Phadke}, {Reuter}, {Spilker}, {Vieira}, \& {Weiss}}]{hill20}
{Hill}, R., {Chapman}, S., {Scott}, D., {et~al.} 2020, arXiv e-prints,
  arXiv:2002.11600

\bibitem[{{Hodge} {et~al.}(2013){Hodge}, {Carilli}, {Walter}, {Daddi}, \&
  {Riechers}}]{hodge13}
{Hodge}, J.~A., {Carilli}, C.~L., {Walter}, F., {Daddi}, E., \& {Riechers}, D.
  2013, \apj, 776, 22

\bibitem[{{Hodge} {et~al.}(2016){Hodge}, {Swinbank}, {Simpson}, {Smail},
  {Walter}, {Alexander}, {Bertoldi}, {Biggs}, {Brandt}, {Chapman}, {Chen},
  {Coppin}, {Cox}, {Dannerbauer}, {Edge}, {Greve}, {Ivison}, {Karim},
  {Knudsen}, {Menten}, {Rix}, {Schinnerer}, {Wardlow}, {Weiss}, \& {van der
  Werf}}]{hodge16}
{Hodge}, J.~A., {Swinbank}, A.~M., {Simpson}, J.~M., {et~al.} 2016, \apj, 833,
  103

\bibitem[{{Ibar} {et~al.}(2010){Ibar}, {Ivison}, {Cava}, {Rodighiero},
  {Buttiglione}, {Temi}, {Frayer}, {Fritz}, {Leeuw}, {Baes}, {Rigby}, {Verma},
  {Serjeant}, {M{\"u}ller}, {Auld}, {Dariush}, {Dunne}, {Eales}, {Maddox},
  {Panuzzo}, {Pascale}, {Pohlen}, {Smith}, {de Zotti}, {Vaccari}, {Hopwood},
  {Cooray}, {Burgarella}, \& {Jarvis}}]{ibar10}
{Ibar}, E., {Ivison}, R.~J., {Cava}, A., {et~al.} 2010, \mnras, 409, 38

\bibitem[{{Ikarashi} {et~al.}(2015){Ikarashi}, {Ivison}, {Caputi}, {Aretxaga},
  {Dunlop}, {Hatsukade}, {Hughes}, {Iono}, {Izumi}, {Kawabe}, {Kohno}, {Lagos},
  {Motohara}, {Nakanishi}, {Ohta}, {Tamura}, {Umehata}, {Wilson}, {Yabe}, \&
  {Yun}}]{ikarashi15}
{Ikarashi}, S., {Ivison}, R.~J., {Caputi}, K.~I., {et~al.} 2015, \apj, 810, 133

\bibitem[{Ikarashi {et~al.}(2017)Ikarashi, Caputi, Ohta, Ivison, Lagos,
  Bisigello, Hatsukade, Aretxaga, Dunlop, Hughes, Iono, Izumi, Kashikawa,
  Koyama, Kawabe, Kohno, Motohara, Nakanishi, Tamura, Umehata, Wilson, Yabe, \&
  Yun}]{ikarashi17}
Ikarashi, S., Caputi, K.~I., Ohta, K., {et~al.} 2017, The Astrophysical
  Journal, 849, L36

\bibitem[{{Ivison} {et~al.}(2013){Ivison}, {Swinbank}, {Smail}, {Harris},
  {Bussmann}, {Cooray}, {Cox}, {Fu}, {Kov{\'a}cs}, {Krips}, {Narayanan},
  {Negrello}, {Neri}, {Pe{\~n}arrubia}, {Richard}, {Riechers}, {Rowlands},
  {Staguhn}, {Targett}, {Amber}, {Baker}, {Bourne}, {Bertoldi}, {Bremer},
  {Calanog}, {Clements}, {Dannerbauer}, {Dariush}, {De Zotti}, {Dunne},
  {Eales}, {Farrah}, {Fleuren}, {Franceschini}, {Geach}, {George}, {Helly},
  {Hopwood}, {Ibar}, {Jarvis}, {Kneib}, {Maddox}, {Omont}, {Scott}, {Serjeant},
  {Smith}, {Thompson}, {Valiante}, {Valtchanov}, {Vieira}, \& {van der
  Werf}}]{ivison13}
{Ivison}, R.~J., {Swinbank}, A.~M., {Smail}, I., {et~al.} 2013, \apj, 772, 137

\bibitem[{{Ivison} {et~al.}(2016){Ivison}, {Lewis}, {Weiss}, {Arumugam},
  {Simpson}, {Holland}, {Maddox}, {Dunne}, {Valiante}, {van der Werf}, {Omont},
  {Dannerbauer}, {Smail}, {Bertoldi}, {Bremer}, {Bussmann}, {Cai}, {Clements},
  {Cooray}, {De Zotti}, {Eales}, {Fuller}, {Gonzalez-Nuevo}, {Ibar},
  {Negrello}, {Oteo}, {P{\'e}rez-Fournon}, {Riechers}, {Stevens}, {Swinbank},
  \& {Wardlow}}]{ivison16}
{Ivison}, R.~J., {Lewis}, A.~J.~R., {Weiss}, A., {et~al.} 2016, \apj, 832, 78

\bibitem[{{Jarugula} {et~al.}(2019){Jarugula}, {Vieira}, {Spilker},
  {Apostolovski}, {Aravena}, {B{\'e}thermin}, {de Breuck}, {Chen},
  {Cunningham}, {Dong}, {Greve}, {Hayward}, {Hezaveh}, {Litke}, {Mangian},
  {Narayanan}, {Phadke}, {Reuter}, {Van der Werf}, \& {Weiss}}]{jarugula19}
{Jarugula}, S., {Vieira}, J.~D., {Spilker}, J.~S., {et~al.} 2019, \apj, 880, 92

\bibitem[{Jin {et~al.}(2019)Jin, Daddi, Magdis, Liu, Schinnerer, Papadopoulos,
  Gu, Gao, \& Calabr{\`{o}}}]{jin19}
Jin, S., Daddi, E., Magdis, G.~E., {et~al.} 2019, The Astrophysical Journal,
  887, 144

\bibitem[{{Klein} {et~al.}(2014){Klein}, {Ciechanowicz}, {Leinz}, {Heyminck},
  {Güsten}, {Kasemann}, {Wunsch}, {Maier}, \& {Sekimoto}}]{klein14}
{Klein}, T., {Ciechanowicz}, M., {Leinz}, C., {et~al.} 2014, IEEE Transactions
  on Terahertz Science and Technology, 4, 588

\bibitem[{{Koprowski} {et~al.}(2014){Koprowski}, {Dunlop}, {Micha{\l}owski},
  {Cirasuolo}, \& {Bowler}}]{koprowski14}
{Koprowski}, M.~P., {Dunlop}, J.~S., {Micha{\l}owski}, M.~J., {Cirasuolo}, M.,
  \& {Bowler}, R.~A.~A. 2014, \mnras, 444, 117

\bibitem[{{Kroupa}(2001)}]{kroupa01}
{Kroupa}, P. 2001, \mnras, 322, 231

\bibitem[{{Lacey} {et~al.}(2010){Lacey}, {Baugh}, {Frenk}, {Benson}, {Orsi},
  {Silva}, {Granato}, \& {Bressan}}]{lacey10}
{Lacey}, C.~G., {Baugh}, C.~M., {Frenk}, C.~S., {et~al.} 2010, \mnras, 405, 2

\bibitem[{{Lagos} {et~al.}(2018){Lagos}, {Tobar}, {Robotham}, {Obreschkow},
  {Mitchell}, {Power}, \& {Elahi}}]{lagos18}
{Lagos}, C. d.~P., {Tobar}, R.~J., {Robotham}, A. S.~G., {et~al.} 2018, \mnras,
  481, 3573

\bibitem[{{Lagos} {et~al.}(2019){Lagos}, {Robotham}, {Trayford}, {Tobar},
  {Bravo}, {Bellstedt}, {Davies}, {Driver}, {Elahi}, {Obreschkow}, \&
  {Power}}]{lagos19}
{Lagos}, C. d.~P., {Robotham}, A. S.~G., {Trayford}, J.~W., {et~al.} 2019,
  \mnras, 489, 4196

\bibitem[{{Leitherer} {et~al.}(1999){Leitherer}, {Schaerer}, {Goldader},
  {Delgado}, {Robert}, {Kune}, {de Mello}, {Devost}, \&
  {Heckman}}]{leitherer99}
{Leitherer}, C., {Schaerer}, D., {Goldader}, J.~D., {et~al.} 1999, \apjs, 123,
  3

\bibitem[{{Lewis} {et~al.}(2018){Lewis}, {Ivison}, {Best}, {Simpson}, {Weiss},
  {Oteo}, {Zhang}, {Arumugam}, {Bremer}, {Chapman}, {Clements}, {Dannerbauer},
  {Dunne}, {Eales}, {Maddox}, {Oliver}, {Omont}, {Riechers}, {Serjeant},
  {Valiante}, {Wardlow}, {van der Werf}, \& {De Zotti}}]{lewis18}
{Lewis}, A.~J.~R., {Ivison}, R.~J., {Best}, P.~N., {et~al.} 2018, \apj, 862, 96

\bibitem[{{Liang} {et~al.}(2019){Liang}, {Feldmann}, {Kere{\v{s}}}, {Scoville},
  {Hayward}, {Faucher-Gigu{\`e}re}, {Schreiber}, {Ma}, {Hopkins}, \&
  {Quataert}}]{liang19}
{Liang}, L., {Feldmann}, R., {Kere{\v{s}}}, D., {et~al.} 2019, \mnras, 489,
  1397

\bibitem[{{Litke} {et~al.}(2019){Litke}, {Marrone}, {Spilker}, {Aravena},
  {B{\'e}thermin}, {Chapman}, {Chen}, {de Breuck}, {Dong}, {Gonzalez}, {Greve},
  {Hayward}, {Hezaveh}, {Jarugula}, {Ma}, {Morningstar}, {Narayanan}, {Phadke},
  {Reuter}, {Vieira}, \& {Weiss}}]{litke19}
{Litke}, K.~C., {Marrone}, D.~P., {Spilker}, J.~S., {et~al.} 2019, \apj, 870,
  80

\bibitem[{{Lovell} {et~al.}(2020){Lovell}, {Geach}, {Dav{\'e}}, {Narayanan}, \&
  {Li}}]{lovell20}
{Lovell}, C.~C., {Geach}, J.~E., {Dav{\'e}}, R., {Narayanan}, D., \& {Li}, Q.
  2020, arXiv e-prints, arXiv:2006.15156

\bibitem[{{Ma} {et~al.}(2016){Ma}, {Gonzalez}, {Vieira}, {Aravena}, {Ashby},
  {B{\'e}thermin}, {Bothwell}, {Brandt}, {de Breuck}, {Carlstrom}, {Chapman},
  {Gullberg}, {Hezaveh}, {Litke}, {Malkan}, {Marrone}, {McDonald}, {Murphy},
  {Spilker}, {Sreevani}, {Stark}, {Strandet}, \& {Wang}}]{ma16}
{Ma}, J., {Gonzalez}, A.~H., {Vieira}, J.~D., {et~al.} 2016, \apj, 832, 114

\bibitem[{{Ma} {et~al.}(2019){Ma}, {Hayward}, {Casey}, {Hopkins}, {Quataert},
  {Liang}, {Faucher-Gigu{\`e}re}, {Feldmann}, \& {Kere{\v{s}}}}]{ma19}
{Ma}, X., {Hayward}, C.~C., {Casey}, C.~M., {et~al.} 2019, \mnras, 487, 1844

\bibitem[{{Magdis} {et~al.}(2012){Magdis}, {Daddi}, {B{\'e}thermin}, {Sargent},
  {Elbaz}, {Pannella}, {Dickinson}, {Dannerbauer}, {da Cunha}, {Walter},
  {Rigopoulou}, {Charmandaris}, {Hwang}, \& {Kartaltepe}}]{magdis12}
{Magdis}, G.~E., {Daddi}, E., {B{\'e}thermin}, M., {et~al.} 2012, \apj, 760, 6

\bibitem[{{Magnelli} {et~al.}(2014){Magnelli}, {Lutz}, {Saintonge}, {Berta},
  {Santini}, {Symeonidis}, {Altieri}, {Andreani}, {Aussel}, {B{\'e}thermin},
  {Bock}, {Bongiovanni}, {Cepa}, {Cimatti}, {Conley}, {Daddi}, {Elbaz},
  {F{\"o}rster Schreiber}, {Genzel}, {Ivison}, {Le Floc'h}, {Magdis},
  {Maiolino}, {Nordon}, {Oliver}, {Page}, {P{\'e}rez Garc{\'\i}a}, {Poglitsch},
  {Popesso}, {Pozzi}, {Riguccini}, {Rodighiero}, {Rosario}, {Roseboom},
  {Sanchez-Portal}, {Scott}, {Sturm}, {Tacconi}, {Valtchanov}, {Wang}, \&
  {Wuyts}}]{magnelli14}
{Magnelli}, B., {Lutz}, D., {Saintonge}, A., {et~al.} 2014, \aap, 561, A86

\bibitem[{{Marrone} {et~al.}(2017){Marrone}, {Spilker}, {Hayward}, {Vieira},
  {Strandet}, {Weiss}, {De Breuck}, {Ashby}, {B{\'e}thermin}, {Bothwell},
  {Bradford}, {Carlstrom}, {Chapman}, {Cunningham}, {Chen}, {Fassnacht},
  {Gonzalez}, {Greve}, {Gullberg}, {Hezaveh}, {Litke}, {Ma}, {Malkan},
  {Menten}, {Miller}, {Murphy}, {Narayanan}, {Phadke}, {Rotermund}, , \&
  {Sreevani}}]{marrone17}
{Marrone}, D.~P., {Spilker}, J.~S., {Hayward}, C.~C., {et~al.} 2017, \nat, 842,
  XX

\bibitem[{{Marrone} {et~al.}(2018){Marrone}, {Spilker}, {Hayward}, {Vieira},
  {Aravena}, {Ashby}, {Bayliss}, {B{\'e}thermin}, {Brodwin}, {Bothwell},
  {Carlstrom}, {Chapman}, {Chen}, {Crawford}, {Cunningham}, {De Breuck},
  {Fassnacht}, {Gonzalez}, {Greve}, {Hezaveh}, {Lacaille}, {Litke}, {Lower},
  {Ma}, {Malkan}, {Miller}, {Morningstar}, {Murphy}, {Narayanan}, {Phadke},
  {Rotermund}, {Sreevani}, {Stalder}, {Stark}, {Strandet}, {Tang}, \&
  {Wei{\ss}}}]{marrone18}
---. 2018, \nat, 553, 51

\bibitem[{Marsden {et~al.}(2014)Marsden, Gralla, Marriage, Switzer, Partridge,
  Massardi, Morales, Addison, Bond, Crichton, Das, Devlin, Dünner, Hajian,
  Hilton, Hincks, Hughes, Irwin, Kosowsky, Menanteau, Moodley, Niemack, Page,
  Reese, Schmitt, Sehgal, Sievers, Staggs, Swetz, Thornton, \&
  Wollack}]{marsden14}
Marsden, D., Gralla, M., Marriage, T.~A., {et~al.} 2014, Monthly Notices of the
  Royal Astronomical Society, 439, 1556

\bibitem[{{McMullin} {et~al.}(2007){McMullin}, {Waters}, {Schiebel}, {Young},
  \& {Golap}}]{mcmullin07}
{McMullin}, J.~P., {Waters}, B., {Schiebel}, D., {Young}, W., \& {Golap}, K.
  2007, in Astronomical Society of the Pacific Conference Series, Vol. 376,
  Astronomical Data Analysis Software and Systems XVI, ed. R.~A. {Shaw},
  F.~{Hill}, \& D.~J. {Bell}, 127

\bibitem[{{Messias} {et~al.}(2014){Messias}, {Dye}, {Nagar}, {Orellana},
  {Bussmann}, {Calanog}, {Dannerbauer}, {Fu}, {Ibar}, {Inohara}, {Ivison},
  {Negrello}, {Riechers}, {Sheen}, {Aguirre}, {Amber}, {Birkinshaw}, {Bourne},
  {Bradford}, {Clements}, {Cooray}, {De Zotti}, {Demarco}, {Dunne}, {Eales},
  {Fleuren}, {Kamenetzky}, {Lupu}, {Maddox}, {Marrone}, {Micha{\l}owski},
  {Murphy}, {Nguyen}, {Omont}, {Rowlands}, {Smith}, {Smith}, {Valiante}, \&
  {Vieira}}]{messias14}
{Messias}, H., {Dye}, S., {Nagar}, N., {et~al.} 2014, \aap, 568, A92

\bibitem[{Micha{\l}owski {et~al.}(2017)Micha{\l}owski, Dunlop, Koprowski,
  Cirasuolo, Geach, Bowler, Mortlock, Caputi, Aretxaga, Arumugam, Chen, McLure,
  Birkinshaw, Bourne, Farrah, Ibar, van~der Werf, \& Zemcov}]{michalowski17}
Micha{\l}owski, M.~J., Dunlop, J.~S., Koprowski, M.~P., {et~al.} 2017, Monthly
  Notices of the Royal Astronomical Society, 469, 492

\bibitem[{{Miettinen} {et~al.}(2015){Miettinen}, {Smol{\v c}i{\'c}}, {Novak},
  {Aravena}, {Karim}, {Masters}, {Riechers}, {Bussmann}, {McCracken}, {Ilbert},
  {Bertoldi}, {Capak}, {Feruglio}, {Halliday}, {Kartaltepe}, {Navarrete},
  {Salvato}, {Sanders}, {Schinnerer}, \& {Sheth}}]{miettinen15}
{Miettinen}, O., {Smol{\v c}i{\'c}}, V., {Novak}, M., {et~al.} 2015, \aap, 577,
  A29

\bibitem[{{Miller} {et~al.}(2018){Miller}, {Chapman}, {Aravena}, {Ashby},
  {Hayward}, {Vieira}, {Marrone}, {Spilker}, {Bayliss}, {B{\'e}thermin},
  {Brodwin}, {Bothwell}, {Carlstrom}, {Chen}, {Crawford}, {Cunningham}, {De
  Breuck}, {Fassnacht}, {Gonzalez}, {Greve}, {Hezaveh}, {Lacaille}, {Litke},
  {Lower}, {Ma}, {Malkan}, {Miller}, {Morningstar}, {Murphy}, {Narayanan},
  {Phadke}, {Rotermund}, {Sreevani}, {Stalder}, {Stark}, {Strandet}, {Tang}, \&
  {Wei{\ss}}}]{miller18}
{Miller}, T.~B., {Chapman}, S.~C., {Aravena}, M., {et~al.} 2018, \nat, 000, XX

\bibitem[{{Mocanu} {et~al.}(2013){Mocanu}, {Crawford}, {Vieira}, {Aird},
  {Aravena}, {Austermann}, {Benson}, {B{\'e}thermin}, {Bleem}, {Bothwell},
  {Carlstrom}, {Chang}, {Chapman}, {Cho}, {Crites}, {de Haan}, {Dobbs},
  {Everett}, {George}, {Halverson}, {Harrington}, {Hezaveh}, {Holder},
  {Holzapfel}, {Hoover}, {Hrubes}, {Keisler}, {Knox}, {Lee}, {Leitch},
  {Lueker}, {Luong-Van}, {Marrone}, {McMahon}, {Mehl}, {Meyer}, {Mohr},
  {Montroy}, {Natoli}, {Padin}, {Plagge}, {Pryke}, {Rest}, {Reichardt}, {Ruhl},
  {Sayre}, {Schaffer}, {Shirokoff}, {Spieler}, {Spilker}, {Stalder},
  {Staniszewski}, {Stark}, {Story}, {Switzer}, {Vanderlinde}, \&
  {Williamson}}]{mocanu13}
{Mocanu}, L.~M., {Crawford}, T.~M., {Vieira}, J.~D., {et~al.} 2013, \apj, 779,
  61

\bibitem[{{Mroczkowski} {et~al.}(2019){Mroczkowski}, {De Breuck}, {Kemper},
  {Phillips}, {Fuller}, {Beltr{\'a}n}, {Laing}, {Marconi}, {Testi}, {Yagoubov},
  {George}, \& {McGenn}}]{mroczkowski19}
{Mroczkowski}, T., {De Breuck}, C., {Kemper}, C., {et~al.} 2019, arXiv
  e-prints, arXiv:1905.09064

\bibitem[{{Murphy} {et~al.}(2011){Murphy}, {Condon}, {Schinnerer}, {Kennicutt},
  {Calzetti}, {Armus}, {Helou}, {Turner}, {Aniano}, {Beir{\~a}o}, {Bolatto},
  {Brandl}, {Croxall}, {Dale}, {Donovan Meyer}, {Draine}, {Engelbracht},
  {Hunt}, {Hao}, {Koda}, {Roussel}, {Skibba}, \& {Smith}}]{murphy11}
{Murphy}, E.~J., {Condon}, J.~J., {Schinnerer}, E., {et~al.} 2011, \apj, 737,
  67

\bibitem[{{Narayanan} {et~al.}(2018){Narayanan}, {Dav{\'e}}, {Johnson},
  {Thompson}, {Conroy}, \& {Geach}}]{narayanan18}
{Narayanan}, D., {Dav{\'e}}, R., {Johnson}, B.~D., {et~al.} 2018, \mnras, 474,
  1718

\bibitem[{{Narayanan} {et~al.}(2010){Narayanan}, {Dey}, {Hayward}, {Cox},
  {Bussmann}, {Brodwin}, {Jonsson}, {Hopkins}, {Groves}, {Younger}, \&
  {Hernquist}}]{narayanan10}
{Narayanan}, D., {Dey}, A., {Hayward}, C.~C., {et~al.} 2010, \mnras, 407, 1701

\bibitem[{{Narayanan} {et~al.}(2015){Narayanan}, {Turk}, {Feldmann},
  {Robitaille}, {Hopkins}, {Thompson}, {Hayward}, {Ball},
  {Faucher-Gigu{\`e}re}, \& {Kere{\v s}}}]{narayanan15}
{Narayanan}, D., {Turk}, M., {Feldmann}, R., {et~al.} 2015, \nat, 525, 496

\bibitem[{{Negrello} {et~al.}(2007){Negrello}, {Perrotta},
  {Gonz{\'a}lez-Nuevo}, {Silva}, {De Zotti}, {Granato}, {Baccigalupi}, \&
  {Danese}}]{negrello07}
{Negrello}, M., {Perrotta}, F., {Gonz{\'a}lez-Nuevo}, J., {et~al.} 2007,
  \mnras, 377, 1557

\bibitem[{{Negrello} {et~al.}(2010){Negrello}, {Hopwood}, {De Zotti}, {Cooray},
  {Verma}, {Bock}, {Frayer}, {Gurwell}, {Omont}, {Neri}, {Dannerbauer},
  {Leeuw}, {Barton}, {Cooke}, {Kim}, {da Cunha}, {Rodighiero}, {Cox},
  {Bonfield}, {Jarvis}, {Serjeant}, {Ivison}, {Dye}, {Aretxaga}, {Hughes},
  {Ibar}, {Bertoldi}, {Valtchanov}, {Eales}, {Dunne}, {Driver}, {Auld},
  {Buttiglione}, {Cava}, {Grady}, {Clements}, {Dariush}, {Fritz}, {Hill},
  {Hornbeck}, {Kelvin}, {Lagache}, {Lopez-Caniego}, {Gonzalez-Nuevo}, {Maddox},
  {Pascale}, {Pohlen}, {Rigby}, {Robotham}, {Simpson}, {Smith}, {Temi},
  {Thompson}, {Woodgate}, {York}, {Aguirre}, {Beelen}, {Blain}, {Baker},
  {Birkinshaw}, {Blundell}, {Bradford}, {Burgarella}, {Danese}, {Dunlop},
  {Fleuren}, {Glenn}, {Harris}, {Kamenetzky}, {Lupu}, {Maddalena}, {Madore},
  {Maloney}, {Matsuhara}, {Michaowski}, {Murphy}, {Naylor}, {Nguyen},
  {Popescu}, {Rawlings}, {Rigopoulou}, {Scott}, {Scott}, {Seibert}, {Smail},
  {Tuffs}, {Vieira}, {van der Werf}, \& {Zmuidzinas}}]{negrello10}
{Negrello}, M., {Hopwood}, R., {De Zotti}, G., {et~al.} 2010, Science, 330, 800

\bibitem[{{Neri} {et~al.}(2020){Neri}, {Cox}, {Omont}, {Beelen}, {Berta},
  {Bakx}, {Lehnert}, {Baker}, {Buat}, {Cooray}, {Dannerbauer}, {Dunne}, {Dye},
  {Eales}, {Gavazzi}, {Harris}, {Herrera}, {Hughes}, {Ivison}, {Jin}, {Krips},
  {Lagache}, {Marchetti}, {Messias}, {Negrello}, {Perez-Fournon}, {Riechers},
  {Serjeant}, {Urquhart}, {Vlahakis}, {Wei{\ss}}, {van der Werf}, {Yang}, \&
  {Young}}]{neri20}
{Neri}, R., {Cox}, P., {Omont}, A., {et~al.} 2020, \aap, 635, A7

\bibitem[{{Nguyen} {et~al.}(2010){Nguyen}, {Schulz}, {Levenson}, {Amblard},
  {Arumugam}, {Aussel}, {Babbedge}, {Blain}, {Bock}, {Boselli}, {Buat},
  {Castro-Rodriguez}, {Cava}, {Chanial}, {Chapin}, {Clements}, {Conley},
  {Conversi}, {Cooray}, {Dowell}, {Dwek}, {Eales}, {Elbaz}, {Fox},
  {Franceschini}, {Gear}, {Glenn}, {Griffin}, {Halpern}, {Hatziminaoglou},
  {Ibar}, {Isaak}, {Ivison}, {Lagache}, {Lu}, {Madden}, {Maffei}, {Mainetti},
  {Marchetti}, {Marsden}, {Marshall}, {O'Halloran}, {Oliver}, {Omont}, {Page},
  {Panuzzo}, {Papageorgiou}, {Pearson}, {Perez Fournon}, {Pohlen}, {Rangwala},
  {Rigopoulou}, {Rizzo}, {Roseboom}, {Rowan-Robinson}, {Scott}, {Seymour},
  {Shupe}, {Smith}, {Stevens}, {Symeonidis}, {Trichas}, {Tugwell}, {Vaccari},
  {Valtchanov}, {Vigroux}, {Wang}, {Ward}, {Wiebe}, {Wright}, {Xu}, \&
  {Zemcov}}]{nguyen10}
{Nguyen}, H.~T., {Schulz}, B., {Levenson}, L., {et~al.} 2010, \aap, 518, L5

\bibitem[{{Oliver} {et~al.}(2012){Oliver}, {Bock}, {Altieri}, {Amblard},
  {Arumugam}, {Aussel}, {Babbedge}, {Beelen}, {B{\'e}thermin}, {Blain},
  {Boselli}, {Bridge}, {Brisbin}, {Buat}, {Burgarella},
  {Castro-Rodr{\'\i}guez}, {Cava}, {Chanial}, {Cirasuolo}, {Clements},
  {Conley}, {Conversi}, {Cooray}, {Dowell}, {Dubois}, {Dwek}, {Dye}, {Eales},
  {Elbaz}, {Farrah}, {Feltre}, {Ferrero}, {Fiolet}, {Fox}, {Franceschini},
  {Gear}, {Giovannoli}, {Glenn}, {Gong}, {Gonz{\'a}lez Solares}, {Griffin},
  {Halpern}, {Harwit}, {Hatziminaoglou}, {Heinis}, {Hurley}, {Hwang}, {Hyde},
  {Ibar}, {Ilbert}, {Isaak}, {Ivison}, {Lagache}, {Le Floc'h}, {Levenson},
  {Faro}, {Lu}, {Madden}, {Maffei}, {Magdis}, {Mainetti}, {Marchetti},
  {Marsden}, {Marshall}, {Mortier}, {Nguyen}, {O'Halloran}, {Omont}, {Page},
  {Panuzzo}, {Papageorgiou}, {Patel}, {Pearson}, {P{\'e}rez-Fournon}, {Pohlen},
  {Rawlings}, {Raymond}, {Rigopoulou}, {Riguccini}, {Rizzo}, {Rodighiero},
  {Roseboom}, {Rowan-Robinson}, {S{\'a}nchez Portal}, {Schulz}, {Scott},
  {Seymour}, {Shupe}, {Smith}, {Stevens}, {Symeonidis}, {Trichas}, {Tugwell},
  {Vaccari}, {Valtchanov}, {Vieira}, {Viero}, {Vigroux}, {Wang}, {Ward},
  {Wardlow}, {Wright}, {Xu}, \& {Zemcov}}]{oliver12}
{Oliver}, S.~J., {Bock}, J., {Altieri}, B., {et~al.} 2012, \mnras, 424, 1614

\bibitem[{{Omont} {et~al.}(2011){Omont}, {Neri}, {Cox}, {Lupu}, {Gu{\'e}lin},
  {van der Werf}, {Wei{\ss}}, {Ivison}, {Negrello}, {Leeuw}, {Lehnert},
  {Smail}, {Verma}, {Baker}, {Beelen}, {Aguirre}, {Baes}, {Bertoldi},
  {Clements}, {Cooray}, {Coppin}, {Dannerbauer}, {de Zotti}, {Dye}, {Fiolet},
  {Frayer}, {Gavazzi}, {Hughes}, {Jarvis}, {Krips}, {Micha{\l}owski}, {Murphy},
  {Riechers}, {Serjeant}, {Swinbank}, {Temi}, {Vaccari}, {Vieira}, {Auld},
  {Buttiglione}, {Cava}, {Dariush}, {Dunne}, {Eales}, {Fritz}, {Gomez}, {Ibar},
  {Maddox}, {Pascale}, {Pohlen}, {Rigby}, {Smith}, {Bock}, {Bradford}, {Glenn},
  {Scott}, \& {Zmuidzinas}}]{omont11}
{Omont}, A., {Neri}, R., {Cox}, P., {et~al.} 2011, \aap, 530, L3+

\bibitem[{{Oteo} {et~al.}(2017){Oteo}, {Ivison}, {Negrello}, {Smail},
  {P{\'e}rez-Fournon}, {Bremer}, {De Zotti}, {Eales}, {Farrah}, {Temi},
  {Clements}, {Cooray}, {Dannerbauer}, {Duivenvoorden}, {Dunne}, {Ibar},
  {Lewis}, {Marques-Chaves}, {Mart{\'\i}nez-Navajas}, {Micha{\l}owski},
  {Omont}, {Oliver}, {Riechers}, {Scott}, \& {van der Werf}}]{oteo17}
{Oteo}, I., {Ivison}, R.~J., {Negrello}, M., {et~al.} 2017, arXiv e-prints,
  arXiv:1709.04191

\bibitem[{Oteo {et~al.}(2018)Oteo, Ivison, Dunne, Manilla-Robles, Maddox,
  Lewis, de~Zotti, Bremer, Clements, Cooray, Dannerbauer, Eales, Greenslade,
  Omont, Perez{\textendash}Fourn{\'{o}}n, Riechers, Scott, van~der Werf, Weiss,
  \& Zhang}]{oteo18}
Oteo, I., Ivison, R.~J., Dunne, L., {et~al.} 2018, The Astrophysical Journal,
  856, 72

\bibitem[{{Ott}(2011)}]{ott11}
{Ott}, S. 2011, in Astronomical Society of the Pacific Conference Series, Vol.
  442, Astronomical Data Analysis Software and Systems XX, ed. I.~N. {Evans},
  A.~{Accomazzi}, D.~J. {Mink}, \& A.~H. {Rots}, 347

\bibitem[{{Overzier}(2016)}]{overzier16}
{Overzier}, R.~A. 2016, \aapr, 24, 14

\bibitem[{Pavesi {et~al.}(2018)Pavesi, Riechers, Sharon, Smol{\v{c}}i{\'{c}},
  Faisst, Schinnerer, Carilli, Capak, Scoville, \& Stacey}]{pavesi18}
Pavesi, R., Riechers, D.~A., Sharon, C.~E., {et~al.} 2018, The Astrophysical
  Journal, 861, 43

\bibitem[{{Petry} {et~al.}(2012)}]{petry12}
{Petry}, D., {et~al.} 2012, ArXiv e-prints, arXiv:1201.3454

\bibitem[{{Planck Collaboration} {et~al.}(2015){Planck Collaboration},
  {Aghanim}, {Altieri}, {Arnaud}, {Ashdown}, {Aumont}, {Baccigalupi}, {Banday},
  {Barreiro}, {Bartolo}, \& et~al.}]{planck15}
{Planck Collaboration}, {Aghanim}, N., {Altieri}, B., {et~al.} 2015, \aap, 582,
  A30

\bibitem[{{Planck Collaboration} {et~al.}(2016){Planck Collaboration}, {Ade},
  {Aghanim}, {Arnaud}, {Ashdown}, {Aumont}, {Baccigalupi}, {Banday},
  {Barreiro}, {Bartlett}, \& et~al.}]{planck16cosmo}
{Planck Collaboration}, {Ade}, P.~A.~R., {Aghanim}, N., {et~al.} 2016, \aap,
  594, A13

\bibitem[{{Rawle} {et~al.}(2014){Rawle}, {Egami}, {Bussmann}, {Gurwell},
  {Ivison}, {Boone}, {Combes}, {Danielson}, {Rex}, {Richard}, {Smail},
  {Swinbank}, {Altieri}, {Blain}, {Clement}, {Dessauges-Zavadsky}, {Edge},
  {Fazio}, {Jones}, {Kneib}, {Omont}, {P{\'e}rez-Gonz{\'a}lez}, {Schaerer},
  {Valtchanov}, {van der Werf}, {Walth}, {Zamojski}, \& {Zemcov}}]{rawle14}
{Rawle}, T.~D., {Egami}, E., {Bussmann}, R.~S., {et~al.} 2014, \apj, 783, 59

\bibitem[{{Riechers} {et~al.}(2013){Riechers}, {Bradford}, {Clements},
  {Dowell}, {P{\'e}rez-Fournon}, {Ivison}, {Bridge}, {Conley}, {Fu}, {Vieira},
  {Wardlow}, {Calanog}, {Cooray}, {Hurley}, {Neri}, {Kamenetzky}, {Aguirre},
  {Altieri}, {Arumugam}, {Benford}, {B{\'e}thermin}, {Bock}, {Burgarella},
  {Cabrera-Lavers}, {Chapman}, {Cox}, {Dunlop}, {Earle}, {Farrah}, {Ferrero},
  {Franceschini}, {Gavazzi}, {Glenn}, {Solares}, {Gurwell}, {Halpern},
  {Hatziminaoglou}, {Hyde}, {Ibar}, {Kov{\'a}cs}, {Krips}, {Lupu}, {Maloney},
  {Martinez-Navajas}, {Matsuhara}, {Murphy}, {Naylor}, {Nguyen}, {Oliver},
  {Omont}, {Page}, {Petitpas}, {Rangwala}, {Roseboom}, {Scott}, {Smith},
  {Staguhn}, {Streblyanska}, {Thomson}, {Valtchanov}, {Viero}, {Wang},
  {Zemcov}, \& {Zmuidzinas}}]{riechers13}
{Riechers}, D.~A., {Bradford}, C.~M., {Clements}, D.~L., {et~al.} 2013, \nat,
  496, 329

\bibitem[{{Riechers} {et~al.}(2014){Riechers}, {Carilli}, {Capak}, {Scoville},
  {Smolcic}, {Schinnerer}, {Yun}, {Cox}, {Bertoldi}, {Karim}, \&
  {Yan}}]{riechers14}
{Riechers}, D.~A., {Carilli}, C.~L., {Capak}, P.~L., {et~al.} 2014, ArXiv
  e-prints, arXiv:1404.7159

\bibitem[{Riechers {et~al.}(2017)Riechers, Leung, Ivison, P{\'{e}}rez-Fournon,
  Lewis, Marques-Chaves, Oteo, Clements, Cooray, Greenslade,
  Mart{\'{\i}}nez-Navajas, Oliver, Rigopoulou, Scott, \& Weiss}]{riechers17}
Riechers, D.~A., Leung, T. K.~D., Ivison, R.~J., {et~al.} 2017, The
  Astrophysical Journal, 850, 1

\bibitem[{{Riechers} {et~al.}(2020){Riechers}, {Hodge}, {Pavesi}, {Daddi},
  {Decarli}, {Ivison}, {Sharon}, {Smail}, {Walter}, {Aravena}, {Capak},
  {Carilli}, {Cox}, {da Cunha}, {Dannerbauer}, {Dickinson}, {Neri}, \&
  {Wagg}}]{riechers20}
{Riechers}, D.~A., {Hodge}, J.~A., {Pavesi}, R., {et~al.} 2020, arXiv e-prints,
  arXiv:2004.10204

\bibitem[{{Saintonge} {et~al.}(2013){Saintonge}, {Lutz}, {Genzel}, {Magnelli},
  {Nordon}, {Tacconi}, {Baker}, {Bandara}, {Berta}, {F{\"o}rster Schreiber},
  {Poglitsch}, {Sturm}, {Wuyts}, \& {Wuyts}}]{saintonge13}
{Saintonge}, A., {Lutz}, D., {Genzel}, R., {et~al.} 2013, \apj, 778, 2

\bibitem[{{Sault} {et~al.}(1995){Sault}, {Teuben}, \& {Wright}}]{sault95}
{Sault}, R.~J., {Teuben}, P.~J., \& {Wright}, M.~C.~H. 1995, in Astronomical
  Society of the Pacific Conference Series, Vol.~77, Astronomical Data Analysis
  Software and Systems IV, ed. R.~A. {Shaw}, H.~E. {Payne}, \& J.~J.~E.
  {Hayes}, 433

\bibitem[{{Schreiber} {et~al.}(2018){Schreiber}, {Elbaz}, {Pannella}, {Ciesla},
  {Wang}, \& {Franco}}]{schreiber18}
{Schreiber}, C., {Elbaz}, D., {Pannella}, M., {et~al.} 2018, \aap, 609, A30

\bibitem[{{Schuller} {et~al.}(2010){Schuller}, {Nord}, {Vlahakis}, {Albrecht},
  {Beelen}, {Bertoldi}, {Mueller}, \& {Schaaf}}]{schuller10}
{Schuller}, F., {Nord}, M., {Vlahakis}, C., {et~al.} 2010, BoA -- The Bolometer
  Data Analysis Software. User and Reference Manual
  (http://www.mpifr-bonn.mpg.de/div/submmtech/software/boa /boa\_main.html)

\bibitem[{{Scott} {et~al.}(2011){Scott}, {Lupu}, {Aguirre}, {Auld}, {Aussel},
  {Baker}, {Beelen}, {Bock}, {Bradford}, {Brisbin}, {Burgarella}, {Carpenter},
  {Chanial}, {Chapman}, {Clements}, {Conley}, {Cooray}, {Cox}, {Dowell},
  {Eales}, {Farrah}, {Franceschini}, {Frayer}, {Gavazzi}, {Glenn}, {Griffin},
  {Harris}, {Ibar}, {Ivison}, {Kamenetzky}, {Kim}, {Krips}, {Maloney},
  {Matsuhara}, {Mortier}, {Murphy}, {Naylor}, {Neri}, {Nguyen}, {Oliver},
  {Omont}, {Page}, {Papageorgiou}, {Pearson}, {P{\'e}rez-Fournon}, {Pohlen},
  {Rawlings}, {Raymond}, {Riechers}, {Rodighiero}, {Roseboom},
  {Rowan-Robinson}, {Scott}, {Seymour}, {Smith}, {Symeonidis}, {Tugwell},
  {Vaccari}, {Vieira}, {Vigroux}, {Wang}, {Wright}, \& {Zmuidzinas}}]{scott11}
{Scott}, K.~S., {Lupu}, R.~E., {Aguirre}, J.~E., {et~al.} 2011, \apj, 733, 29

\bibitem[{{Silva} {et~al.}(1998){Silva}, {Granato}, {Bressan}, \&
  {Danese}}]{silva98}
{Silva}, L., {Granato}, G.~L., {Bressan}, A., \& {Danese}, L. 1998, \apj, 509,
  103

\bibitem[{{Simpson} {et~al.}(2014){Simpson}, {Swinbank}, {Smail}, {Alexander},
  {Brandt}, {Bertoldi}, {de Breuck}, {Chapman}, {Coppin}, {da Cunha},
  {Danielson}, {Dannerbauer}, {Greve}, {Hodge}, {Ivison}, {Karim}, {Knudsen},
  {Poggianti}, {Schinnerer}, {Thomson}, {Walter}, {Wardlow}, {Wei{\ss}}, \&
  {van der Werf}}]{simpson14}
{Simpson}, J.~M., {Swinbank}, A.~M., {Smail}, I., {et~al.} 2014, \apj, 788, 125

\bibitem[{{Simpson} {et~al.}(2015){Simpson}, {Smail}, {Swinbank}, {Chapman},
  {Geach}, {Ivison}, {Thomson}, {Aretxaga}, {Blain}, {Cowley}, {Chen},
  {Coppin}, {Dunlop}, {Edge}, {Farrah}, {Ibar}, {Karim}, {Knudsen},
  {Meijerink}, {Micha{\l}owski}, {Scott}, {Spaans}, \& {van der
  Werf}}]{simpson15}
{Simpson}, J.~M., {Smail}, I., {Swinbank}, A.~M., {et~al.} 2015, \apj, 807, 128

\bibitem[{Simpson {et~al.}(2017)Simpson, Smail, Swinbank, Ivison, Dunlop,
  Geach, Almaini, Arumugam, Bremer, Chen, \& et~al.}]{simpson17}
Simpson, J.~M., Smail, I., Swinbank, A.~M., {et~al.} 2017, The Astrophysical
  Journal, 839, 58

\bibitem[{{Siringo} {et~al.}(2009){Siringo}, {Kreysa}, {Kov{\'a}cs},
  {Schuller}, {Wei{\ss}}, {Esch}, {Gem{\"u}nd}, {Jethava}, {Lundershausen},
  {Colin}, {G{\"u}sten}, {Menten}, {Beelen}, {Bertoldi}, {Beeman}, \&
  {Haller}}]{siringo09}
{Siringo}, G., {Kreysa}, E., {Kov{\'a}cs}, A., {et~al.} 2009, \aap, 497, 945

\bibitem[{{Spilker} {et~al.}(2014){Spilker}, {Marrone}, {Aguirre}, {Aravena},
  {Ashby}, {B{\'e}thermin}, {Bradford}, {Bothwell}, {Brodwin}, {Carlstrom},
  {Chapman}, {Crawford}, {de Breuck}, {Fassnacht}, {Gonzalez}, {Greve},
  {Gullberg}, {Hezaveh}, {Holzapfel}, {Husband}, {Ma}, {Malkan}, {Murphy},
  {Reichardt}, {Rotermund}, {Stalder}, {Stark}, {Strandet}, {Vieira},
  {Wei{\ss}}, \& {Welikala}}]{spilker14}
{Spilker}, J.~S., {Marrone}, D.~P., {Aguirre}, J.~E., {et~al.} 2014, \apj, 785,
  149

\bibitem[{{Spilker} {et~al.}(2015){Spilker}, {Aravena}, {Marrone},
  {B{\'e}thermin}, {Bothwell}, {Carlstrom}, {Chapman}, {Collier}, {de Breuck},
  {Fassnacht}, {Galvin}, {Gonzalez}, {Gonz{\'a}lez-L{\'o}pez}, {Grieve},
  {Hezaveh}, {Ma}, {Malkan}, {O'Brien}, {Rotermund}, {Strandet}, {Vieira},
  {Weiss}, \& {Wong}}]{spilker15}
{Spilker}, J.~S., {Aravena}, M., {Marrone}, D.~P., {et~al.} 2015, \apj, 811,
  124

\bibitem[{{Spilker} {et~al.}(2016){Spilker}, {Marrone}, {Aravena},
  {B{\'e}thermin}, {Bothwell}, {Carlstrom}, {Chapman}, {Crawford}, {de Breuck},
  {Fassnacht}, {Gonzalez}, {Greve}, {Hezaveh}, {Litke}, {Ma}, {Malkan},
  {Rotermund}, {Strandet}, {Vieira}, {Weiss}, \& {Welikala}}]{spilker16}
{Spilker}, J.~S., {Marrone}, D.~P., {Aravena}, M., {et~al.} 2016, \apj, 826,
  112

\bibitem[{{Spilker} {et~al.}(2018){Spilker}, {Aravena}, {B{\'e}thermin},
  {Chapman}, {Chen}, {Cunningham}, {De Breuck}, {Dong}, {Gonzalez}, {Hayward},
  {Hezaveh}, {Litke}, {Ma}, {Malkan}, {Marrone}, {Miller}, {Morningstar},
  {Narayanan}, {Phadke}, {Sreevani}, {Stark}, {Vieira}, \&
  {Wei{\ss}}}]{spilker18}
{Spilker}, J.~S., {Aravena}, M., {B{\'e}thermin}, M., {et~al.} 2018, Science,
  361, 1016

\bibitem[{{Staguhn} {et~al.}(2014){Staguhn}, {Kov{\'a}cs}, {Arendt}, {Benford},
  {Decarli}, {Dwek}, {Fixsen}, {Hilton}, {Irwin}, {Jhabvala}, {Karim},
  {Leclercq}, {Maher}, {Miller}, {Moseley}, {Sharp}, {Walter}, \&
  {Wollack}}]{staguhn14}
{Staguhn}, J.~G., {Kov{\'a}cs}, A., {Arendt}, R.~G., {et~al.} 2014, \apj, 790,
  77

\bibitem[{{Strandet} {et~al.}(2016){Strandet}, {Weiss}, {Vieira}, {de Breuck},
  {Aguirre}, {Aravena}, {Ashby}, {B{\'e}thermin}, {Bradford}, {Carlstrom},
  {Chapman}, {Crawford}, {Everett}, {Fassnacht}, {Furstenau}, {Gonzalez},
  {Greve}, {Gullberg}, {Hezaveh}, {Kamenetzky}, {Litke}, {Ma}, {Malkan},
  {Marrone}, {Menten}, {Murphy}, {Nadolski}, {Rotermund}, {Spilker}, {Stark},
  \& {Welikala}}]{strandet16}
{Strandet}, M.~L., {Weiss}, A., {Vieira}, J.~D., {et~al.} 2016, \apj, 822, 80

\bibitem[{{Strandet} {et~al.}(2017){Strandet}, {Weiss}, {De Breuck}, {Marrone},
  {Vieira}, {Aravena}, {Ashby}, {B{\'e}thermin}, {Bothwell}, {Bradford},
  {Carlstrom}, {Chapman}, {Cunningham}, {Chen}, {Fassnacht}, {Gonzalez},
  {Greve}, {Gullberg}, {Hayward}, {Hezaveh}, {Litke}, {Ma}, {Malkan}, {Menten},
  {Miller}, {Murphy}, {Narayanan}, {Phadke}, {Rotermund}, {Spilker}, \&
  {Sreevani}}]{strandet17}
{Strandet}, M.~L., {Weiss}, A., {De Breuck}, C., {et~al.} 2017, \apjl, 842, L15

\bibitem[{{Swinbank} {et~al.}(2014){Swinbank}, {Simpson}, {Smail}, {Harrison},
  {Hodge}, {Karim}, {Walter}, {Alexander}, {Brandt}, {de Breuck}, {da Cunha},
  {Chapman}, {Coppin}, {Danielson}, {Dannerbauer}, {Decarli}, {Greve},
  {Ivison}, {Knudsen}, {Lagos}, {Schinnerer}, {Thomson}, {Wardlow}, {Wei{\ss}},
  \& {van der Werf}}]{swinbank14}
{Swinbank}, A.~M., {Simpson}, J.~M., {Smail}, I., {et~al.} 2014, \mnras, 438,
  1267

\bibitem[{{Tacconi} {et~al.}(2008){Tacconi}, {Genzel}, {Smail}, {Neri},
  {Chapman}, {Ivison}, {Blain}, {Cox}, {Omont}, {Bertoldi}, {Greve},
  {F{\"o}rster Schreiber}, {Genel}, {Lutz}, {Swinbank}, {Shapley}, {Erb},
  {Cimatti}, {Daddi}, \& {Baker}}]{tacconi08}
{Tacconi}, L.~J., {Genzel}, R., {Smail}, I., {et~al.} 2008, \apj, 680, 246

\bibitem[{{Tacconi} {et~al.}(2018){Tacconi}, {Genzel}, {Saintonge}, {Combes},
  {Garc{\'\i}a-Burillo}, {Neri}, {Bolatto}, {Contini}, {F{\"o}rster Schreiber},
  {Lilly}, {Lutz}, {Wuyts}, {Accurso}, {Boissier}, {Boone}, {Bouch{\'e}},
  {Bournaud}, {Burkert}, {Carollo}, {Cooper}, {Cox}, {Feruglio}, {Freundlich},
  {Herrera-Camus}, {Juneau}, {Lippa}, {Naab}, {Renzini}, {Salome}, {Sternberg},
  {Tadaki}, {{\"U}bler}, {Walter}, {Weiner}, \& {Weiss}}]{tacconi18}
{Tacconi}, L.~J., {Genzel}, R., {Saintonge}, A., {et~al.} 2018, \apj, 853, 179

\bibitem[{{Vieira} {et~al.}(2010){Vieira}, {Crawford}, {Switzer}, {Ade},
  {Aird}, {Ashby}, {Benson}, {Bleem}, {Brodwin}, {Carlstrom}, {Chang}, {Cho},
  {Crites}, {de Haan}, {Dobbs}, {Everett}, {George}, {Gladders}, {Hall},
  {Halverson}, {High}, {Holder}, {Holzapfel}, {Hrubes}, {Joy}, {Keisler},
  {Knox}, {Lee}, {Leitch}, {Lueker}, {Marrone}, {McIntyre}, {McMahon}, {Mehl},
  {Meyer}, {Mohr}, {Montroy}, {Padin}, {Plagge}, {Pryke}, {Reichardt}, {Ruhl},
  {Schaffer}, {Shaw}, {Shirokoff}, {Spieler}, {Stalder}, {Staniszewski},
  {Stark}, {Vanderlinde}, {Walsh}, {Williamson}, {Yang}, {Zahn}, \&
  {Zenteno}}]{vieira10}
{Vieira}, J.~D., {Crawford}, T.~M., {Switzer}, E.~R., {et~al.} 2010, \apj, 719,
  763

\bibitem[{{Vieira} {et~al.}(2013){Vieira}, {Marrone}, {Chapman}, {De Breuck},
  {Hezaveh}, {Wei{$\beta$}}, {Aguirre}, {Aird}, {Aravena}, {Ashby}, {Bayliss},
  {Benson}, {Biggs}, {Bleem}, {Bock}, {Bothwell}, {Bradford}, {Brodwin},
  {Carlstrom}, {Chang}, {Crawford}, {Crites}, {de Haan}, {Dobbs}, {Fomalont},
  {Fassnacht}, {George}, {Gladders}, {Gonzalez}, {Greve}, {Gullberg},
  {Halverson}, {High}, {Holder}, {Holzapfel}, {Hoover}, {Hrubes}, {Hunter},
  {Keisler}, {Lee}, {Leitch}, {Lueker}, {Luong-van}, {Malkan}, {McIntyre},
  {McMahon}, {Mehl}, {Menten}, {Meyer}, {Mocanu}, {Murphy}, {Natoli}, {Padin},
  {Plagge}, {Reichardt}, {Rest}, {Ruel}, {Ruhl}, {Sharon}, {Schaffer}, {Shaw},
  {Shirokoff}, {Spilker}, {Stalder}, {Staniszewski}, {Stark}, {Story},
  {Vanderlinde}, {Welikala}, \& {Williamson}}]{vieira13}
{Vieira}, J.~D., {Marrone}, D.~P., {Chapman}, S.~C., {et~al.} 2013, \nat, 495,
  344

\bibitem[{{Walter} {et~al.}(2012){Walter}, {Decarli}, {Carilli}, {Bertoldi},
  {Cox}, {da Cunha}, {Daddi}, {Dickinson}, {Downes}, {Elbaz}, {Ellis}, {Hodge},
  {Neri}, {Riechers}, {Weiss}, {Bell}, {Dannerbauer}, {Krips}, {Krumholz},
  {Lentati}, {Maiolino}, {Menten}, {Rix}, {Robertson}, {Spinrad}, {Stark}, \&
  {Stern}}]{walter12}
{Walter}, F., {Decarli}, R., {Carilli}, C., {et~al.} 2012, \nat, 486, 233

\bibitem[{{Wardlow} {et~al.}(2012){Wardlow}, {Cooray}, {De Bernardis},
  {Amblard}, {Arumugam}, {Aussel}, {Baker}, {B{\'e}thermin}, {Blundell},
  {Bock}, {Boselli}, {Bridge}, {Buat}, {Burgarella}, {Bussmann}, {Calanog},
  {Carpenter}, {Casey}, {Castro-Rodr{\'{\i}}guez}, {Cava}, {Chanial},
  {Chapman}, {Clements}, {Conley}, {Cox}, {Dowell}, {Dye}, {Eales}, {Farrah},
  {Franceschini}, {Frayer}, {Frazer}, {Fu}, {Gavazzi}, {Glenn}, {Griffin},
  {Gurwell}, {Harris}, {Hatziminaoglou}, {Hopwood}, {Ibar}, {Ivison}, {Kim},
  {Lagache}, {Levenson}, {Marchetti}, {Marsden}, {Negrello}, {Neri}, {Nguyen},
  {O'Halloran}, {Oliver}, {Omont}, {Page}, {Panuzzo}, {Papageorgiou},
  {Pearson}, {P{\'e}rez-Fournon}, {Pohlen}, {Riechers}, {Rigopoulou},
  {Roseboom}, {Rowan-Robinson}, {Schulz}, {Scott}, {Scoville}, {Seymour},
  {Shupe}, {Smith}, {Symeonidis}, {Trichas}, {Vaccari}, {Vieira}, {Viero},
  {Wang}, {Xu}, {Yan}, \& {Zemcov}}]{wardlow12}
{Wardlow}, J.~L., {Cooray}, A., {De Bernardis}, F., {et~al.} 2012, ArXiv
  e-prints, arXiv:1205.3778

\bibitem[{{Wardlow} {et~al.}(2013){Wardlow}, {Cooray}, {De Bernardis},
  {Amblard}, {Arumugam}, {Aussel}, {Baker}, {B{\'e}thermin}, {Blundell},
  {Bock}, {Boselli}, {Bridge}, {Buat}, {Burgarella}, {Bussmann},
  {Cabrera-Lavers}, {Calanog}, {Carpenter}, {Casey}, {Castro-Rodr{\'{\i}}guez},
  {Cava}, {Chanial}, {Chapin}, {Chapman}, {Clements}, {Conley}, {Cox},
  {Dowell}, {Dye}, {Eales}, {Farrah}, {Ferrero}, {Franceschini}, {Frayer},
  {Frazer}, {Fu}, {Gavazzi}, {Glenn}, {Gonz{\'a}lez Solares}, {Griffin},
  {Gurwell}, {Harris}, {Hatziminaoglou}, {Hopwood}, {Hyde}, {Ibar}, {Ivison},
  {Kim}, {Lagache}, {Levenson}, {Marchetti}, {Marsden}, {Martinez-Navajas},
  {Negrello}, {Neri}, {Nguyen}, {O'Halloran}, {Oliver}, {Omont}, {Page},
  {Panuzzo}, {Papageorgiou}, {Pearson}, {P{\'e}rez-Fournon}, {Pohlen},
  {Riechers}, {Rigopoulou}, {Roseboom}, {Rowan-Robinson}, {Schulz}, {Scott},
  {Scoville}, {Seymour}, {Shupe}, {Smith}, {Streblyanska}, {Strom},
  {Symeonidis}, {Trichas}, {Vaccari}, {Vieira}, {Viero}, {Wang}, {Xu}, {Yan},
  \& {Zemcov}}]{wardlow13}
---. 2013, \apj, 762, 59

\bibitem[{{Weingartner} \& {Draine}(2001)}]{weingartner01}
{Weingartner}, J.~C., \& {Draine}, B.~T. 2001, \apj, 548, 296

\bibitem[{{Wei{\ss}} {et~al.}(2009){Wei{\ss}}, {Kov{\'a}cs}, {Coppin}, {Greve},
  {Walter}, {Smail}, {Dunlop}, {Knudsen}, {Alexander}, {Bertoldi}, {Brandt},
  {Chapman}, {Cox}, {Dannerbauer}, {De Breuck}, {Gawiser}, {Ivison}, {Lutz},
  {Menten}, {Koekemoer}, {Kreysa}, {Kurczynski}, {Rix}, {Schinnerer}, \& {van
  der Werf}}]{weiss09}
{Wei{\ss}}, A., {Kov{\'a}cs}, A., {Coppin}, K., {et~al.} 2009, \apj, 707, 1201

\bibitem[{{Wei{\ss}} {et~al.}(2013){Wei{\ss}}, {De Breuck}, {Marrone},
  {Vieira}, {Aguirre}, {Aird}, {Aravena}, {Ashby}, {Bayliss}, {Benson},
  {B{\'e}thermin}, {Biggs}, {Bleem}, {Bock}, {Bothwell}, {Bradford}, {Brodwin},
  {Carlstrom}, {Chang}, {Chapman}, {Crawford}, {Crites}, {de Haan}, {Dobbs},
  {Downes}, {Fassnacht}, {George}, {Gladders}, {Gonzalez}, {Greve},
  {Halverson}, {Hezaveh}, {High}, {Holder}, {Holzapfel}, {Hoover}, {Hrubes},
  {Husband}, {Keisler}, {Lee}, {Leitch}, {Lueker}, {Luong-Van}, {Malkan},
  {McIntyre}, {McMahon}, {Mehl}, {Menten}, {Meyer}, {Murphy}, {Padin},
  {Plagge}, {Reichardt}, {Rest}, {Rosenman}, {Ruel}, {Ruhl}, {Schaffer},
  {Shirokoff}, {Spilker}, {Stalder}, {Staniszewski}, {Stark}, {Story},
  {Vanderlinde}, {Welikala}, \& {Williamson}}]{weiss13}
{Wei{\ss}}, A., {De Breuck}, C., {Marrone}, D.~P., {et~al.} 2013, \apj, 767, 88

\bibitem[{{Wilson} {et~al.}(2011){Wilson}, {Ferris}, {Axtens}, {Brown},
  {Davis}, {Hampson}, {Leach}, {Roberts}, {Saunders}, {Koribalski}, {Caswell},
  {Lenc}, {Stevens}, {Voronkov}, {Wieringa}, {Brooks}, {Edwards}, {Ekers},
  {Emonts}, {Hindson}, {Johnston}, {Maddison}, {Mahony}, {Malu}, {Massardi},
  {Mao}, {McConnell}, {Norris}, {Schnitzeler}, {Subrahmanyan}, {Urquhart},
  {Thompson}, \& {Wark}}]{wilson11}
{Wilson}, W.~E., {Ferris}, R.~H., {Axtens}, P., {et~al.} 2011, \mnras, 416, 832

\bibitem[{{Zavala} {et~al.}(2018){Zavala}, {Monta{\~n}a}, {Hughes}, {Yun},
  {Ivison}, {Valiante}, {Wilner}, {Spilker}, {Aretxaga}, {Eales},
  {Avila-Reese}, {Ch{\'a}vez}, {Cooray}, {Dannerbauer}, {Dunlop}, {Dunne},
  {G{\'o}mez-Ruiz}, {Micha{\l}owski}, {Narayanan}, {Nayyeri}, {Oteo}, {Rosa
  Gonz{\'a}lez}, {S{\'a}nchez-Arg{\"u}elles}, {Schloerb}, {Serjeant}, {Smith},
  {Terlevich}, {Vega}, {Villalba}, {van der Werf}, {Wilson}, \&
  {Zeballos}}]{zavala18}
{Zavala}, J.~A., {Monta{\~n}a}, A., {Hughes}, D.~H., {et~al.} 2018, Nature
  Astronomy, 2, 56

\bibitem[{{Zhang} {et~al.}(2018){Zhang}, {Romano}, {Ivison}, {Papadopoulos}, \&
  {Matteucci}}]{zhang18}
{Zhang}, Z.-Y., {Romano}, D., {Ivison}, R.~J., {Papadopoulos}, P.~P., \&
  {Matteucci}, F. 2018, \nat, 558, 260

\end{thebibliography}

\end{document}